\PassOptionsToPackage{numbers,sort&compress}{natbib}
\documentclass[preprint,11pt,a4paper,number]{elsarticle}
\usepackage{iftex}
\usepackage{lmodern}
\usepackage[a4paper,left=1.7cm,right=1.7cm,top=1.7cm,bottom=1.75cm,footskip=0.85cm]{geometry}
\usepackage{amsmath,amssymb,mathtools,bm}
\ifPDFTeX
  \usepackage[T1]{fontenc}
  \IfFileExists{newtxtext.sty}{\usepackage{newtxtext}}{\usepackage{mathptmx}}
  \IfFileExists{newtxmath.sty}{\usepackage{newtxmath}}{}
\else
  \usepackage{fontspec}
  \IfFileExists{newtxmath.sty}{\usepackage{newtxmath}}{}
  \IfFontExistsTF{Times New Roman}{\setmainfont{Times New Roman}}{%
    \IfFontExistsTF{Tinos}{\setmainfont{Tinos}}{%
      \IfFontExistsTF{TeX Gyre Termes}{\setmainfont{TeX Gyre Termes}}{\setmainfont{Latin Modern Roman}}}}
\fi
\usepackage{microtype,titlesec}
\usepackage{caption}
\usepackage[hidelinks]{hyperref}
\hypersetup{colorlinks=false,pdfborder={0 0 0},pdftitle={Operational Characterizations and Resource Structure of the Interactive Weight of Quantum Instruments},pdfsubject={Interactive quantum instruments}}
\newcommand{\doi}[1]{\href{https://doi.org/#1}{\nolinkurl{https://doi.org/#1}}}
\allowdisplaybreaks[2]
\renewcommand{\thesection}{\arabic{section}}
\renewcommand{\thesubsection}{\thesection.\arabic{subsection}}
\titleformat{\section}{\centering\bfseries\normalsize}{\thesection.}{0.45em}{}
\titleformat{\subsection}{\bfseries\normalsize}{\thesubsection}{0.45em}{}
\titlespacing{\section}{0pt}{0.85em}{0.50em}
\titlespacing{\subsection}{0pt}{0.65em}{0.32em}

\newcommand{\ManuscriptFigure}[2]{\makebox[\linewidth][c]{\includegraphics[width=#2,keepaspectratio]{#1}}}
\begin{document}
\begin{frontmatter}
\title{Operational Characterizations and Resource Structure of the Interactive Weight of Quantum Instruments}
\author[a]{Yongxi He}
\author[b]{Xin Hong}
\author[c]{Yu Luo\corref{cor1}}
\author[a,d]{Xiangzhen Zhou\corref{cor1}}
\author[b]{Shenggang Ying}
\cortext[cor1]{Corresponding authors.}
\address[a]{Arclight Quantum Computing Inc., Beijing, China.}
\address[b]{Key Laboratory of System Software (Chinese Academy of Sciences), Institute of Software, Chinese Academy of Sciences, Beijing, China.}
\address[c]{School of Artificial Intelligence and Computer Science, Shaanxi Normal University, No. 199 Chang'an South Road, Yanta District, Xi'an 710062, China.}
\address[d]{College of Artificial Intelligence, Nanjing Tech University, No. 30 Puzhu South Road, Pukou District, Nanjing 211800, China.}
\begin{abstract}
Quantifying the interactivity of quantum instruments is fundamental to understanding their resource structure and operational capabilities. Here, we introduce the interactive weight of quantum instruments as a resource quantifier that measures the minimum fraction of an instrument that cannot be explained by non-interactive components. Through quantum-information tasks including quantum-state recovery, positive linear loss games, and environment-assisted input-blind extraction, we establish quantitative bounds on the interactive weight of quantum instruments and exact operational characterizations of its complementary quantity, the maximal non-interactive content. In particular, we extend environment-assisted input-blind extraction to the setting of restricted environmental access and develop a corresponding framework. Within the allowed-operations framework introduced by Hsieh \textit{et al.}, we establish fundamental resource-theoretic properties of the interactive weight of quantum instruments, including faithfulness, convexity, and monotonicity under deterministic allowed operations. Furthermore, we prove that the maximal non-interactive content satisfies exact multiplicativity under independent parallel composition, yielding necessary constraints on multicopy resource conversion and establishing the finite-copy maximal-weight distillation no-go result under deterministic allowed operations. These results establish a unified theoretical framework for the interactive weight of quantum instruments from the perspectives of resource quantification, operational interpretation, and resource conversion, providing systematic theoretical tools for the quantitative comparison and conversion of interactive quantum instruments.
\end{abstract}
\end{frontmatter}

\section{INTRODUCTION}
Quantum measurement is a fundamental operation for extracting classical information from quantum systems. It also plays a central role in information-processing tasks such as quantum-state discrimination~\cite{barnett2009,conlon2023,hirche2017detector}, parameter estimation~\cite{paris2009,liu2024}, and quantum control~\cite{zhang2017}. As the statistical description of a quantum measurement, a positive operator-valued measure (POVM) specifies the probability distribution of measurement outcomes for a given input state~\cite{zhu2022measurements}. Moreover, when considered at the level of measurement statistics, quantum measurements can also exhibit a variety of nontrivial features, including the inability to be simulated by projective measurements~\cite{oszmaniec2017,brinster2026nonprojective}, measurement incompatibility~\cite{guhne2023,skrzypczyk2019,tendick2023distributed,porto2026steering,tendick2023distance,tendick2025correlations}, and coherence properties of quantum measurements~\cite{baek2020,cimini2019}. All these features can give rise to quantifiable operational advantages in quantum-information tasks~\cite{oszmaniec2019,skrzypczyk2019robustness,plavala2025nonlocality}.

However, in a physical quantum measurement process, measurement does not merely produce a classical outcome but is also typically accompanied by a change in the state of the measured quantum system~\cite{stricker2022,lee2021balance}. The post-measurement quantum state is equally important, as it may affect subsequent information extraction, conditional control, and quantum-state recovery~\cite{leppajarvi2021,han2018residual}. A POVM primarily specifies the probabilities of different measurement outcomes. For measurement processes that yield identical measurement statistics but different post-measurement states, a POVM does not provide a complete mathematical description~\cite{leppajarvi2021}. Therefore, when both the classical measurement outcome and the corresponding quantum output need to be described, a more general mathematical framework is required, namely a quantum instrument~\cite{mitra2023instruments,ji2024programmable,lee2026functionals}. A quantum instrument specifies not only the probability of each measurement outcome but also the state transformation conditioned on that outcome~\cite{dressel2013}. Therefore, it provides a more complete description of how quantum information is acquired, preserved, and transmitted during a measurement process~\cite{davies1970,ozawa1984,khandelwal2025}.

The coexistence of classical outcomes and quantum outputs gives quantum instruments a richer information structure than POVMs. This naturally raises a fundamental question: to what extent do the outputs of a quantum instrument genuinely depend on the input quantum state? If both the classical and quantum outputs can be generated independently of the input, the measurement process does not make use of the information encoded in the input state. Genuine interactivity arises only when the generated outputs depend nontrivially on the input. Quantifying this input-output interaction is therefore essential for comparing the resource properties of quantum instruments and for understanding their information-processing capabilities.

Recently, Hsieh \textit{et al.} developed a resource theory of interactive quantum instruments based on this perspective~\cite{hsieh2026}. Instruments whose classical and quantum outputs are both independent of the input state were identified as non-interactive instruments, thereby defining the corresponding free set and a natural class of allowed operations. Interactive instrument robustness was further introduced to quantify interactivity from a noise-mixing perspective. This quantity was related to quantum-information recovery tasks and to resource conversion between instruments. Meanwhile, weight-based resource quantifiers have also been investigated in multiple resource theories, including those of entanglement, coherence, quantum measurements, and quantum channels~\cite{chitambar2019,lewenstein1998,bu2018,ducuara2020,uola2020,luo2022,kuroiwa2024,zhangluo2025,fu2026dynamic}. Unlike robustness, a weight-based resource quantifier is defined through a convex decomposition of the target object into free and resourceful components. It measures the minimum fraction that cannot be explained by free components and admits concrete operational interpretations in exclusion, discrimination, and other related quantum-information tasks~\cite{ducuara2020,uola2020}. However, existing studies on the resource theory of interactive quantum instruments have primarily focused on the robustness perspective, leaving a research gap in the systematic weight-based characterization of interactivity in terms of non-interactive components. In particular, a weight-based resource theory capable of systematically characterizing the maximal non-interactive content of quantum instruments, elucidating its operational significance and composition properties, and revealing its implications for resource conversion remains to be established.

To address this problem, the interactive weight of quantum instruments is introduced as a resource quantifier. It is defined through convex decompositions of a quantum instrument relative to the free set of non-interactive instruments. The quantity measures the minimum fraction that cannot be accounted for by non-interactive components, whereas its complementary quantity represents the maximal non-interactive content that can be embedded in the target instrument. By exploiting the special Choi structure of non-interactive instruments, primal and dual semidefinite programming (SDP) formulations of the interactive weight of quantum instruments are derived. Strong duality is also established, yielding a computable convex-optimization characterization of the resource. The interactive weight of quantum instruments is further shown to be faithful and convex. Moreover, the maximal non-interactive content satisfies exact multiplicativity under independent parallel composition.

The operational significance of the interactive weight of quantum instruments is then investigated through several distinct quantum-information tasks. In maximally entangled-state recovery and Haar-averaged pure-state recovery, quantitative bounds are established between the recovery performance and the interactive weight of quantum instruments, which further yield a unified quantitative relation between the interactive weight of quantum instruments and the previously introduced interactive instrument robustness. In addition, the maximal non-interactive content is shown to admit an exact characterization through a class of positive linear loss games that retain the native classical outcome of the instrument.

More importantly, we introduce an environment-assisted input-blind extraction scenario, in which an assistant performs postselection by measuring the minimal Stinespring environment associated with the quantum instrument, thereby extracting a fully input-blind subprocess without additional processing of the original classical--quantum output. We prove that the maximum success probability of this task is exactly equal to the maximal non-interactive content, providing a direct probabilistic operational interpretation of this resource quantity. Furthermore, we show that the maximal non-interactive content can be exactly characterized by the positive quantum Doeblin coefficient of the associated classical--quantum channel retaining the instrument's native classical outcome~\cite{george2026doeblin}. We further extend the environment-assisted input-blind extraction framework to restricted environmental access, obtaining a universal upper bound on the extraction probability and an access data-processing inequality.

Finally, within the allowed-operations framework introduced in Ref.~\cite{hsieh2026}, the interactive weight of quantum instruments is shown to be monotonic under deterministic allowed operations. Moreover, we find that the exact multiplicativity of the maximal non-interactive content further induces a strictly additive logarithmic interactivity measure. The resulting additive mathematical structure imposes necessary constraints on exact multicopy resource conversion and reveals a finite-copy maximal-weight distillation no-go result. In particular, a finite number of copies of an instrument for which the interactive weight of quantum instruments is not maximal cannot be converted exactly, under deterministic allowed operations, into an instrument with a maximal value of the interactive weight of quantum instruments. The interactive weight of quantum instruments therefore provides a resource quantification complementary to robustness and connects the convex-decomposition structure of quantum instruments with operational tasks, parallel composition, and multicopy resource conversion within a unified framework.

This paper is organized as follows. In \mbox{\hyperref[sec:preliminaries]{Sec.\,2}}, we introduce the necessary notation and definitions of quantum instruments and non-interactive instruments. In \mbox{\hyperref[sec:interactive-weight]{Sec.\,3}}, we define the interactive weight of quantum instruments, derive its semidefinite programming characterization, and establish its fundamental structural properties, including exact multiplicativity of the maximal non-interactive content under parallel composition. In \mbox{\hyperref[sec:operational-interpretations]{Sec.\,4}}, we investigate its operational interpretations through quantum-state recovery, positive linear loss games, and environment-assisted input-blind extraction, including restricted environmental access and the connection to the positive quantum Doeblin coefficient. In \mbox{\hyperref[sec:resource-conversion]{Sec.\,5}}, we establish monotonicity under deterministic allowed operations and derive constraints on multicopy resource conversion and establish the finite-copy maximal-weight distillation no-go result. Finally, we summarize our results in \mbox{\hyperref[sec:conclusion]{Sec.\,6}}.

\section{PRELIMINARIES}\label{sec:preliminaries}
\subsection{Notation}
Throughout this work, all Hilbert spaces are finite-dimensional, and all classical outcome sets are finite. We use capital Roman letters \(A,B,C,D,E,R,Z\) to denote physical systems and their corresponding Hilbert spaces. An isomorphic reference copy of system \(A\) is denoted by \(A'\simeq A\). The same convention applies to composite systems, with \(A_1'A_2'\simeq A_1A_2\) denoting a copy of \(A_1A_2\). We write \(d_A:=\dim A\), abbreviate \(A\otimes B\) as \(AB\), and order tensor factors according to their system labels.
For normalized quantum states on these systems, we use the lowercase Greek letter \(\rho\). In the notation \(\rho_\alpha^S\), the superscript \(S\) specifies the underlying system, while the optional subscript \(\alpha\) distinguishes different states or classical conditions. Other lowercase Greek letters, such as \(\tau,\kappa,\eta\), are used for positive operators that need not be normalized. General operators are denoted by capital Roman letters such as \(X,Y\), and the space of linear operators on \(A\) is denoted by \(\mathcal B(A)\).
For quantum measurements, we use positive operator-valued measures (POVMs), denoted by \(\{M_j^S\}_j\) on system \(S\), satisfying \(M_j^S\succeq0\) and \(\sum_j M_j^S=I_S\). A binary POVM is written as \(\{M_0^S,\ I_S-M_0^S\}\), with \(0\preceq M_0^S\preceq I_S\), where \(I_S\) denotes the identity operator on system \(S\).
We use calligraphic letters such as \(\mathcal M,\mathcal N,\mathcal P\) to denote linear maps between operator spaces, with their specific properties stated in the relevant context. The action of a map is written in parentheses, as in \(\mathcal M(X)\). In particular, completely positive and trace-preserving (CPTP) maps are called quantum channels.
Following Ref.~\cite{hsieh2026}, a quantum instrument from system \(A\) to system \(B\) is denoted by \(\bm{\mathcal E}^{A\to B}=\{\mathcal E_a^{A\to B}\}_a,\) where each branch \(\mathcal E_a\) is completely positive (CP) and trace-nonincreasing, while \(\sum_a\mathcal E_a\) is CPTP. For CP maps \(\mathcal S\) and \(\mathcal T\), the notation \(\mathcal S\preceq_{\mathrm{CP}}\mathcal T\) means that \(\mathcal T-\mathcal S\) is completely positive.
Throughout, we adopt the normalized Choi representation~\cite{choi1975,jamiolkowski1972,watrous2018}. For \(A'\simeq A\), the normalized maximally entangled state is defined as \(\Phi^+_{A'A}:=|\Phi^+\rangle\langle\Phi^+|_{A'A}=\frac{1}{d_A}\sum_{i,j=0}^{d_A-1}|ii\rangle\langle jj|_{A'A},\) where \(|\Phi^+\rangle_{A'A}=\frac{1}{\sqrt{d_A}}\sum_{i=0}^{d_A-1}|ii\rangle_{A'A}.\) For any linear map \(\mathcal M^{A\to B}\), its normalized Choi operator is defined by
\begin{equation}
J^{\mathcal M}_{A'B}:=(\operatorname{id}_{A'}\otimes\mathcal M)(\Phi^+_{A'A}),
\label{eq:notation-choi}
\end{equation}
where \(\operatorname{id}_{A'}\) denotes the identity channel on \(A'\).
Under this normalized convention, for any \(X\in\mathcal B(A)\), the map \(\mathcal M^{A\to B}\) can be recovered from its Choi operator through
\begin{equation}
\mathcal M(X)=d_A\operatorname{Tr}_{A'}\left[(X^T\otimes I_B)J^{\mathcal M}_{A'B}\right].
\label{eq:notation-inverse-choi}
\end{equation}
In particular, the Choi operator of the \(a\)-th instrument branch is written as \(J_a^{\bm{\mathcal E}}:=J^{\mathcal E_a}.\) For a fixed instrument, we use the shorthands \(d:=d_A\) and \(J_a:=J_a^{\bm{\mathcal E}}\). Transposes are taken with respect to the chosen bases, and system labels are omitted when no ambiguity arises.

\subsection{Interactive quantum instruments}
Quantum instruments describe quantum measurement processes that produce both classical measurement outcomes and corresponding quantum output states~\cite{davies1970,ozawa1984}. To characterize the dependence of these joint classical--quantum outputs on the input state, and thereby quantitatively assess the degree of interactivity of quantum instruments, we adopt the resource theory of interactive quantum instruments introduced in Ref.~\cite{hsieh2026}. We denote by \(\mathcal I\) the set of all valid quantum instruments with the same input system, output system, and classical outcome set.

\begin{figure}[!htb]
\centering
\ManuscriptFigure{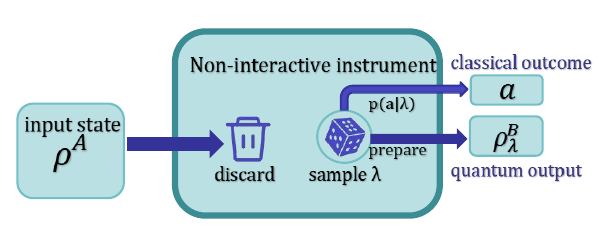}{9cm}
\caption{\textbf{Schematic of a non-interactive quantum instrument.} The instrument first discards the input quantum state and then samples an internal classical variable \(\lambda\). Based on \(\lambda\), it generates the classical outcome \(a\) with conditional probability \(p(a|\lambda)\) and prepares the corresponding quantum output state \(\rho_\lambda^B\).}
\label{fig:ni-instrument}
\end{figure}

Within this framework, an instrument is said to be non-interactive if its joint classical--quantum output is completely independent of the input state. Such an instrument can be written as
\begin{equation}
\mathcal L_a(\,\cdot\,)
=\sum_\lambda q_\lambda p(a|\lambda)\rho_\lambda^B\operatorname{Tr}(\,\cdot\,).
\label{eq:ni-decomposition}
\end{equation}
Here, \(\lambda\) denotes an internal classical random variable of the instrument, \(\{q_\lambda\}_\lambda\) is a probability distribution, \(p(a|\lambda)\) is the conditional probability of generating the classical outcome \(a\), and \(\rho_\lambda^B\) is the corresponding prepared quantum output state. As illustrated in Fig.~\ref{fig:ni-instrument}, the input state is first discarded. The subsequent classical outcome and quantum output are determined solely by the internal random variable of the instrument and therefore contain no information about the original input.

\par\indent Define
\begin{equation}
r(a):=\sum_\lambda q_\lambda p(a|\lambda).
\label{eq:r-def}
\end{equation}
For \(r(a)>0\), we further define \(\rho_a^B:=\frac{\sum_\lambda q_\lambda p(a|\lambda)\rho_\lambda^B}{r(a)}.\) When \(r(a)=0\), \(\rho_a^B\) may be chosen as an arbitrary quantum state. A non-interactive instrument can therefore be equivalently written as \(\mathcal L_a(X)=r(a)\rho_a^B\operatorname{Tr}X.\) Introducing the unnormalized positive operator \(\tau_a^B:=r(a)\rho_a^B,\) this form simplifies to \(\mathcal L_a(X)=\tau_a^B\operatorname{Tr}X,\) with \(\tau_a^B\succeq0\) and \(\sum_a\operatorname{Tr}\tau_a^B=1\). The set of all non-interactive instruments is denoted by \(\mathbf{NI}\subseteq\mathcal I\), and any instrument \(\bm{\mathcal E}\notin\mathbf{NI}\) is referred to as an interactive instrument.

According to the normalized Choi representation introduced above, the Choi operators of the instrument branches satisfy
\begin{equation}
J_a^{\bm{\mathcal E}}\succeq0,\qquad
\sum_a\operatorname{Tr}_B J_a^{\bm{\mathcal E}}=\frac{I_{A'}}{d_A}.
\label{eq:instrument-choi-normalization}
\end{equation}
For a non-interactive instrument, the above form directly gives
\begin{equation}
J_a^{\bm{\mathcal L}}=\frac{I_{A'}}{d_A}\otimes\tau_a^B.
\label{eq:ni-choi-product}
\end{equation}
Hence, non-interactive instruments possess a simple product structure in the Choi representation. This structure provides the direct basis for the semidefinite programming formulation of the interactive weight of quantum instruments introduced below.

Having characterized the structure of non-interactive instruments, a natural next question is how much of a given quantum instrument can be explained by non-interactive components. We therefore introduce a weight-based resource quantifier below to quantify this contribution.

\section{INTERACTIVE WEIGHT OF QUANTUM INSTRUMENTS AND STRUCTURAL PROPERTIES}\label{sec:interactive-weight}
\subsection{Interactive weight of quantum instruments and SDP characterization}
Since the set \(\mathbf{NI}\) is convex~\cite{hsieh2026}, a corresponding resource measure can be constructed following the standard weight-based prescription for convex quantum resource theories~\cite{chitambar2019,ducuara2020,kuroiwa2024}. Specifically, the interactive weight of a quantum instrument \(\bm{\mathcal E}\) is defined as

\begin{equation}
W_{\mathrm{int}}(\bm{\mathcal E}):=\min\bigl\{s\in[0,1]\ \big|\;
\bm{\mathcal E}=(1-s)\bm{\mathcal L}+s\bm{\mathcal G},\quad
\bm{\mathcal L}\in\mathbf{NI},\;\bm{\mathcal G}\in\mathcal I\bigr\}.
\label{eq:interactive-weight-definition}
\end{equation}

In this decomposition, \(\bm{\mathcal L}\in\mathbf{NI}\) denotes the non-interactive component, while \(\bm{\mathcal G}=\{\mathcal G_a^{A\to B}\}_a\in\mathcal I\) denotes an arbitrary valid quantum instrument, with \(\mathcal G_a\) representing its \(a\)-th branch. Both instruments have the same input system, output system, and classical outcome set as \(\bm{\mathcal E}\). The coefficient \(1-s\) is the fraction associated with the non-interactive component. Hence, \(1-W_{\mathrm{int}}(\bm{\mathcal E})\) represents the maximal non-interactive content that can be embedded in the target instrument, whereas \(W_{\mathrm{int}}(\bm{\mathcal E})\) quantifies the minimum remaining fraction that cannot be explained by non-interactive components.

Owing to the special Choi structure of non-interactive instruments established above, the interactive weight of quantum instruments can be formulated as a semidefinite program. Introducing the variable substitution \(\kappa_a^B:=(1-s)\tau_a^B\) converts the original decomposition condition into a branchwise linear matrix inequality. Applying SDP duality~\cite{boyd2004,skrzypczyk2023} yields the following characterization. The complete equivalence transformation, dual construction, and proof of strong duality are given in \hyperref[app:A]{Appendix~A}.

\phantomsection\label{prop:1}\noindent \textbf{Proposition 1.} For any quantum instrument \(\bm{\mathcal E}^{A\to B}=\{\mathcal E_a^{A\to B}\}_a\), the interactive weight of the instrument is given by the primal SDP

\begin{equation}
W_{\mathrm{int}}(\bm{\mathcal E})
=
\min_{\{\kappa_a^B\succeq0\}_a}
\left[
1-\sum_a\operatorname{Tr}\kappa_a^B
\right]
\label{eq:weight-primal-objective}
\end{equation}
subject to
\begin{equation}
J_a^{\bm{\mathcal E}}
-
\frac{I_{A'}}{d_A}\otimes\kappa_a^B
\succeq0,
\qquad
\forall a.
\label{eq:weight-primal-constraints}
\end{equation}
The corresponding dual SDP is
\begin{equation}
W_{\mathrm{int}}(\bm{\mathcal E})
=
\sup_{\{\Omega_a^{A'B}\succeq0\}_a}
\left[
1-
\sum_a
\operatorname{Tr}
\left(
\Omega_a^{A'B}J_a^{\bm{\mathcal E}}
\right)
\right]
\label{eq:weight-dual-objective}
\end{equation}
subject to
\begin{equation}
\operatorname{Tr}_{A'}
\Omega_a^{A'B}
\succeq
d_A I_B,
\qquad
\forall a.
\label{eq:weight-dual-constraints}
\end{equation}
The primal and dual problems satisfy strong duality and therefore have the same optimal value.

To directly characterize the non-interactive component that can be retained within the target instrument, we introduce the complementary quantity
\begin{equation}
\lambda_{\mathbf{NI}}(\bm{\mathcal E})
:=
1-W_{\mathrm{int}}(\bm{\mathcal E}).
\label{eq:ni-content-definition}
\end{equation}
Whereas \(W_{\mathrm{int}}(\bm{\mathcal E})\) quantifies the minimum fraction that cannot be explained by non-interactive components, \(\lambda_{\mathbf{NI}}(\bm{\mathcal E})\) describes the maximal non-interactive content that can be embedded in the target instrument. Equations \eqref{eq:weight-primal-objective} and \eqref{eq:weight-primal-constraints} are therefore equivalent to
\begin{equation}
\lambda_{\mathbf{NI}}(\bm{\mathcal E})
=
\max_{\{\kappa_a^B\succeq0\}_a}
\sum_a\operatorname{Tr}\kappa_a^B
\label{eq:ni-content-primal-objective}
\end{equation}
subject to
\begin{equation}
\frac{I_{A'}}{d_A}
\otimes\kappa_a^B
\preceq
J_a^{\bm{\mathcal E}},
\qquad
\forall a.
\label{eq:ni-content-primal-constraints}
\end{equation}
Here, \(\kappa_a^B\) represents the unnormalized non-interactive component that can be embedded in the target branch \(J_a^{\bm{\mathcal E}}\), while \(\sum_a\operatorname{Tr}\kappa_a^B\) gives the total weight of these components. Hence, \(\lambda_{\mathbf{NI}}(\bm{\mathcal E})\) is precisely the maximal non-interactive content of the target instrument, whereas \(W_{\mathrm{int}}(\bm{\mathcal E})=1-\lambda_{\mathbf{NI}}(\bm{\mathcal E})\) is the minimum residual fraction that cannot be explained by any non-interactive component.

The SDP characterization also provides a convenient basis for examining the intrinsic structural properties of the interactive weight of quantum instruments.

\subsection{Basic structural properties}\label{sec:structural}
The interactive weight of quantum instruments satisfies the natural normalization and faithfulness conditions expected of a weight-based resource quantifier. For every finite-dimensional quantum instrument \(\bm{\mathcal E}\),
\begin{equation}
0\le W_{\mathrm{int}}(\bm{\mathcal E})\le1,
\qquad
W_{\mathrm{int}}(\bm{\mathcal E})=0
\iff
\bm{\mathcal E}\in\mathbf{NI},
\label{eq:weight-range-faithfulness}
\end{equation}
and the minimum in its definition is attained.

The first relation fixes the natural scale of the resource measure, while faithfulness guarantees that the interactive weight of quantum instruments vanishes exactly on the free set of non-interactive instruments. Thus, no genuinely interactive instrument is assigned zero resource. The detailed proof of faithfulness follows from the attained primal SDP and is given in \hyperref[app:A]{Appendix~A}.

The interactive weight of quantum instruments is also convex under probabilistic mixtures of quantum instruments. For any probability distribution \(\{p_i\}_i\) and any family of instruments \(\{\bm{\mathcal E}^{(i)}\}_i\) with the same input system, output system, and classical outcome set,
\begin{equation}
W_{\mathrm{int}}
\!\left(
\sum_i p_i\bm{\mathcal E}^{(i)}
\right)
\le
\sum_i p_i
W_{\mathrm{int}}\!\left(\bm{\mathcal E}^{(i)}\right).
\label{eq:weight-convexity}
\end{equation}
Correspondingly,
\begin{equation}
\lambda_{\mathbf{NI}}
\!\left(
\sum_i p_i\bm{\mathcal E}^{(i)}
\right)
\ge
\sum_i p_i
\lambda_{\mathbf{NI}}\!\left(\bm{\mathcal E}^{(i)}\right),
\label{eq:ni-content-concavity}
\end{equation}
so the maximal non-interactive content is concave.

Convexity has a direct resource-theoretic interpretation. Classical randomization among different quantum instruments cannot increase the average value of the interactive weight of quantum instruments. Equivalently, probabilistic mixing cannot reduce the maximal non-interactive content below the corresponding average value. Together with faithfulness and the normalization range, this establishes the expected basic behavior of \(W_{\mathrm{int}}\) as a convex weight-based quantifier of instrument interactivity. A proof of convexity is provided in \hyperref[app:A]{Appendix~A}.

\subsection{Exact multiplicativity under parallel composition}
Beyond these single-instrument properties, the complementary quantity \(\lambda_{\mathbf{NI}}\) exhibits a stronger compositional structure under independent parallel use. Consider two finite-dimensional quantum instruments \(\bm{\mathcal E}^{A_1\to B_1}=\{\mathcal E_a\}_a\) and \(\bm{\mathcal F}^{A_2\to B_2}=\{\mathcal F_b\}_b\), with parallel composition defined branchwise by \((\bm{\mathcal E}\otimes\bm{\mathcal F})_{a,b}:=\mathcal E_a\otimes\mathcal F_b\). The primal and dual SDP characterizations above imply the following exact composition law.

\phantomsection\label{prop:2}\noindent \textbf{Proposition 2.} For any finite-dimensional quantum instruments \(\bm{\mathcal E}\) and \(\bm{\mathcal F}\),
\begin{equation}
\lambda_{\mathbf{NI}}
(\bm{\mathcal E}\otimes\bm{\mathcal F})
=
\lambda_{\mathbf{NI}}(\bm{\mathcal E})
\lambda_{\mathbf{NI}}(\bm{\mathcal F}).
\label{eq:ni-content-multiplicativity}
\end{equation}
Equivalently,
\begin{equation}
1-W_{\mathrm{int}}(\bm{\mathcal E}\otimes\bm{\mathcal F})
=
\bigl[1-W_{\mathrm{int}}(\bm{\mathcal E})\bigr]
\bigl[1-W_{\mathrm{int}}(\bm{\mathcal F})\bigr],
\label{eq:complementary-weight-product}
\end{equation}
or
\begin{equation}
W_{\mathrm{int}}(\bm{\mathcal E}\otimes\bm{\mathcal F})
=
W_{\mathrm{int}}(\bm{\mathcal E})
+
W_{\mathrm{int}}(\bm{\mathcal F})
-
W_{\mathrm{int}}(\bm{\mathcal E})W_{\mathrm{int}}(\bm{\mathcal F}).
\label{eq:weight-parallel-composition}
\end{equation}
A complete proof based on the primal and dual SDPs, including the tensor-factor ordering of the corresponding Choi operators, is provided in \hyperref[app:F]{Appendix~F}.

Equation~\eqref{eq:ni-content-multiplicativity} shows that the maximal non-interactive content obeys an exact multiplicative law under the parallel composition of two independent quantum instruments, rather than merely satisfying a one-sided bound. For multiple copies of the same instrument, by writing the \(n\)-fold tensor product as \(\bm{\mathcal E}^{\otimes n}\) and repeatedly applying Eq.~\eqref{eq:ni-content-multiplicativity}, this multiplicative law extends to any positive integer \(n\), yielding
\begingroup
\begin{equation}
\lambda_{\mathbf{NI}}(\bm{\mathcal E}^{\otimes n})
=
\left[\lambda_{\mathbf{NI}}(\bm{\mathcal E})\right]^n.
\label{eq:ni-content-tensor-power}
\end{equation}
\endgroup
Thus, the maximal non-interactive content of independent copies factorizes exactly. For an instrument satisfying \(0<\lambda_{\mathbf{NI}}(\bm{\mathcal E})<1\), this content decreases exponentially with the number of copies but remains strictly positive for every finite \(n\). This exact multicopy structure will later provide the basis for introducing an additive logarithmic interactivity measure and for deriving constraints on multicopy resource conversion under deterministic allowed operations.

Having established the definition, computable SDP characterization, and intrinsic structural properties of the interactive weight of quantum instruments, we next turn to its operational significance in concrete quantum-information tasks.

\section{OPERATIONAL INTERPRETATIONS}\label{sec:operational-interpretations}
In this section, we explore the operational significance of the interactive weight of quantum instruments and its complementary maximal non-interactive content through four complementary settings. We begin with \hyperref[interpretation:1]{Interpretation 1} and \hyperref[interpretation:2]{Interpretation 2}, which establish quantitative bounds through maximally entangled-state recovery and Haar-averaged pure-state recovery, respectively. We then turn to \hyperref[interpretation:3]{Interpretation 3}, where positive linear loss games provide an exact characterization of the maximal non-interactive content. Finally, \hyperref[interpretation:4]{Interpretation 4} introduces a new operational scenario of environment-assisted input-blind extraction, where we prove that the maximum success probability of extracting a fully input-blind subprocess is exactly equal to the maximal non-interactive content, and further extend this framework to restricted environmental access.

\subsection{Interpretation 1}\label{interpretation:1}
The SDP formulation above shows that the interactive weight of quantum instruments can be understood in terms of the maximal non-interactive component that can be contained in an instrument. We now show that this quantity also directly constrains the performance of the instrument in a maximally entangled-state recovery task.

\begin{figure}[!htb]
\centering
\ManuscriptFigure{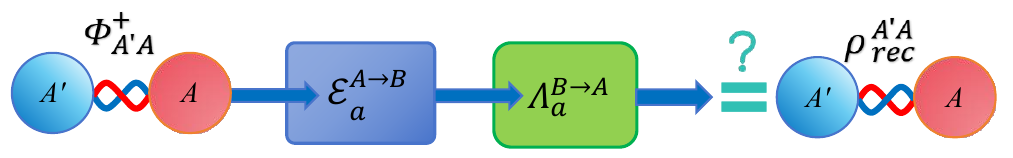}{9.8cm}
\caption{\textbf{Schematic of maximally entangled-state recovery.} Starting from the initial maximally entangled state \(\Phi^+_{A'A}\), the reference system \(A'\) is retained while the other subsystem \(A\) is sent through the quantum instrument \(\bm{\mathcal E}^{A\to B}\), followed by the recovery channel \(\Lambda_a^{B\to A}\) conditioned on the classical outcome \(a\). Finally, the recovery performance is quantified by the fidelity between the recovered joint state \(\rho_{\mathrm{rec}}^{A'A}\) and the initial maximally entangled state.}
\label{fig:maxent-recovery}
\end{figure}

Consider a maximally entangled state \(|\Phi^{+}\rangle_{A'A}\) shared between a reference system \(A'\) and the input system \(A\). System \(A\) is sent through the quantum instrument \(\bm{\mathcal E}^{A\to B} = \{\mathcal{E}^{A\to B}_a\}_a,\) while the reference system \(A'\) remains unchanged. The instrument produces a classical outcome \(a\) and a quantum output \(B\). Conditioned on the classical outcome \(a\), a recovery channel \(\Lambda_a^{B\to A}\) is applied with the aim of restoring the final state as close as possible to the initial maximally entangled state. This recovery process is illustrated in Fig.~\ref{fig:maxent-recovery}.

For a fixed family of recovery channels \(\bm\Lambda=\{\Lambda_a\}_a\), the recovered state, averaged over all classical outcomes, is given by
\begin{equation}
\begin{aligned}
\rho_{\mathrm{rec}}^{A'A}&:=\sum_a
\bigl[\operatorname{id}_{A'}\otimes(\Lambda_a\circ\mathcal E_a)\bigr]
(\Phi^+_{A'A}).
\end{aligned}
\label{eq:maxent-recovered-state}
\end{equation}
The corresponding average maximally entangled-state recovery performance is defined as

\begin{equation}
F_{+}(\bm{\mathcal E};\bm\Lambda)
:=
\sum_a
\langle\Phi^{+}|
\bigl[
\operatorname{id}_{A'}
\otimes
(\Lambda_a\circ\mathcal{E}_a)
\bigr]
(\Phi^{+})
|\Phi^{+}\rangle .
\label{eq:maxent-recovery-performance}
\end{equation}
Optimizing over all recovery channels gives

\begin{equation}
F_{+}(\bm{\mathcal E})
:=
\max_{\{\Lambda_a^{B\to A}\}_a}
F_{+}(\bm{\mathcal E};\bm\Lambda).
\label{eq:maxent-recovery-optimum}
\end{equation}
For any non-interactive instrument \(\mathcal{L}_a(X)=\tau_a\operatorname{Tr}X\), this recovery performance is independent of the particular recovery strategy and satisfies

\begin{equation}
F_{+}(\bm{\mathcal L};\bm\Lambda)
=
\frac{1}{d_A^2}.
\label{eq:maxent-ni-benchmark}
\end{equation}
Hence, \(1/d_A^2\) provides the benchmark performance for all non-interactive instruments in this task.

The interactive weight of quantum instruments is related to the recovery performance through the following bound.

\phantomsection\label{prop:3}\noindent \textbf{Proposition 3.} For any quantum instrument \(\bm{\mathcal E}^{A\to B}\) with input dimension \(d_A\),

\begin{equation}
\frac{1}{d_A^2}
\le
F_{+}(\bm{\mathcal E})
\le
W_{\mathrm{int}}(\bm{\mathcal E})
+
\frac{1-W_{\mathrm{int}}(\bm{\mathcal E})}{d_A^2}.
\label{eq:maxent-recovery-bounds}
\end{equation}
For \(d_A\ge2\),

\begin{equation}
W_{\mathrm{int}}(\bm{\mathcal E})
\ge
\frac{d_A^2F_{+}(\bm{\mathcal E})-1}
{d_A^2-1}.
\label{eq:maxent-weight-bound}
\end{equation}
\noindent \textbf{Proof.} Let \(w:=W_{\mathrm{int}}(\bm{\mathcal E}).\) An optimal decomposition of the interactive weight of quantum instruments can be written as \(\bm{\mathcal E} = (1-w)\bm{\mathcal L} + w\bm{\mathcal G},\) where \(\bm{\mathcal L}\in\mathbf{NI}\) and \(\bm{\mathcal G}\) is a valid quantum instrument. For any fixed recovery family \(\bm\Lambda=\{\Lambda_a\}_a\), the linearity of \(F_{+}\) with respect to the instrument branches gives

\begin{equation}
F_{+}(\bm{\mathcal E};\bm\Lambda)
=
(1-w)F_{+}(\bm{\mathcal L};\bm\Lambda)
+
wF_{+}(\bm{\mathcal G};\bm\Lambda).
\label{eq:maxent-recovery-linearity}
\end{equation}
For a non-interactive instrument, \(F_{+}(\bm{\mathcal L};\bm\Lambda) = \frac{1}{d_A^2},\) whereas \(F_{+}(\bm{\mathcal G};\bm\Lambda)\le1\) for any valid instrument \(\bm{\mathcal G}\) and recovery strategy. Therefore, \(F_{+}(\bm{\mathcal E};\bm\Lambda) \le \frac{1-w}{d_A^2}+w.\) The right-hand side is independent of the particular recovery strategy. Maximizing over all \(\Lambda_a\) therefore yields \(F_{+}(\bm{\mathcal E}) \le W_{\mathrm{int}}(\bm{\mathcal E}) + \frac{1-W_{\mathrm{int}}(\bm{\mathcal E})}{d_A^2}.\) On the other hand, a fixed-state preparation channel can always be chosen as the recovery operation for any quantum instrument. This gives \(F_{+}(\bm{\mathcal E}) \ge \frac{1}{d_A^2}.\) Combining the two inequalities proves \mbox{Eq.~\eqref{eq:maxent-recovery-bounds}}. Rearranging its upper bound gives \mbox{Eq.~\eqref{eq:maxent-weight-bound}}.

The complete calculation of the non-interactive benchmark and a detailed proof of the above bounds are provided in \hyperref[app:B]{Appendix~B}.

The operational meaning of this result is direct. A completely non-interactive instrument cannot achieve a recovery performance above \(1/d_A^2\) in the maximally entangled-state recovery task. Therefore, any experimentally observed value satisfying \(F_{+}(\bm{\mathcal E}) > \frac{1}{d_A^2}\) directly certifies a nonzero value of the interactive weight of quantum instruments. Moreover, \mbox{Eq.~\eqref{eq:maxent-weight-bound}} converts the measured recovery performance into a quantitative lower bound on \(W_{\mathrm{int}}(\bm{\mathcal E})\).

\subsection{Interpretation 2}\label{interpretation:2}
The interactive weight of quantum instruments can also be operationally constrained through a recovery task that does not require entanglement assistance.

\begin{figure}[!htb]
\centering
\ManuscriptFigure{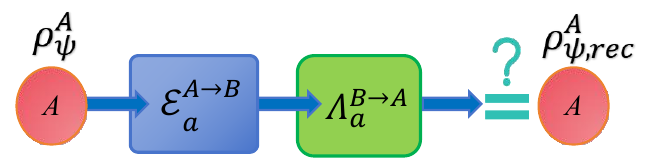}{8.1cm}
\caption{\textbf{Schematic of pure-state recovery.} An unknown input pure state \(\rho_\psi^A\) is randomly sampled according to the Haar measure and sent through the quantum instrument \(\bm{\mathcal E}^{A\to B}\), followed by the recovery channel \(\Lambda_a^{B\to A}\) conditioned on the classical outcome \(a\). The average recovery performance is quantified by the fidelity between the recovered state \(\rho_{\psi,\mathrm{rec}}^A\) and the original input state, averaged over Haar-random pure states.}
\label{fig:pure-recovery}
\end{figure}

Consider an unknown pure state \(\rho_\psi^A = |\psi\rangle\langle\psi|^A\) randomly sampled from a \(d_A\)-dimensional Hilbert space according to the normalized Haar measure. The state is sent through the quantum instrument \(\bm{\mathcal E}^{A\to B}\), which produces a classical outcome \(a\) and a quantum output \(B\). Conditioned on \(a\), a recovery channel \(\Lambda_a^{B\to A}\) is applied in an attempt to restore the original unknown state. This recovery process is illustrated in Fig.~\ref{fig:pure-recovery}. The recovery performance is quantified by the fidelity between the recovered state and the original input and is then averaged over all Haar-random pure states.

For a fixed recovery family \(\bm\Lambda=\{\Lambda_a\}_a\), the recovered state corresponding to an input \(\rho_\psi^A\), averaged over all classical outcomes, is given by
\begin{equation}
\rho_{\psi,\mathrm{rec}}^{A}
:=\sum_a(\Lambda_a\circ\mathcal E_a)(\rho_\psi^A).
\label{eq:pure-recovered-state}
\end{equation}
The average recovery fidelity is defined as

\begin{equation}
F_{\mathrm{ave}}(\bm{\mathcal E};\bm\Lambda)
:=
\int d\psi
\sum_a
\langle\psi|
(\Lambda_a\circ\mathcal{E}_a)(\rho_\psi)
|\psi\rangle .
\label{eq:haar-recovery-performance}
\end{equation}
The optimal average recovery performance is defined as

\begin{equation}
F_{\mathrm{ave}}(\bm{\mathcal E})
:=
\max_{\{\Lambda_a^{B\to A}\}_a}
F_{\mathrm{ave}}(\bm{\mathcal E};\bm\Lambda).
\label{eq:haar-recovery-optimum}
\end{equation}
Unlike the benchmark \(1/d_A^2\) in the maximally entangled-state recovery task, every non-interactive instrument satisfies

\begin{equation}
F_{\mathrm{ave}}(\bm{\mathcal L};\bm\Lambda)
=
\frac{1}{d_A},
\qquad
\forall\,\bm{\mathcal L}\in\mathbf{NI},
\label{eq:haar-ni-benchmark}
\end{equation}
independently of the recovery strategy. A non-interactive instrument completely discards the unknown input state. Consequently, after any conditional recovery operation, the overall output remains a fixed state independent of \(\rho_\psi\). The Haar-averaged overlap between any fixed quantum state and a random pure state is exactly \(1/d_A\).

The interactive weight of quantum instruments therefore constrains the average recovery performance as follows.

\phantomsection\label{prop:4}\noindent \textbf{Proposition 4.} For any quantum instrument \(\bm{\mathcal E}^{A\to B}\) with input dimension \(d_A\),

\begin{equation}
\frac{1}{d_A}
\le
F_{\mathrm{ave}}(\bm{\mathcal E})
\le
W_{\mathrm{int}}(\bm{\mathcal E})
+
\frac{1-W_{\mathrm{int}}(\bm{\mathcal E})}{d_A}.
\label{eq:haar-recovery-bounds}
\end{equation}
Hence, for \(d_A\ge2\),

\begin{equation}
W_{\mathrm{int}}(\bm{\mathcal E})
\ge
\frac{d_AF_{\mathrm{ave}}(\bm{\mathcal E})-1}
{d_A-1}.
\label{eq:haar-weight-bound}
\end{equation}
\noindent \textbf{Proof.} Let \(w:=W_{\mathrm{int}}(\bm{\mathcal E}),\) and take an optimal weight decomposition \(\bm{\mathcal E} = (1-w)\bm{\mathcal L} + w\bm{\mathcal G},\allowbreak\; \bm{\mathcal L}\in\mathbf{NI},\) where \(\bm{\mathcal G}\) is a valid quantum instrument. For any fixed recovery family \(\bm\Lambda=\{\Lambda_a\}_a\), the linearity of the average recovery fidelity with respect to the instrument gives

\begin{equation}
F_{\mathrm{ave}}(\bm{\mathcal E};\bm\Lambda)
=
(1-w)F_{\mathrm{ave}}(\bm{\mathcal L};\bm\Lambda)
+
wF_{\mathrm{ave}}(\bm{\mathcal G};\bm\Lambda).
\label{eq:haar-recovery-linearity}
\end{equation}
The non-interactive component satisfies \(F_{\mathrm{ave}}(\bm{\mathcal L};\bm\Lambda) = \frac{1}{d_A},\) whereas \(F_{\mathrm{ave}}(\bm{\mathcal G};\bm\Lambda)\le1\) for any valid instrument \(\bm{\mathcal G}\) and recovery strategy. Therefore, \(F_{\mathrm{ave}}(\bm{\mathcal E};\bm\Lambda) \le \frac{1-w}{d_A}+w.\) Since the right-hand side is independent of the recovery strategy, maximization over all recovery channels yields \(F_{\mathrm{ave}}(\bm{\mathcal E}) \le W_{\mathrm{int}}(\bm{\mathcal E}) + \frac{1-W_{\mathrm{int}}(\bm{\mathcal E})}{d_A}.\) Conversely, a fixed-state preparation channel can always be used as the recovery strategy for any instrument, giving \(F_{\mathrm{ave}}(\bm{\mathcal E}) \ge \frac{1}{d_A}.\) This proves \mbox{Eq.~\eqref{eq:haar-recovery-bounds}}, while rearranging the upper bound gives \mbox{Eq.~\eqref{eq:haar-weight-bound}}.

The complete proof and the calculation of the Haar average are provided in \hyperref[app:C]{Appendix~C}.

Thus, \(1/d_A\) is the exact benchmark for all non-interactive instruments in this Haar-averaged pure-state recovery task. Any recovery performance satisfying \(F_{\mathrm{ave}}(\bm{\mathcal E}) > \frac{1}{d_A}\) certifies a nonzero value of the interactive weight of quantum instruments. Equation \eqref{eq:haar-weight-bound} further provides a quantitative lower bound on \(W_{\mathrm{int}}(\bm{\mathcal E})\).

Before establishing the quantitative relation between the interactive weight of quantum instruments and the previously introduced interactive instrument robustness, we first recall the definition of the interactive instrument robustness introduced in Ref.~\cite{hsieh2026}. For any quantum instrument \(\bm{\mathcal E}\in\mathcal I\), the interactive instrument robustness is defined as
\begin{equation}
R_{\mathrm{int}}(\bm{\mathcal E}):=\min_{\substack{\bm{\mathcal N}\in\mathcal I\\ t\ge0}}\left\{t:\frac{\bm{\mathcal E}+t\bm{\mathcal N}}{1+t}\in\mathbf{NI}\right\}.
\label{eq:robustness-def}
\end{equation}
Here, \(\bm{\mathcal N}=\{\mathcal N_a^{A\to B}\}_a\in\mathcal I\) is an arbitrary quantum instrument with the same input system, output system, and classical outcome set as \(\bm{\mathcal E}\), while \(t\ge0\) quantifies the amount of instrument noise mixed with \(\bm{\mathcal E}\). In the finite-dimensional setting considered here, the minimum is attained. The quantity \(R_{\mathrm{int}}(\bm{\mathcal E})\) therefore characterizes instrument interactivity from the perspective of the amount of noise required to eliminate the resource.

\phantomsection\label{cor:1}\noindent \textbf{Corollary 1.} Combining the two recovery tasks above yields a unified quantitative relation between the interactive weight of quantum instruments and the interactive instrument robustness. For any quantum instrument \(\bm{\mathcal E}^{A\to B}\) with \(d_A\ge2\),
\begin{equation}
W_{\mathrm{int}}(\bm{\mathcal E})\ge\frac{R_{\mathrm{int}}(\bm{\mathcal E})}{d_A^2-1}.
\label{eq:weight-robustness-relation}
\end{equation}
This result follows directly from the recovery bounds above together with the recovery relations for the interactive instrument robustness established in Ref.~\cite{hsieh2026}. The maximally entangled-state recovery task gives
\begin{equation}
W_{\mathrm{int}}(\bm{\mathcal E})\ge\frac{d_A^2F_+(\bm{\mathcal E})-1}{d_A^2-1},
\label{eq:corollary-maxent-weight}
\end{equation}
while
\begin{equation}
1+R_{\mathrm{int}}(\bm{\mathcal E})=d_A^2F_+(\bm{\mathcal E})
\label{eq:corollary-maxent-robustness}
\end{equation}
was established in Ref.~\cite{hsieh2026}. Substitution of the latter relation into the former immediately gives \mbox{Eq.~\eqref{eq:weight-robustness-relation}}.

Likewise, the Haar-averaged pure-state recovery task gives
\begin{equation}
W_{\mathrm{int}}(\bm{\mathcal E})\ge\frac{d_AF_{\mathrm{ave}}(\bm{\mathcal E})-1}{d_A-1},
\label{eq:corollary-haar-weight}
\end{equation}
whereas Ref.~\cite{hsieh2026} gives
\begin{equation}
F_{\mathrm{ave}}(\bm{\mathcal E})=\frac{1}{d_A}\left[1+\frac{R_{\mathrm{int}}(\bm{\mathcal E})}{d_A+1}\right].
\label{eq:corollary-haar-robustness}
\end{equation}
Substituting this expression into the preceding bound again yields \mbox{Eq.~\eqref{eq:weight-robustness-relation}}.

Although maximally entangled-state recovery and Haar-averaged pure-state recovery correspond to two distinct operational tasks, they lead to exactly the same quantitative constraint between \(W_{\mathrm{int}}\) and \(R_{\mathrm{int}}\).

This relation also reflects the complementary nature of the two resource quantifiers. The interactive instrument robustness \(R_{\mathrm{int}}\) measures the amount of noise required to eliminate instrument interactivity. By contrast, the interactive weight of quantum instruments \(W_{\mathrm{int}}\) quantifies the fraction of interactivity that must remain after the largest possible non-interactive component has been extracted. Equation~\eqref{eq:weight-robustness-relation} shows that an instrument with large interactive instrument robustness cannot have an arbitrarily small value of the interactive weight of quantum instruments.

Equivalently, \mbox{Eq.~\eqref{eq:weight-robustness-relation}} can be written as
\begin{equation}
R_{\mathrm{int}}(\bm{\mathcal E})\le(d_A^2-1)W_{\mathrm{int}}(\bm{\mathcal E}).
\label{eq:robustness-weight-upper-bound}
\end{equation}
Thus, the interactive weight of quantum instruments not only admits its own operational characterization but also constrains the previously introduced interactive instrument robustness. More importantly, the same bound is obtained independently from two distinct recovery tasks. This indicates that the relation is not tied to a particular experiment, but instead reflects a more general structural connection between the two interactivity resource quantifiers.

\subsection{Interpretation 3}\label{interpretation:3}
We now present an exact operational characterization of the maximal non-interactive content, with respect to the non-interactive free set \(\mathbf{NI}\), through positive linear loss games. Unlike the recovery tasks considered above, the present setting involves a finite Alice--Bob prepare--instrument--measure game in which the native classical outcome generated by the quantum instrument is retained.

Suppose that, in each experimental round, Alice randomly selects an input state \(\rho_i^A\) from a finite set \(\{\rho_i^A\}_i\) according to a probability distribution \(\{p_i\}_i\), and sends it through a quantum instrument \(\bm{\mathcal E}^{A\to B}=\{\mathcal E_a^{A\to B}\}_a\). The instrument produces a classical outcome \(a\) and a corresponding quantum output in system \(B\). The classical outcome is recorded, while the quantum output is delivered to Bob. Bob then performs a fixed POVM \(\{M_j^B\}_j\), satisfying \(M_j^B\succeq0\) and \(\sum_j M_j^B=I_B\), and records the measurement outcome \(j\). The same POVM is used in every round, independently of the instrument outcome \(a\).

\begin{figure}[!htb]
\centering
\ManuscriptFigure{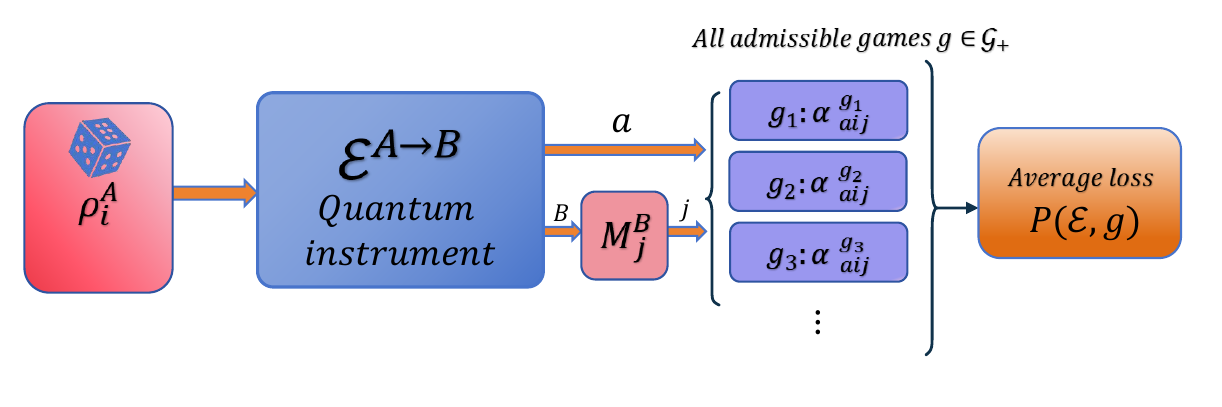}{13cm}
\caption{\textbf{Schematic of positive linear loss games.} Alice first selects and prepares an input state \(\rho_i^A\) with probability \(p_i\) and sends it through the quantum instrument \(\bm{\mathcal E}^{A\to B}\). The instrument produces its native classical outcome \(a\) and delivers the quantum output system \(B\) to Bob, who performs a fixed POVM \(\{M_j^B\}_j\) and obtains outcome \(j\). Each experimental round thereby generates a complete record \((i,a,j)\), to which a loss coefficient \(\alpha_{aij}\) is assigned. Finally, the probability-weighted average over all possible records yields the instrument's average loss \(P(\bm{\mathcal E},g)\) in game \(g\).}
\label{fig:positive-loss}
\end{figure}

Each experimental round therefore generates a complete record \((i,a,j)\), with joint probability \(p_i\allowbreak\operatorname{Tr}[\mathcal E_a(\rho_i)M_j]\). To evaluate the performance of the instrument, a referee assigns a real-valued loss coefficient \(\alpha_{aij}\) to each possible record. The dependence on \(a\) allows the game to distinguish different branches of the instrument rather than evaluating only its outcome-averaged quantum channel. A game is thus specified by

\begin{equation}
g=
\bigl(
\{p_i,\rho_i\}_i,
\{M_j\}_j,
\{\alpha_{aij}\}_{a,i,j}
\bigr),
\label{eq:game-def}
\end{equation}
and the average loss of the instrument \(\bm{\mathcal E}\) in this game is defined as

\begin{equation}
P(\bm{\mathcal E},g)
:=
\sum_{a,i,j}
p_i\alpha_{aij}
\operatorname{Tr}
\!\left[
\mathcal E_a(\rho_i)M_j
\right].
\label{eq:game-average-loss}
\end{equation}
The prepare--instrument--measure game and its loss evaluation are illustrated in Fig.~\ref{fig:positive-loss}.

The aim is to characterize the maximal non-interactive content by comparing the loss of the target instrument with the minimum loss achievable by non-interactive instruments in the same game.

To establish this connection, we express the average loss in terms of the Choi operators of the instrument. Using the normalized inverse Choi formula,

\begin{equation}
\mathcal E_a(\rho)
=
d_A
\operatorname{Tr}_{A'}
\!\left[
(\rho^T\otimes I_B)
J_a^{\bm{\mathcal E}}
\right].
\label{eq:instrument-inverse-choi}
\end{equation}

we obtain \(\operatorname{Tr}[\mathcal E_a(\rho_i)M_j]=d_A\operatorname{Tr}[(\rho_i^T\otimes M_j)J_a^{\bm{\mathcal E}}]\). For each classical branch \(a\), define the corresponding Choi loss operator as

\begin{equation}
H_a^g
:=
d_A
\sum_{i,j}
p_i\alpha_{aij}
\rho_i^T\otimes M_j .
\label{eq:game-choi-loss-operators}
\end{equation}

\noindent Substituting this expression into \mbox{Eq.~\eqref{eq:game-average-loss}}, the average loss becomes

\begin{equation}
P(\bm{\mathcal E},g)
=
\sum_a
\operatorname{Tr}
\!\left[
H_a^gJ_a^{\bm{\mathcal E}}
\right].
\label{eq:game-choi-loss-representation}
\end{equation}

Thus, the average loss is represented as a linear functional of the branchwise Choi operators, matching the trace structure appearing in the dual SDP characterization of the maximal non-interactive content.

To obtain a well-defined relative performance measure, we restrict attention to games satisfying \(H_a^g\succeq0\) for every \(a\), with a strictly positive minimum loss over non-interactive instruments. The positivity condition ensures that the average loss is nonnegative for every valid instrument, while the positive non-interactive benchmark permits normalization. Specifically, define

\begin{equation}
b_{\mathbf{NI}}(g)
:=
\min_{\bm{\mathcal L}\in\mathbf{NI}}
P(\bm{\mathcal L},g)
>
0,
\label{eq:game-positive-free-benchmark}
\end{equation}
and denote the set of all such games by \(\mathcal G_+\).

Using the non-interactive Choi structure established in Eq.~\eqref{eq:ni-choi-product}, the average loss in Eq.~\eqref{eq:game-choi-loss-representation} reduces, for any non-interactive instrument \(\bm{\mathcal L}\), to \(P(\bm{\mathcal L},g)=\allowbreak\sum_a\operatorname{Tr}\left[\left(\frac{\operatorname{Tr}_{A'}H_a^g}{d_A}\right)\tau_a^B\right]\), where \(\tau_a^B\succeq0\) and \(\sum_a\operatorname{Tr}\tau_a^B=1\).

The evaluation of the non-interactive benchmark therefore reduces to an optimization over the classical branches and their corresponding quantum output states. \hyperref[lem:D.1]{Lemma D.1} in \hyperref[app:D]{Appendix~D} proves that the minimum is attained by selecting the branch with the smallest eigenvalue and preparing a corresponding eigenstate as the quantum output. Consequently, the non-interactive benchmark admits the closed-form expression

\begin{equation}
b_{\mathbf{NI}}(g)
=
\min_a
\lambda_{\min}
\!\left(
\frac{\operatorname{Tr}_{A'}H_a^g}{d_A}
\right).
\label{eq:game-free-benchmark-spectrum}
\end{equation}

Hence, the non-interactive benchmark can be calculated directly from the loss operators. Comparing \(P(\bm{\mathcal E},g)\) with this benchmark yields a normalized loss ratio, whose optimization over all admissible games will be shown below to characterize the maximal non-interactive content exactly.

\phantomsection\label{prop:5}\noindent \textbf{Proposition 5.} For any finite-dimensional quantum instrument \(\bm{\mathcal E}\),

\begin{equation}
1-W_{\mathrm{int}}(\bm{\mathcal E})
=
\inf_{g\in\mathcal G_+}
\frac{
P(\bm{\mathcal E},g)
}{
\displaystyle
\min_{\bm{\mathcal L}\in\mathbf{NI}}
P(\bm{\mathcal L},g)
}.
\label{eq:game-exact-characterization}
\end{equation}
Thus, the maximal non-interactive content \(1-W_{\mathrm{int}}(\bm{\mathcal E})\) is exactly the infimum of the normalized loss ratio of the target instrument relative to the non-interactive benchmark over all admissible positive linear loss games.

\noindent \textbf{Proof.} Let \(\lambda_{\mathbf{NI}}(\bm{\mathcal E}) = 1-W_{\mathrm{int}}(\bm{\mathcal E}).\) From the dual SDP in \hyperref[prop:1]{Proposition 1},

\begin{equation}
\begin{aligned}
\lambda_{\mathbf{NI}}(\bm{\mathcal E})
&=\inf_{\substack{\Omega_a\succeq0\\\operatorname{Tr}_{A'}\Omega_a\succeq d_AI_B}}\;\sum_a\operatorname{Tr}(\Omega_aJ_a^{\bm{\mathcal E}}).
\end{aligned}
\label{eq:ni-content-dual-for-games}
\end{equation}
First, fix an arbitrary \(g\in\mathcal G_+\), and define \(\Omega_a:=H_a^g/b_{\mathbf{NI}}(g).\) Equation \eqref{eq:game-free-benchmark-spectrum} gives \(\operatorname{Tr}_{A'}H_a^g \succeq d_A b_{\mathbf{NI}}(g)I_B,\) and therefore \(\operatorname{Tr}_{A'}\Omega_a \succeq d_AI_B.\) Since \(H_a^g\succeq0\), the family \(\{\Omega_a\}_a\) is feasible for the dual problem in \mbox{Eq.~\eqref{eq:ni-content-dual-for-games}}. Hence,

\begin{equation}
\frac{P(\bm{\mathcal E},g)}
{b_{\mathbf{NI}}(g)}
=
\sum_a
\operatorname{Tr}
\!\left(
\Omega_aJ_a^{\bm{\mathcal E}}
\right)
\ge
\lambda_{\mathbf{NI}}(\bm{\mathcal E}).
\label{eq:game-normalized-loss-bound}
\end{equation}
Taking the infimum over all \(g\in\mathcal G_+\) gives

\begin{equation}
\inf_{g\in\mathcal G_+}
\frac{P(\bm{\mathcal E},g)}
{b_{\mathbf{NI}}(g)}
\ge
\lambda_{\mathbf{NI}}(\bm{\mathcal E}).
\label{eq:game-infimum-lower-bound}
\end{equation}
It remains to establish the reverse inequality. Fix an arbitrary \(\epsilon>0\), and choose an \(\epsilon\)-optimal dual-feasible family \(\{\Omega_a\}_a\) for \mbox{Eq.~\eqref{eq:ni-content-dual-for-games}} such that

\begin{equation}
\Omega_a\succeq0,\quad\operatorname{Tr}_{A'}\Omega_a\succeq d_AI_B,\quad\sum_a\operatorname{Tr}(\Omega_aJ_a^{\bm{\mathcal E}})\le\lambda_{\mathbf{NI}}(\bm{\mathcal E})+\epsilon.
\label{eq:game-epsilon-optimal-witness}
\end{equation}
The finite-dimensional Hermitian operator space can be spanned by tensor products of a finite set of transposed physical input states \(\{\rho_i^T\}_i\) and an informationally complete POVM \(\{M_j\}_j\). Therefore, finite \(\{p_i,\rho_i\}_i\), a common POVM \(\{M_j\}_j\), and real coefficients \(\alpha_{aij}\) can be chosen such that the corresponding game satisfies

\begin{equation}
H_a^g
=
d_A
\sum_{i,j}
p_i\alpha_{aij}
\rho_i^T\otimes M_j
=
\Omega_a,
\qquad
\forall a.
\label{eq:game-witness-realization}
\end{equation}
Since \(\Omega_a\succeq0\), one has \(H_a^g\succeq0\). Furthermore, the dual-feasibility conditions in Eq.~\eqref{eq:game-epsilon-optimal-witness}, together with \(H_a^g=\Omega_a\), imply \(\operatorname{Tr}_{A'}H_a^g/d_A\succeq I_B\) for every \(a\). Hence, Eq.~\eqref{eq:game-free-benchmark-spectrum} gives \(b_{\mathbf{NI}}(g)\ge1\).
The constructed game therefore belongs to \(\mathcal G_+\). Since \(H_a^g\succeq0\) and \(J_a^{\bm{\mathcal E}}\succeq0\), one also has \(P(\bm{\mathcal E},g)\ge0\). Consequently,

\begin{equation}
\frac{P(\bm{\mathcal E},g)}
{b_{\mathbf{NI}}(g)}
\le
\lambda_{\mathbf{NI}}(\bm{\mathcal E})+\epsilon.
\label{eq:game-epsilon-upper-bound}
\end{equation}
Since \(\epsilon>0\) is arbitrary, taking \(\epsilon\to0\) yields \(\inf_{g\in\mathcal G_+} \frac{P(\bm{\mathcal E},g)} {b_{\mathbf{NI}}(g)} \le \lambda_{\mathbf{NI}}(\bm{\mathcal E}).\) Together with \mbox{Eq.~\eqref{eq:game-infimum-lower-bound}}, this proves \mbox{Eq.~\eqref{eq:game-exact-characterization}}. \hfill\ensuremath{\square}

A finite prepare-and-measure realization of the dual operator family, together with explicit constructions of the informationally complete input states and POVM, is provided in \hyperref[app:D]{Appendix~D}.

Unlike the bounds obtained from the two recovery tasks above, \mbox{Eq.~\eqref{eq:game-exact-characterization}} is an exact equality. For each game, the denominator gives the minimum loss attainable by a non-interactive instrument, namely the non-interactive benchmark. The numerator is the loss of the target instrument in the same task. Taking the infimum of the normalized loss ratio over all admissible games yields exactly the maximal non-interactive content \(1-W_{\mathrm{int}}(\bm{\mathcal E})\). Hence, this quantity can be characterized exactly through an experimental task involving a finite set of quantum-state preparations, the native classical outcome of the instrument, and a fixed POVM on Bob's side.

\subsection{Interpretation 4}\label{interpretation:4}
We now present another operational interpretation of the interactive weight of quantum instruments. Unlike the recovery tasks considered above, the input state is not restored here. Instead, access to the full environment generated during the operation of the instrument is allowed, and a physical, fully input-blind subprocess carrying no information about the input is extracted from within the original instrument.

We assume that Alice holds an arbitrary joint input state \(\rho^{RA}\), where \(R\) is a reference system retained by Alice and system \(A\) is sent through the quantum instrument \(\bm{\mathcal E}^{A\to B}=\{\mathcal E_a^{A\to B}\}_a\). Here, the joint state \(\rho^{RA}\) need not be separable across \(R\) and \(A\). Thus, system \(A\) may initially share arbitrary correlations with the external reference \(R\), including quantum entanglement. After the input system \(A\) passes through the quantum instrument, the instrument produces its native classical outcome \(a\) and a quantum output system \(B\).

To retain both the classical outcome and the quantum output of the instrument, we introduce a classical register \(C\) and define the corresponding classical--quantum channel (cq channel)~\cite{choi1975,watrous2018}
\begin{equation}
\widetilde{\mathcal E}^{A\to CB}(X)
=\sum_a |a\rangle\langle a|^C\otimes\mathcal E_a(X).
\label{eq:associated-cq-channel}
\end{equation}
Furthermore, to describe this process, we fix a minimal Stinespring realization of the associated channel \(\widetilde{\mathcal E}\)~\cite{watrous2018,stinespring1955}, \(V:A\longrightarrow C\otimes B\otimes E\), such that
\begin{equation}
\widetilde{\mathcal E}(X)=\operatorname{Tr}_E\!\left[VXV^\dagger\right].
\label{eq:minimal-stinespring-realization}
\end{equation}
Here, \(C\) records the native classical outcome of the instrument, \(B\) denotes its quantum output, and the environment system \(E\) contains the quantum degrees of freedom from a single run that are not retained in the public output \(CB\).

Based on the associated channel and its Stinespring realization introduced above, we consider a scenario in which an environmental assistant, Eve, is allowed to access the environment system \(E\) after the instrument has acted. Eve's task is to measure the environment and, through postselection, extract a subprocess whose retained public output carries no information about the input.

To ensure that the extracted subprocess is physically contained in the original quantum instrument rather than constructed through additional operations, we restrict Eve's admissible interventions. Specifically, Eve is allowed only to measure the environment system \(E\) and perform postselection based on the measurement outcome. She is not allowed to alter the unknown input or apply additional processing to the original public output \(CB\). Moreover, for a given quantum instrument, Eve's environmental measurement effect is chosen in advance, independently of the native classical outcome \(a\) in each run. She has no additional access to the classical register \(C\) or separate readout of \(a\), although she may exploit information about \(a\) contained in \(E\). In this model, any explicit readout of \(C\) occurs only after the environmental success decision.

Specifically, Eve performs a binary POVM on the environment system \(E\), \(\{M_{\mathrm{succ}}^E,I_E-M_{\mathrm{succ}}^E\}\), where \(0\preceq M_{\mathrm{succ}}^E\preceq I_E\). For the chosen environmental measurement effect \(M_{\mathrm{succ}}^E\), we denote the completely positive (CP) submap induced on the public output \(CB\) by \(\mathcal S_M^{A\to CB}\), where the subscript \(M\) indicates its dependence on the chosen environmental effect. Explicitly,
\begin{equation}
\mathcal S_M^{A\to CB}(X)
:=\operatorname{Tr}_E\Big[
\left(I_{CB}\otimes\sqrt{M_{\mathrm{succ}}^E}\right)VXV^\dagger
\left(I_{CB}\otimes\sqrt{M_{\mathrm{succ}}^E}\right)\Big].
\label{eq:environmental-success-submap}
\end{equation}
In this binary measurement, the outcome associated with \(M_{\mathrm{succ}}^E\) is designated as a successful event, and the original classical outcome \(a\) and quantum output \(B\) of that run are retained. The complementary outcome associated with \(I_E-M_{\mathrm{succ}}^E\) is designated as a failure, and the corresponding run is discarded. This postselection procedure does not involve any additional measurement or dephasing of the classical register \(C\).

To ensure that the retained subprocess not only has a success probability independent of the input state but also produces a joint classical--quantum public output containing no information about the original input, we impose an additional structural condition on the CP submap defined in Eq.~\eqref{eq:environmental-success-submap}. Specifically, a successful event is called fully input-blind if there exists a family of positive operators \(\{\kappa_a^B\}_a\) such that
\begin{equation}
\mathcal S_M(X)
=\sum_a |a\rangle\langle a|^C\otimes\kappa_a^B\operatorname{Tr}X,
\qquad\forall X.
\label{eq:fully-input-blind-condition}
\end{equation}
For Eq.~\eqref{eq:fully-input-blind-condition}, since every normalized input state satisfies \(\operatorname{Tr}\rho=1\), the output of the successful subprocess is determined solely by the fixed family of positive operators \(\{\kappa_a^B\}_a\). Consequently, both the joint probability of success and each native classical outcome \(a\), and the corresponding unnormalized quantum output \(\kappa_a^B\), are independent of the input state. Thus, the entire retained public output \(CB\) carries no information about the original input. It is important to emphasize that the fully input-blind condition is imposed only on the successful public submap \(\mathcal S_M^{A\to CB}\), rather than on the environment or the original instrument as a whole.

Moreover, the resulting replacement map satisfies a complete-decoupling property. For any reference system \(R\) and any joint input state \(\rho^{RA}\),
\begin{equation}
(\operatorname{id}_R\otimes\mathcal S_M)(\rho^{RA})
=\operatorname{Tr}_A\rho^{RA}\otimes
\sum_a |a\rangle\langle a|^C\otimes\kappa_a^B.
\label{eq:input-blind-reference-decoupling}
\end{equation}
Thus, even if the input system \(A\) initially shares arbitrary correlations with the reference system \(R\), including quantum entanglement, the public output \(CB\) of the successful subprocess remains completely decoupled from \(R\), and the initial correlations are not transmitted to the public output. A detailed proof of this complete-decoupling property is provided in \hyperref[app:E]{Appendix~E}.

To further establish that the CP subprocesses considered above are not merely abstract mathematical descriptions but can be physically implemented by measuring the original environment, we establish a correspondence between CP submaps of the original quantum instrument and environmental measurement effects. We use a standard structural property of a minimal Stinespring representation. For the fixed minimal realization \(V\) introduced above, every CP submap satisfying
\begin{equation}
0\preceq_{\mathrm{CP}}\mathcal S\preceq_{\mathrm{CP}}\widetilde{\mathcal E}
\label{eq:cq-submap-cp-order}
\end{equation}
can be implemented by an effect \(0\preceq M_{\mathrm{succ}}^E\preceq I_E\) on the same environment system \(E\). Conversely, every such environmental effect induces a CP submap satisfying Eq.~\eqref{eq:cq-submap-cp-order}. This is precisely the CP Radon--Nikodym correspondence for a minimal Stinespring representation~\cite{belavkin1986,raginsky2003}. This correspondence ensures that subprocesses satisfying the corresponding CP-domination condition can be realized through environmental measurements without additional processing of the original public output. A complete proof of this correspondence, together with the role played by minimality, is provided in \hyperref[app:E]{Appendix~E}.

Next, we further quantify the extraction performance of fully input-blind subprocesses by considering the probability of a successful event. For any normalized input state \(\rho\), let \(p(M,a|\rho)\) denote the joint probability of obtaining the successful environmental measurement outcome associated with \(M_{\mathrm{succ}}^E\) and the instrument's native classical outcome \(a\). By Eq.~\eqref{eq:fully-input-blind-condition}, this probability is given by \(p(M,a|\rho) =\operatorname{Tr}\left[ \left(|a\rangle\langle a|^C\otimes I_B\right)\mathcal S_M(\rho)\right] =\operatorname{Tr}\kappa_a^B.\) Summing over all classical outcomes \(a\), the total success probability, denoted by \(p_M\), is \(p_M=\sum_a p(M,a|\rho) =\sum_a\operatorname{Tr}\kappa_a^B.\) Thus, under the fully input-blind condition, the total success probability is also independent of the unknown input state.

\begin{figure}[!htb]
\centering
\ManuscriptFigure{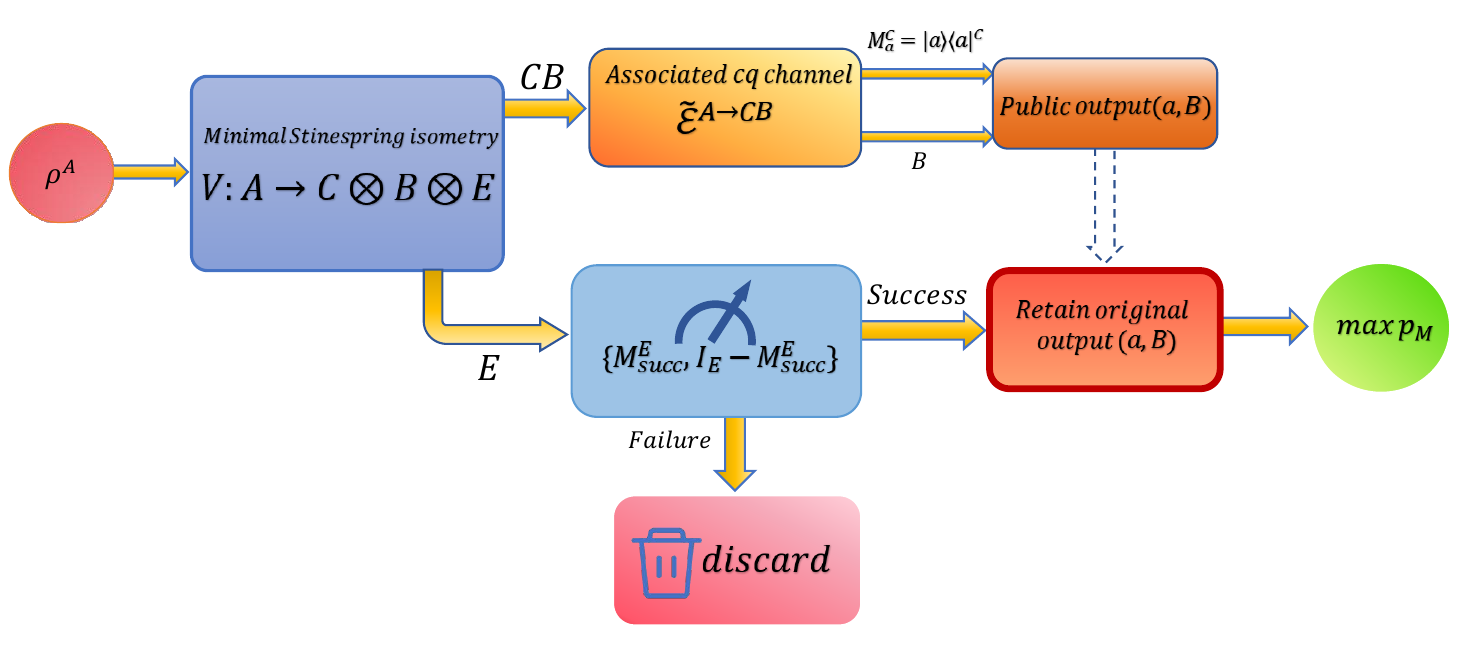}{13cm}
\caption{\textbf{Schematic of environment-assisted input-blind extraction.} An input quantum state \(\rho^A\) passes through a minimal Stinespring isometry \(V:A\to C\otimes B\otimes E\), producing the public output \(CB\) and the environment system \(E\). The public output \(CB\) is described by the associated classical--quantum channel \(\widetilde{\mathcal E}^{A\to CB}(X)=\sum_a |a\rangle\langle a|^C\otimes\mathcal E_a(X)\), where \(C\) is the register that records the instrument's native classical outcome \(a\), which can be read out after successful environmental postselection through a projective measurement \(M_a^C=|a\rangle\langle a|^C\) on \(C\), while \(B\) is the corresponding quantum output system, which can be retained without additional measurement. Eve performs a predetermined binary positive operator-valued measurement (POVM) \(\{M_{\mathrm{succ}}^E,I_E-M_{\mathrm{succ}}^E\}\) only on the environment system \(E\) and postselects based on the measurement outcome. Upon success, the original public output \((a,B)\) associated with the classical--quantum channel is retained without additional processing; otherwise, the run is discarded. The environmental effect is chosen such that the induced successful submap \(\mathcal S_M\) is fully input-blind. Optimizing over all admissible environmental measurement effects satisfying this condition yields the maximum success probability \(P_{\mathrm{blind}}(\bm{\mathcal E})=\max p_M\).}
\label{fig:blind-extraction}
\end{figure}

To determine the optimal success probability for extracting a fully input-blind subprocess from the original quantum instrument under full environmental access, we optimize over all environmental measurement effects satisfying the above conditions and their corresponding input-blind submaps. We define
\begin{equation}
P_{\mathrm{blind}}(\bm{\mathcal E})
:=\sup\!\Bigl\{\sum_a\operatorname{Tr}\kappa_a^B\;\Big|\;
0\preceq M_{\mathrm{succ}}^E\preceq I_E,\;
\mathcal S_M(X)=\sum_a|a\rangle\langle a|^C\otimes\kappa_a^B\operatorname{Tr}X,\;\forall X,\;
\kappa_a^B\succeq0,\;\forall a\Bigr\}.
\label{eq:full-access-extraction-probability}
\end{equation}
The environment-assisted postselection procedure is illustrated in Fig.~\ref{fig:blind-extraction}.

This leads to the following exact operational characterization.

\phantomsection\label{prop:6}\noindent \textbf{Proposition 6.} For any finite-dimensional quantum instrument \(\bm{\mathcal E}\), if the environmental assistant has access to the minimal Stinespring environment of the complete cq channel \(\widetilde{\mathcal E}\), then

\begin{equation}
P_{\mathrm{blind}}(\bm{\mathcal E})
=
1-W_{\mathrm{int}}(\bm{\mathcal E}).
\label{eq:full-access-extraction-equality}
\end{equation}
Moreover, the optimal success probability is attained by an actual environmental effect \(M_{\mathrm{succ}}^{E,\star}\).

\noindent \textbf{Proof.} The proof proceeds in two directions.

First, consider an arbitrary feasible environmental effect \(M_{\mathrm{succ}}^E\). The fully input-blind condition requires the \(a\)-th successful branch to have the replacement form \(X\mapsto\kappa_a\operatorname{Tr}X,\) whose Choi block is \(\frac{I_{A'}}{d_A}\otimes\kappa_a.\) Since the successful subprocess is a CP submap of the original cq channel, \(0 \preceq_{\mathrm{CP}} \mathcal S_M \preceq_{\mathrm{CP}} \widetilde{\mathcal E},\) the branchwise relation

\begin{equation}
\frac{I_{A'}}{d_A}
\otimes
\kappa_a
\preceq
J_a^{\bm{\mathcal E}}
\label{eq:extraction-branch-feasibility}
\end{equation}
must hold. Thus, \(\{\kappa_a\}_a\) is feasible for the primal SDP of the interactive weight of quantum instruments. Consequently,

\begin{equation}
p_M
=
\sum_a
\operatorname{Tr}\kappa_a
\le
1-W_{\mathrm{int}}(\bm{\mathcal E}).
\label{eq:extraction-success-upper-bound}
\end{equation}
Conversely, the optimum of the primal SDP is attained. Let \(\{\kappa_a^\star\}_a\) be an optimal family and define

\begin{equation}
\mathcal S^\star(X)
=
\sum_a
|a\rangle\langle a|^C
\otimes
\kappa_a^\star
\operatorname{Tr}X.
\label{eq:optimal-input-blind-submap}
\end{equation}
Since \(\kappa_a^\star\succeq0\), the map \(\mathcal S^\star\) is completely positive. The branchwise constraints of the primal SDP further guarantee \(0 \preceq_{\mathrm{CP}} \mathcal S^\star \preceq_{\mathrm{CP}} \widetilde{\mathcal E}.\) By the CP Radon--Nikodym correspondence on the minimal Stinespring environment, there exists an environmental effect \(M_{\mathrm{succ}}^{E,\star}\) that realizes \(\mathcal S^\star\) exactly. Hence,

\begin{equation}
P_{\mathrm{blind}}(\bm{\mathcal E})
\ge
\sum_a
\operatorname{Tr}\kappa_a^\star
=
1-W_{\mathrm{int}}(\bm{\mathcal E}).
\label{eq:extraction-success-lower-bound}
\end{equation}
Combining \mbox{Eqs.~\eqref{eq:extraction-success-upper-bound}} and \eqref{eq:extraction-success-lower-bound} gives \mbox{Eq.~\eqref{eq:full-access-extraction-equality}}. The construction of \(M_{\mathrm{succ}}^{E,\star}\) also shows that the optimal success probability is attained by an actual environmental effect rather than merely approached as a supremum. \hfill\ensuremath{\square}

Complete proofs of the decoupling between an input-blind replacement map and an arbitrary reference system, together with the CP Radon--Nikodym correspondence and the construction of the corresponding environmental effect on the minimal Stinespring environment, are provided in \hyperref[app:E]{Appendix~E}.

Although the success probability \(p_M\) depends on the chosen environmental effect, \(P_{\mathrm{blind}}(\bm{\mathcal E})\) is optimized over all effects inducing fully input-blind submaps. By the CP Radon--Nikodym correspondence, this optimization is equivalent to the primal SDP for the maximal non-interactive content, so its optimal value depends only on the instrument, not on any particular environmental measurement.

Equation \eqref{eq:full-access-extraction-equality} provides a direct probabilistic interpretation of the interactive weight of quantum instruments. For the fixed minimal Stinespring realization, when the assistant can measure the complete environment without modifying the original public output \(CB\), \(1-W_{\mathrm{int}}(\bm{\mathcal E})\) is exactly the maximum success probability for extracting a fully input-blind subprocess from within the instrument. Correspondingly, \(W_{\mathrm{int}}(\bm{\mathcal E})\) is the minimum residual fraction that cannot be interpreted as a fully input-blind subprocess through any such environmental postselection.

It is important to emphasize that the operators \(\kappa_a\) are not artificial optimization parameters introduced into the experiment. They are the unnormalized output operators that necessarily characterize any fully input-blind successful branch. The corresponding map must take the form \(X\mapsto\kappa_a\operatorname{Tr}X\), and its Choi block must therefore be \(I_{A'}/d_A\otimes\kappa_a\). The exact correspondence between the maximal success probability of the environmental postselection task and the maximal non-interactive content of the instrument follows from the fact that this structure coincides precisely with the branchwise constraints of the primal SDP.


\phantomsection\label{rem:1}\noindent \textbf{Remark 1.} In this channel representation of \mbox{Eq.~(\ref{eq:associated-cq-channel})}, the maximal non-interactive content can be directly related to the positive quantum Doeblin coefficient. For a quantum channel \(\mathcal M\), this coefficient is defined as~\cite{george2026doeblin}
\begin{equation}
\alpha_+(\mathcal M)
:=\sup_{X\succeq0}\left\{\operatorname{Tr}X:
I_{A'}\otimes X\preceq\Gamma^{\mathcal M}\right\},
\label{eq:doeblin-positive}
\end{equation}
where \(\Gamma^{\mathcal M}:=d_AJ^{\mathcal M}\) denotes the unnormalized Choi operator of \(\mathcal M\). Here, the positivity constraint \(X\succeq0\) distinguishes \(\alpha_+\) from the quantum Doeblin coefficient \(\alpha\), which allows optimization over arbitrary Hermitian operators \(X\).
For the associated channel \(\widetilde{\mathcal E}\), its Choi operator is block diagonal in the classical register \(C\). Pinching on \(C\) therefore allows the optimization to be restricted, without changing feasibility or the objective value, to operators of the form \(X^{CB}=\sum_a|a\rangle\langle a|^C\otimes\kappa_a^B\). The Doeblin constraint then reduces to \(\frac{I_{A'}}{d_A}\otimes\kappa_a^B\preceq J_a^{\bm{\mathcal E}}\) for all \(a\), which has the same mathematical form as the primal optimization defining the maximal non-interactive content. Consequently,
\begin{equation}
\lambda_{\mathbf{NI}}(\bm{\mathcal E})
=
\alpha_+\!\left(\widetilde{\mathcal E}\right).
\label{eq:doeblin-instrument}
\end{equation}
Thus, when the instrument is represented by the associated channel retaining its native classical outcome, \(\lambda_{\mathbf{NI}}\) and \(\alpha_+\) share the same positive-replacement optimization structure.

\phantomsection\label{rem:2}\noindent \textbf{Remark 2.}
\hyperref[prop:6]{Proposition~6} considers the case in which the environmental assistant has access to the complete minimal Stinespring environment. In a physical device, however, the assistant may have access to only partial environmental information because some environmental degrees of freedom are inaccessible or environmental information is lost.

To describe this situation, fix the Stinespring realization \(V\) of the quantum instrument and introduce a completely positive and trace-preserving (CPTP) environmental access channel \(\mathcal A^{E\to Z}\), where \(Z\) denotes the system actually accessible to the assistant. The assistant can measure only system \(Z\), and a success effect \(M_{\mathrm{succ}}^Z\) corresponds to the effective measurement effect on the original environment
\begin{equation}
M_{\mathrm{eff}}^E=\mathcal A^\dagger(M_{\mathrm{succ}}^Z),
\label{eq:restricted-effective-effect}
\end{equation}
where \(\mathcal A^\dagger\) denotes the adjoint of the access channel \(\mathcal A\).

Let \(P_{\mathrm{blind}}(\bm{\mathcal E};V,\mathcal A)\) denote the maximum probability of extracting a fully input-blind subprocess under this access restriction. Since restricted environmental access permits only a subset of the strategies available with full environmental access, one has
\begin{equation}
0\le P_{\mathrm{blind}}(\bm{\mathcal E};V,\mathcal A)
\le\lambda_{\mathbf{NI}}(\bm{\mathcal E})
=1-W_{\mathrm{int}}(\bm{\mathcal E}).
\label{eq:restricted-extraction-bound}
\end{equation}
Moreover, if two environmental access channels satisfy
\begin{equation}
\mathcal A_2=\mathcal Q\circ\mathcal A_1,
\label{eq:access-degradation}
\end{equation}
where \(\mathcal Q\) is CPTP, then the following access data-processing inequality holds:
\begin{equation}
\begin{aligned}
P_{\mathrm{blind}}(\bm{\mathcal E};V,\mathcal A_2)
\le P_{\mathrm{blind}}(\bm{\mathcal E};V,\mathcal A_1).
\end{aligned}
\label{eq:access-data-processing}
\end{equation}
These results show that the maximal non-interactive content characterizes the non-interactive component intrinsically contained in a quantum instrument, whereas the actual extraction probability under restricted environmental access also depends on the accessibility of environmental information. Further restricting environmental access cannot increase the maximum probability of extracting a fully input-blind subprocess. The relevant definitions, semidefinite programming formulation, and detailed proofs of these inequalities are given in \hyperref[app:restricted-access]{Appendix~\ref*{app:restricted-access}}.

\section{RESOURCE THEORY UNDER ALLOWED OPERATIONS AND RESOURCE CONVERSION}\label{sec:resource-conversion}
\subsection{Allowed operations and monotonicity}
Having established the intrinsic structural properties and operational interpretations of the interactive weight of quantum instruments, we now investigate its behavior under free processing and the resulting constraints on resource conversion. We adopt the allowed-operations framework introduced in Ref.~\cite{hsieh2026}. A deterministic allowed operation \(\Pi\in\mathrm{AO}\) consists of quantum preprocessing, classical processing of the native outcome of the instrument, and conditional quantum postprocessing. Its action can be written as
\begin{equation}
[\Pi(\bm{\mathcal E})]_a
=
\sum_{\lambda,k}
q(\lambda)\,
p(a|k,\lambda)\,
\mathcal D_{k|\lambda}
\circ
\mathcal E_k
\circ
\mathcal P_\lambda .
\label{eq:deterministic-allowed-operation}
\end{equation}
Here, \(\mathcal P_\lambda\) and \(\mathcal D_{k|\lambda}\) are CPTP channels, \(q(\lambda)\) is a probability distribution, and \(p(a|k,\lambda)\) is a conditional probability distribution. As shown in Ref.~\cite{hsieh2026}, these operations map valid quantum instruments to valid quantum instruments and preserve the free set of non-interactive instruments,
\begin{equation}
\Pi(\bm{\mathcal L})\in\mathbf{NI},
\qquad
\forall\,\bm{\mathcal L}\in\mathbf{NI},
\quad
\Pi\in\mathrm{AO}.
\label{eq:allowed-operation-free-preservation}
\end{equation}
Thus, allowed operations cannot generate instrument interactivity from non-interactive instruments.

\phantomsection\label{prop:7}\noindent \textbf{Proposition 7.} For every finite-dimensional quantum instrument \(\bm{\mathcal E}\) and every deterministic allowed operation \(\Pi\in\mathrm{AO}\),
\begin{equation}
W_{\mathrm{int}}[\Pi(\bm{\mathcal E})]
\le
W_{\mathrm{int}}(\bm{\mathcal E}).
\label{eq:weight-ao-monotonicity}
\end{equation}
Equivalently, the maximal non-interactive content satisfies
\begin{equation}
\lambda_{\mathbf{NI}}[\Pi(\bm{\mathcal E})]
\ge
\lambda_{\mathbf{NI}}(\bm{\mathcal E}).
\label{eq:ni-content-ao-monotonicity}
\end{equation}
\noindent \textbf{Proof.} Let \(w:=W_{\mathrm{int}}(\bm{\mathcal E})\), and choose an optimal weight decomposition \(\bm{\mathcal E}=(1-w)\bm{\mathcal L}+w\bm{\mathcal G}\), with \(\bm{\mathcal L}\in\mathbf{NI}\) and \(\bm{\mathcal G}\) a valid quantum instrument. By linearity of \(\Pi\),
\begin{equation}
\Pi(\bm{\mathcal E})
=
(1-w)\Pi(\bm{\mathcal L})
+
w\Pi(\bm{\mathcal G}).
\label{eq:processed-weight-decomposition}
\end{equation}
Since allowed operations preserve the free set, \(\Pi(\bm{\mathcal L})\in\mathbf{NI}\), while \(\Pi(\bm{\mathcal G})\) remains a valid quantum instrument. Hence, \mbox{Eq.~\eqref{eq:processed-weight-decomposition}} is a valid weight decomposition of \(\Pi(\bm{\mathcal E})\) with resource coefficient \(w\). By the minimizing definition of the interactive weight of quantum instruments, \(W_{\mathrm{int}}[\Pi(\bm{\mathcal E})]\le w=W_{\mathrm{int}}(\bm{\mathcal E})\), which proves \mbox{Eq.~\eqref{eq:weight-ao-monotonicity}}. Equation~\eqref{eq:ni-content-ao-monotonicity} follows immediately from \(\lambda_{\mathbf{NI}}=1-W_{\mathrm{int}}\). \hfill\ensuremath{\square}

This result establishes the interactive weight of quantum instruments as a resource monotone under the adopted class of allowed operations. Physically, deterministic free processing consisting only of quantum preprocessing, classical processing of the native outcome, and conditional quantum postprocessing cannot increase the minimum fraction of an instrument that cannot be explained by non-interactive components. Equivalently, such free processing cannot decrease the maximal non-interactive content of the instrument. The interactive weight of quantum instruments therefore respects the resource ordering induced by deterministic allowed operations.

As an immediate consequence of this monotonicity, the interactive weight of quantum instruments is invariant under reversible deterministic allowed operations.

\phantomsection\label{cor:2}\noindent \textbf{Corollary 2.} If there exist \(\Pi_1,\Pi_2\in\mathrm{AO}\) such that
\begin{equation}
\Pi_1(\bm{\mathcal E})=\bm{\mathcal N},
\qquad
\Pi_2(\bm{\mathcal N})=\bm{\mathcal E},
\label{eq:reversible-allowed-operations}
\end{equation}
then
\begin{equation}
W_{\mathrm{int}}(\bm{\mathcal E})
=
W_{\mathrm{int}}(\bm{\mathcal N}).
\label{eq:reversible-weight-invariance}
\end{equation}
Indeed, applying \mbox{Eq.~\eqref{eq:weight-ao-monotonicity}} separately to \(\Pi_1\) and \(\Pi_2\) gives both inequalities and therefore equality.

\subsection{Logarithmic interactivity measure and multicopy resource conversion}
The exact multiplicativity established in \hyperref[prop:2]{Proposition~2} implies that the maximal non-interactive content factorizes under independent parallel composition. This multiplicative structure can be converted into an additive one by defining the logarithmic interactivity measure
\begin{equation}
I_{\mathrm{int}}(\bm{\mathcal E})
:=
-\log_2\lambda_{\mathbf{NI}}(\bm{\mathcal E})
=
-\log_2\bigl[1-W_{\mathrm{int}}(\bm{\mathcal E})\bigr],
\label{eq:logarithmic-interactivity-definition}
\end{equation}
with the convention \(-\log_2 0:=+\infty\). The deterministic-AO monotonicity in \mbox{Eq.~\eqref{eq:ni-content-ao-monotonicity}} and exact multiplicativity immediately imply
\begin{equation}
I_{\mathrm{int}}[\Pi(\bm{\mathcal E})]
\le
I_{\mathrm{int}}(\bm{\mathcal E}),
\label{eq:logarithmic-ao-monotonicity}
\end{equation}
and
\begin{equation}
I_{\mathrm{int}}(\bm{\mathcal E}\otimes\bm{\mathcal F})
=
I_{\mathrm{int}}(\bm{\mathcal E})
+
I_{\mathrm{int}}(\bm{\mathcal F}).
\label{eq:logarithmic-additivity}
\end{equation}
Similarly, by writing the tensor product of \(n\) copies as \(\bm{\mathcal E}^{\otimes n}=\bm{\mathcal E}^{\otimes(n-1)}\otimes\bm{\mathcal E}\) and repeatedly applying the additivity relation in Eq.~\eqref{eq:logarithmic-additivity}, this relation extends to any positive integer \(n\), yielding
\begingroup
\begin{equation}
I_{\mathrm{int}}(\bm{\mathcal E}^{\otimes n})
=
nI_{\mathrm{int}}(\bm{\mathcal E}).
\label{eq:logarithmic-tensor-power}
\end{equation}
\endgroup
Consequently, \(I_{\mathrm{int}}\) is a strictly additive resource monotone under deterministic allowed operations, making it directly suitable for constraining multicopy resource conversion.

\phantomsection\label{prop:8}\noindent \textbf{Proposition 8.} Let \(n\) and \(m\) be arbitrary positive integers. If a deterministic allowed operation \(\Pi\in\mathrm{AO}\) realizes the exact conversion
\begin{equation}
\Pi\!\left(\bm{\mathcal E}^{\otimes n}\right)
=
\bm{\mathcal N}^{\otimes m},
\label{eq:exact-multicopy-conversion}
\end{equation}
then
\begin{equation}
\bigl[\lambda_{\mathbf{NI}}(\bm{\mathcal E})\bigr]^n
\le
\bigl[\lambda_{\mathbf{NI}}(\bm{\mathcal N})\bigr]^m.
\label{eq:multicopy-ni-content-bound}
\end{equation}
Equivalently,
\begin{equation}
nI_{\mathrm{int}}(\bm{\mathcal E})
\ge
mI_{\mathrm{int}}(\bm{\mathcal N}).
\label{eq:multicopy-logarithmic-bound}
\end{equation}
When
\begin{equation}
0<\lambda_{\mathbf{NI}}(\bm{\mathcal E})<1,
\qquad
0<\lambda_{\mathbf{NI}}(\bm{\mathcal N})<1,
\label{eq:finite-positive-conversion-regime}
\end{equation}
every exact deterministic multicopy conversion must satisfy
\begin{equation}
\frac{m}{n}
\le
\frac{I_{\mathrm{int}}(\bm{\mathcal E})}{I_{\mathrm{int}}(\bm{\mathcal N})}.
\label{eq:exact-conversion-rate-bound}
\end{equation}
Furthermore, if \(W_{\mathrm{int}}(\bm{\mathcal E})<1\) and \(W_{\mathrm{int}}(\bm{\mathcal N})=1\), then no deterministic allowed operation can realize
\begin{equation}
\bm{\mathcal E}^{\otimes n}
\longrightarrow
\bm{\mathcal N}^{\otimes m}
\label{eq:finite-copy-distillation-conversion}
\end{equation}
for any finite positive integers \(n\) and \(m\).

The physical meaning of this result is naturally expressed as a multicopy resource-budget constraint. Because \(I_{\mathrm{int}}\) is strictly additive, \(n\) copies of the source instrument have a total value \(nI_{\mathrm{int}}(\bm{\mathcal E})\) of the logarithmic interactivity measure, whereas \(m\) copies of the target have a total value \(mI_{\mathrm{int}}(\bm{\mathcal N})\) of the same measure. Deterministic allowed operations cannot increase this total resource. Hence, an exact conversion is possible only if the source provides at least as large a value of the logarithmic interactivity measure as required by the target.

The maximal-weight case is particularly distinctive. An instrument satisfying \(W_{\mathrm{int}}=1\) has \(\lambda_{\mathbf{NI}}=0\), and therefore contains no nonzero non-interactive component. By contrast, every source instrument satisfying \(W_{\mathrm{int}}<1\) retains a strictly positive non-interactive content. Exact multiplicativity implies that this content remains strictly positive for every finite number of copies. Consequently, finite-copy deterministic processing cannot eliminate it exactly, which yields the finite-copy maximal-weight distillation no-go result above.

\noindent \textbf{Proof.} If \mbox{Eq.~\eqref{eq:exact-multicopy-conversion}} is achievable, monotonicity of the maximal non-interactive content and exact multiplicativity give
\begin{equation}
\bigl[\lambda_{\mathbf{NI}}(\bm{\mathcal N})\bigr]^m
=
\lambda_{\mathbf{NI}}\!\left[\Pi\!\left(\bm{\mathcal E}^{\otimes n}\right)\right]
\ge
\lambda_{\mathbf{NI}}(\bm{\mathcal E}^{\otimes n})
=
\bigl[\lambda_{\mathbf{NI}}(\bm{\mathcal E})\bigr]^n,
\label{eq:conversion-short-proof}
\end{equation}
which proves \mbox{Eq.~\eqref{eq:multicopy-ni-content-bound}}. Applying \(-\log_2\) gives \mbox{Eq.~\eqref{eq:multicopy-logarithmic-bound}}, with the endpoint cases covered by the convention \(-\log_2 0=+\infty\); under the conditions in \mbox{Eq.~\eqref{eq:finite-positive-conversion-regime}}, division by the strictly positive \(I_{\mathrm{int}}(\bm{\mathcal N})\) yields \mbox{Eq.~\eqref{eq:exact-conversion-rate-bound}}. Finally, if \(W_{\mathrm{int}}(\bm{\mathcal E})<1\), then \(\lambda_{\mathbf{NI}}(\bm{\mathcal E})>0\), so \([\lambda_{\mathbf{NI}}(\bm{\mathcal E})]^n>0\) for every finite \(n\). If \(W_{\mathrm{int}}(\bm{\mathcal N})=1\), then \(\lambda_{\mathbf{NI}}(\bm{\mathcal N})=0\) and hence \([\lambda_{\mathbf{NI}}(\bm{\mathcal N})]^m=0\). This contradicts the necessary condition in \mbox{Eq.~\eqref{eq:multicopy-ni-content-bound}}, so the exact deterministic conversion in \mbox{Eq.~\eqref{eq:finite-copy-distillation-conversion}} is impossible for any finite positive integers \(n\) and \(m\). \hfill\ensuremath{\square}

The bound in \mbox{Eq.~\eqref{eq:exact-conversion-rate-bound}} is a necessary condition for exact convertibility and is not, in general, sufficient. Likewise, the finite-copy maximal-weight distillation no-go result concerns finite-copy exact conversion only; it does not exclude approximate transformations or asymptotic conversion in the infinite-copy limit.

\section{CONCLUSION}\label{sec:conclusion}
This work introduces a weight-based approach to quantifying the resource of interactive quantum instruments. The interactive weight of quantum instruments and its complementary quantity are introduced to quantify the degree of instrument interactivity. Using the Choi operators of quantum instruments, the evaluation of the interactive weight of quantum instruments is further formulated as a semidefinite programming (SDP) problem. We then investigate the basic properties of the interactive weight of quantum instruments, including faithfulness, convexity, and monotonicity under deterministic allowed operations, and prove the exact multiplicativity of the maximal non-interactive content under parallel composition. These results yield necessary constraints on exact deterministic multicopy resource conversion. In particular, under deterministic allowed operations, a finite number of copies of an instrument with a nonmaximal value of the interactive weight of quantum instruments cannot be exactly distilled into an instrument with a maximal value of the interactive weight of quantum instruments.

Regarding operational interpretations, the tasks of maximally entangled-state recovery and Haar-averaged pure-state recovery yield quantitative lower bounds on the interactive weight of quantum instruments and a universal inequality relating it to the interactive instrument robustness~\cite{hsieh2026}. By introducing positive linear loss games, we further obtain an exact operational characterization of the maximal non-interactive content. Moreover, with access to the complete minimal Stinespring environment, we prove that the maximal non-interactive content equals the maximum success probability for extracting a fully input-blind subprocess, revealing its physical meaning in environment-assisted subprocess extraction. We also establish an exact correspondence between the maximal non-interactive content and the positive quantum Doeblin coefficient~\cite{george2026doeblin}, and extend environment-assisted input-blind extraction to restricted environmental access, obtaining a universal upper bound on the extraction probability and an access data-processing inequality.

Since the present framework for restricted environmental access mainly concerns exact environment-assisted input-blind extraction in a single use, its operational properties in more general settings remain to be investigated. A natural open question is whether environment-assisted input-blind extraction under restricted environmental access can be extended to approximate and multicopy settings, and how different environmental access restrictions affect the optimal extraction performance. In addition, the attainability conditions for the recovery bounds, as well as the resource-conversion laws for probabilistic, approximate, and asymptotic multicopy settings, remain topics for further investigation. Overall, the results provide a theoretical basis for a deeper understanding of the operational meanings and resource-conversion properties of interactive quantum instruments.

\subsection*{AI statement}
The ideas, methodology, proof strategies and the project scope were designed by the authors. The numerical results and proof calculations were assisted with AI (OpenAI's ChatGPT Pro, GPT-6 Astra), and validated by the authors. All figures were drawn by the authors. The article was entirely written by the authors, with minor refinements with AI to correct spelling and condense some of the sentences.

\subsection*{Acknowledgments}
This research was funded by the Young Scientists Fund of the Natural Science Foundation of Jiangsu Province (No. BK20240536) and the National Natural Science Foundation of China (No. 62502203).

\renewcommand{\refname}{REFERENCES}

\par
\appendix
\renewcommand{\thesection}{\Alph{section}}
\renewcommand{\thesubsection}{\thesection.\arabic{subsection}}
\titleformat{\section}{\centering\bfseries\fontsize{11}{12.5}\selectfont}{Appendix~\thesection.}{0.5em}{}
\titleformat{\subsection}{\bfseries\normalsize}{\thesubsection}{0.45em}{}
\renewcommand{\theequation}{\thesection.\arabic{equation}}
\section{SDP and basic properties of the interactive weight of quantum instruments}\label{app:A}
This appendix proves \hyperref[prop:1]{Proposition 1} and supplies the detailed arguments for the range, attainment, faithfulness, and convexity statements used in \hyperref[sec:structural]{Sec.~3.2}. The normalized Choi convention introduced in the main text is adopted throughout. First, consider an arbitrary non-interactive instrument \(\bm{\mathcal L}^{A\to B}\). Each branch can be written as \(\mathcal L_a^{A\to B}(\,\cdot\,) = r(a)\rho_a^B\operatorname{Tr}(\,\cdot\,),\) where \(r(a)\) is a probability distribution and \(\rho_a^B\) is a quantum state. Defining \(\tau_a^B:=r(a)\rho_a^B,\) gives \(\tau_a^B\succeq0,\allowbreak\; \sum_a\operatorname{Tr}\tau_a^B=1,\) and the corresponding Choi representation is

\begin{equation}
J_a^{\bm{\mathcal L}}
=
(\operatorname{id}_{A'}\otimes\mathcal L_a)(\Phi^+)
=
\frac{I_{A'}}{d}\otimes\tau_a^B .
\label{eq:ni-choi-product-proof}
\end{equation}
From the original convex-decomposition definition \(\bm{\mathcal E}=(1-s)\bm{\mathcal L}+s\bm{\mathcal G},\allowbreak\; \bm{\mathcal L}\in\mathbf{NI}\), each branch satisfies \(\mathcal E_a-(1-s)\mathcal L_a=s\mathcal G_a\succeq_{\mathrm{CP}}0\). Hence, the decomposition condition is equivalently expressed as \(\mathcal E_a\succeq_{\mathrm{CP}}(1-s)\mathcal L_a\) for all \(a\). By the Choi--Jamiołkowski isomorphism~\cite{choi1975,jamiolkowski1972,watrous2018}, \(\mathcal E_a\succeq_{\mathrm{CP}}(1-s)\mathcal L_a\Longleftrightarrow J_a\succeq(1-s)J_a^{\bm{\mathcal L}}\). Using \mbox{Eq.~\eqref{eq:ni-choi-product-proof}}, this becomes
\begin{equation}
J_a\succeq(1-s)\frac{I_{A'}}{d}\otimes\tau_a.
\label{eq:weight-choi-domination}
\end{equation}
The interactive weight of quantum instruments can therefore be written as

\begin{equation}
\begin{aligned}
W_{\mathrm{int}}(\bm{\mathcal E})
=
\min_{s,\{\tau_a\}}
\quad & s\\
\text{s.t.}\quad
&
J_a
-
(1-s)
\frac{I_{A'}}{d}
\otimes\tau_a
\succeq0,
\qquad
\forall a,
\\
&
\tau_a\succeq0,
\qquad
\forall a,
\\
&
\sum_a\operatorname{Tr}\tau_a=1,
\\
&
0\le s\le1.
\end{aligned}
\label{eq:weight-choi-optimization}
\end{equation}
Introduce the variable substitution \(\kappa_a := (1-s)\tau_a.\) Then

\begin{equation}
\kappa_a\succeq0,
\qquad
\sum_a\operatorname{Tr}\kappa_a
=
1-s,
\label{eq:scaled-free-operators}
\end{equation}
and \mbox{Eq.~\eqref{eq:weight-choi-domination}} becomes \(J_a - \frac{I_{A'}}{d} \otimes\kappa_a \succeq0.\) The problem can therefore be written in the standard SDP form

\begin{equation}
\begin{aligned}
W_{\mathrm{int}}(\bm{\mathcal E})
=
\min_{s,\{\kappa_a\}}
\quad & s\\
\text{s.t.}\quad
&
J_a
-
\frac{I_{A'}}{d}
\otimes\kappa_a
\succeq0,
\qquad
\forall a,
\\
&
\kappa_a\succeq0,
\qquad
\forall a,
\\
&
s+\sum_a\operatorname{Tr}\kappa_a=1.
\end{aligned}
\label{eq:weight-scalar-sdp}
\end{equation}
Taking the trace of the first positive-semidefinite constraint and summing over \(a\) gives \(\sum_a\operatorname{Tr}\kappa_a \le \sum_a\operatorname{Tr}J_a = 1.\) On the other hand, \(\kappa_a\succeq0\) implies \(\sum_a\operatorname{Tr}\kappa_a\ge0.\) Together with \mbox{Eq.~\eqref{eq:scaled-free-operators}}, \(s = 1- \sum_a\operatorname{Tr}\kappa_a \in[0,1].\) Hence, the explicit constraint \(0\le s\le1\) is redundant and need not be imposed separately in \mbox{Eq.~\eqref{eq:weight-scalar-sdp}}.

Eliminating \(s\) yields the primal SDP given in the main text:

\begin{equation}
W_{\mathrm{int}}(\bm{\mathcal E})
=
\min_{\{\kappa_a\succeq0\}}
\left[
1-
\sum_a\operatorname{Tr}\kappa_a
\right]
\label{eq:weight-primal-objective-proof}
\end{equation}
subject to

\begin{equation}
J_a
-
\frac{I_{A'}}{d}
\otimes\kappa_a
\succeq0,
\qquad
\forall a.
\label{eq:weight-primal-constraints-proof}
\end{equation}
Equivalently,

\begin{equation}
\lambda_{\mathbf{NI}}(\bm{\mathcal E})
=
1-W_{\mathrm{int}}(\bm{\mathcal E})
=
\max_{\{\kappa_a\succeq0\}}
\sum_a\operatorname{Tr}\kappa_a
\label{eq:ni-content-primal-objective-proof}
\end{equation}
subject to

\begin{equation}
\frac{I_{A'}}{d}
\otimes\kappa_a
\preceq
J_a,
\qquad
\forall a.
\label{eq:ni-content-primal-constraints-proof}
\end{equation}
The zero-operator family is always feasible. The trace bound above further gives \(0 \le \sum_a\operatorname{Tr}\kappa_a \le 1,\) so all \(\kappa_a\) are bounded. In finite dimensions, the primal feasible set is therefore closed and bounded, and hence compact. Consequently, the optimum of the primal problem is attained.

\noindent The Lagrange dual of \mbox{Eq.~\eqref{eq:weight-primal-objective-proof}} is derived next~\cite{boyd2004,skrzypczyk2023}. Positive-semidefinite multipliers \(\Omega_a^{A'B}\succeq0,\allowbreak\; Y_a^B\succeq0\) are introduced for the constraints \(J_a-\frac{I_{A'}}{d}\otimes\kappa_a\succeq0\) and \(\kappa_a\succeq0,\) respectively. The Lagrangian is
{\small
\begin{equation}
\begin{aligned}
L(\kappa,\Omega,Y)
&=1-\sum_a\operatorname{Tr}\kappa_a
-\sum_a\operatorname{Tr}\!\left[\Omega_a\left(J_a-\frac{I_{A'}}{d}\otimes\kappa_a\right)\right]
-\sum_a\operatorname{Tr}(Y_a\kappa_a)\\
&=1-\sum_a\operatorname{Tr}(\Omega_aJ_a)-\sum_a\operatorname{Tr}\kappa_a
+\frac{1}{d}\sum_a\operatorname{Tr}[\Omega_a(I_{A'}\otimes\kappa_a)]
-\sum_a\operatorname{Tr}(Y_a\kappa_a)\\
&=1-\sum_a\operatorname{Tr}(\Omega_aJ_a)-\sum_a\operatorname{Tr}\kappa_a
+\frac{1}{d}\sum_a\operatorname{Tr}[(\operatorname{Tr}_{A'}\Omega_a)\kappa_a]
-\sum_a\operatorname{Tr}(Y_a\kappa_a)\\
&=1-\sum_a\operatorname{Tr}(\Omega_aJ_a)
+\sum_a\operatorname{Tr}\!\left[\kappa_a\left(\frac{\operatorname{Tr}_{A'}\Omega_a}{d}-I_B-Y_a\right)\right].
\end{aligned}
\label{eq:weight-lagrangian}
\end{equation}
}
The dual function is obtained by taking the infimum over \(\{\kappa_a\}_a\). If, for some \(a\), \(\frac{\operatorname{Tr}_{A'}\Omega_a}{d} -I_B-Y_a \neq0,\) then \(\kappa_a\) can be scaled along a direction for which the trace pairing is negative, driving \(L(\kappa,\Omega,Y)\to-\infty.\) Thus, the dual function is finite if and only if

\begin{equation}
\frac{\operatorname{Tr}_{A'}\Omega_a}{d}
-I_B-Y_a
=
0,
\qquad
\forall a.
\label{eq:dual-stationarity}
\end{equation}
Under this condition, \(g(\Omega,Y) = 1- \sum_a \operatorname{Tr}(\Omega_aJ_a).\) Since \(Y_a\succeq0\), \mbox{Eq.~\eqref{eq:dual-stationarity}} is equivalent to \(\operatorname{Tr}_{A'}\Omega_a \succeq dI_B,\allowbreak\; \forall a.\) After eliminating the auxiliary variables \(Y_a\), the dual SDP becomes

\begin{equation}
W_{\mathrm{int}}(\bm{\mathcal E})
=
\sup_{\{\Omega_a\succeq0\}}
\left[
1-
\sum_a
\operatorname{Tr}(\Omega_aJ_a)
\right]
\label{eq:weight-dual-objective-proof}
\end{equation}
subject to

\begin{equation}
\operatorname{Tr}_{A'}\Omega_a
\succeq
dI_B,
\qquad
\forall a.
\label{eq:weight-dual-constraints-proof}
\end{equation}
Strong duality is established next. The primal problem is feasible because the choice \(\kappa_a=0\) for every \(a\) satisfies its constraints, and its optimal value is finite.

For the dual problem, fix any \(\epsilon>0\) and choose \(\Omega_a=(1+\epsilon)I_{A'B}\) and \(Y_a=\epsilon I_B\) for every \(a\). These operators are positive definite and satisfy \(\frac{\operatorname{Tr}_{A'}\Omega_a}{d}-I_B-Y_a=0.\) Equivalently, after eliminating the auxiliary variables \(Y_a\), \(\operatorname{Tr}_{A'}\Omega_a=(1+\epsilon)dI_B\succ dI_B.\) Thus, the dual problem has a strictly feasible point. By Slater's condition~\cite{boyd2004,skrzypczyk2023}, strong duality holds, so the primal and dual SDPs in \hyperref[prop:1]{Proposition~1} have the same optimal value.

Finally, the primal SDP directly gives \(0\le W_{\mathrm{int}}(\bm{\mathcal E})\le1.\) If \(\bm{\mathcal E}\in\mathbf{NI}\), choosing \(\kappa_a=\tau_a\) gives \(W_{\mathrm{int}}(\bm{\mathcal E})=0\). Conversely, suppose that \(W_{\mathrm{int}}(\bm{\mathcal E})=0\). Since the primal optimum is attained, there exists a feasible family satisfying \(\sum_a\operatorname{Tr}\kappa_a=1.\) The positive-semidefinite residual operators \(J_a-\frac{I_{A'}}{d}\otimes\kappa_a\) then have total trace zero, so every residual operator must vanish. Consequently, for all \(a\), \(J_a=\frac{I_{A'}}{d}\otimes\kappa_a\) and \(\sum_a\operatorname{Tr}\kappa_a=1\). This implies \(\bm{\mathcal E}\in\mathbf{NI}\). Therefore,
\begin{equation}
W_{\mathrm{int}}(\bm{\mathcal E})=0
\iff\bm{\mathcal E}\in\mathbf{NI}.
\label{eq:weight-faithfulness-proof}
\end{equation}
This completes the proof of \hyperref[prop:1]{Proposition~1} and the faithfulness statement in \hyperref[sec:structural]{Sec.~3.2}.

\subsection{Convexity of the interactive weight of quantum instruments}
For completeness, we prove the convexity relation stated in \mbox{Eq.~\eqref{eq:weight-convexity}}. Let \(\{p_i\}_i\) be a probability distribution and let \(\bm{\mathcal E}^{(i)}\) be instruments of the same type. Write \(w_i:=W_{\mathrm{int}}(\bm{\mathcal E}^{(i)})\), and choose optimal decompositions
\begin{equation}
\bm{\mathcal E}^{(i)}=(1-w_i)\bm{\mathcal L}^{(i)}+w_i\bm{\mathcal G}^{(i)},
\qquad
\bm{\mathcal L}^{(i)}\in\mathbf{NI}.
\label{eq:convexity-optimal-decompositions}
\end{equation}
Set \(w:=\sum_i p_iw_i\). For \(0<w<1\), define
\begin{equation}
\bm{\mathcal L}
:=
\frac{\sum_i p_i(1-w_i)\bm{\mathcal L}^{(i)}}{1-w},
\qquad
\bm{\mathcal G}
:=
\frac{\sum_i p_iw_i\bm{\mathcal G}^{(i)}}{w}.
\label{eq:convexity-mixed-components}
\end{equation}
Because \(\mathbf{NI}\) is convex and valid quantum instruments are closed under convex combinations, \(\bm{\mathcal L}\in\mathbf{NI}\) and \(\bm{\mathcal G}\) is a valid quantum instrument. Therefore,
\begin{equation}
\sum_i p_i\bm{\mathcal E}^{(i)}
=
(1-w)\bm{\mathcal L}+w\bm{\mathcal G},
\label{eq:convexity-feasible-decomposition}
\end{equation}
which is a feasible weight decomposition with resource coefficient \(w\). Hence,
\begin{equation}
W_{\mathrm{int}}\!\left(\sum_i p_i\bm{\mathcal E}^{(i)}\right)
\le
w
=
\sum_i p_iW_{\mathrm{int}}\!\left(\bm{\mathcal E}^{(i)}\right).
\label{eq:weight-convexity-proof}
\end{equation}
The endpoint cases \(w=0\) and \(w=1\) follow directly from faithfulness and \(0\le W_{\mathrm{int}}\le1\). Equation~\eqref{eq:ni-content-concavity} follows immediately from \(\lambda_{\mathbf{NI}}=1-W_{\mathrm{int}}\). \hfill\ensuremath{\square}

\section{Proof of Proposition 3}\label{app:B}

The recovery performance \(F_+(\bm{\mathcal E};\bm\Lambda)\) and its optimal value \(F_+(\bm{\mathcal E})\) are defined in \mbox{Eqs.~\eqref{eq:maxent-recovery-performance}} and \eqref{eq:maxent-recovery-optimum} of the main text. For any quantum instrument \(\bm{\mathcal X}=\{\mathcal X_a\}_a\) and recovery family \(\bm\Lambda=\{\Lambda_a\}_a\), define the total recovered map as \(\mathcal T_{\bm{\mathcal X},\bm\Lambda}:=\sum_a\Lambda_a\circ\mathcal X_a\). Since each \(\Lambda_a\) is CPTP,
\begin{equation}
\operatorname{Tr}\mathcal T_{\bm{\mathcal X},\bm\Lambda}(X) =\sum_a\operatorname{Tr}\Lambda_a[\mathcal X_a(X)]=\sum_a\operatorname{Tr}\mathcal X_a(X)=\operatorname{Tr}X.
\label{eq:recovered-channel-trace}
\end{equation}
Thus, \(\mathcal T_{\bm{\mathcal X},\bm\Lambda}\) is a CPTP channel, and
\begin{equation}
\begin{aligned}
F_+(\bm{\mathcal X};\bm\Lambda)
&=\langle\Phi^+|
(\operatorname{id}_{A'}\otimes
\mathcal T_{\bm{\mathcal X},\bm\Lambda})(\Phi^+)
|\Phi^+\rangle.
\end{aligned}
\label{eq:recovered-channel-fidelity}
\end{equation}
Let \(\bm{\mathcal L}\in\mathbf{NI}\), with \(\mathcal L_a(X)=\tau_a\operatorname{Tr}X\), where \(\tau_a\succeq0\) and \(\sum_a\operatorname{Tr}\tau_a=1\). Define
\begin{equation}
\rho_{\bm\Lambda}^A:=\sum_a\Lambda_a(\tau_a).
\label{eq:recovered-replacement-state-definition}
\end{equation}
Since each \(\Lambda_a\) is CPTP,
\begin{equation}
\rho_{\bm\Lambda}^A\succeq0,\operatorname{Tr}\rho_{\bm\Lambda}^A =\sum_a\operatorname{Tr}\Lambda_a(\tau_a)=\sum_a\operatorname{Tr}\tau_a=1.
\label{eq:recovered-replacement-state}
\end{equation}
Hence, \(\rho_{\bm\Lambda}^A\) is a normalized quantum state. For any input operator \(X\),
\begin{equation}
\mathcal T_{\bm{\mathcal L},\bm\Lambda}(X) =\sum_a\Lambda_a[\mathcal L_a(X)]=\sum_a\Lambda_a(\tau_a\operatorname{Tr}X)=\left(\sum_a\Lambda_a(\tau_a)\right)\operatorname{Tr}X=\rho_{\bm\Lambda}^A\operatorname{Tr}X.
\label{eq:recovered-ni-channel}
\end{equation}
Thus, the recovered channel is a replacement channel with normalized Choi operator
\begin{equation}
J^{\mathcal T_{\bm{\mathcal L},\bm\Lambda}}=\frac{I_{A'}}{d}\otimes\rho_{\bm\Lambda}^A.
\label{eq:recovered-ni-choi-operator}
\end{equation}
Using the maximally entangled-state identity
\begin{equation}
\langle\Phi^+|(I_{A'}\otimes Y)|\Phi^+\rangle
=\frac{\operatorname{Tr}Y}{d},
\label{eq:maxent-trace-identity}
\end{equation}
we obtain
\begin{equation}
F_+(\bm{\mathcal L};\bm\Lambda) =\left\langle\Phi^+\middle| \frac{I_{A'}}{d}\otimes\rho_{\bm\Lambda}^A \middle|\Phi^+\right\rangle=\frac1d\langle\Phi^+| (I_{A'}\otimes\rho_{\bm\Lambda}^A)|\Phi^+\rangle=\frac{\operatorname{Tr}\rho_{\bm\Lambda}^A}{d^2}=\frac1{d^2}.
\label{eq:maxent-ni-benchmark-calculation}
\end{equation}
Therefore,
\begin{equation}
F_+(\bm{\mathcal L};\bm\Lambda)=\frac1{d^2}
\label{eq:maxent-ni-benchmark-proof}
\end{equation}
for every non-interactive instrument and every recovery strategy.

We first establish the lower bound. Choose an arbitrary normalized state \(\rho_{\mathrm{fix}}^A\) and the constant recovery channels \(\Lambda_a(Y)=\rho_{\mathrm{fix}}^A\operatorname{Tr}Y\) for all \(a\). Since the total instrument map is trace preserving,
\begin{equation}
\sum_a(\Lambda_a\circ\mathcal E_a)(X) =\sum_a\rho_{\mathrm{fix}}^A\operatorname{Tr}[\mathcal E_a(X)]=\rho_{\mathrm{fix}}^A\operatorname{Tr}\left[ \left(\sum_a\mathcal E_a\right)(X)\right]=\rho_{\mathrm{fix}}^A\operatorname{Tr}X.
\label{eq:maxent-constant-recovery}
\end{equation}
By \mbox{Eq.~\eqref{eq:maxent-ni-benchmark-calculation}}, this recovery strategy achieves \(F_+(\bm{\mathcal E};\bm\Lambda)=1/d^2\). Hence,
\begin{equation}
F_+(\bm{\mathcal E})\ge\frac1{d^2}.
\label{eq:maxent-recovery-lower-bound-proof}
\end{equation}
For the upper bound, let \(w:=W_{\mathrm{int}}(\bm{\mathcal E})\) and choose an optimal weight decomposition \(\bm{\mathcal E}=(1-w)\bm{\mathcal L}+w\bm{\mathcal G}\), where \(\bm{\mathcal L}\in\mathbf{NI}\) and \(\bm{\mathcal G}\) is a valid instrument. For any fixed recovery family \(\bm\Lambda\), linearity of the recovery performance gives
{\small
\begin{equation}
\begin{aligned}
F_+(\bm{\mathcal E};\bm\Lambda)&=\sum_a\langle\Phi^+|\bigl[\operatorname{id}_{A'}\otimes
\bigl(\Lambda_a\circ((1-w)\mathcal L_a+w\mathcal G_a)\bigr)\bigr](\Phi^+)|\Phi^+\rangle\\
&=\sum_a\langle\Phi^+|\bigl[(1-w)\operatorname{id}_{A'}\otimes(\Lambda_a\circ\mathcal L_a)
+w\operatorname{id}_{A'}\otimes(\Lambda_a\circ\mathcal G_a)\bigr](\Phi^+)|\Phi^+\rangle\\
&=(1-w)\sum_a\langle\Phi^+|[\operatorname{id}_{A'}\otimes(\Lambda_a\circ\mathcal L_a)](\Phi^+)|\Phi^+\rangle+w\sum_a\langle\Phi^+|[\operatorname{id}_{A'}\otimes(\Lambda_a\circ\mathcal G_a)](\Phi^+)|\Phi^+\rangle\\
&=(1-w)F_+(\bm{\mathcal L};\bm\Lambda)+wF_+(\bm{\mathcal G};\bm\Lambda).
\end{aligned}
\label{eq:maxent-recovery-linearity-proof}
\end{equation}
}
From \mbox{Eq.~\eqref{eq:maxent-ni-benchmark-proof}},
\begin{equation}
F_+(\bm{\mathcal L};\bm\Lambda)=\frac1{d^2}.
\label{eq:maxent-free-term}
\end{equation}
Next, define \(\mathcal T_{\bm{\mathcal G},\bm\Lambda}:=\sum_a\Lambda_a\circ\mathcal G_a\). By \mbox{Eq.~\eqref{eq:recovered-channel-trace}}, this map is CPTP. Hence, \(\rho_{\bm{\mathcal G},\bm\Lambda}^{A'A}:=(\operatorname{id}_{A'}\otimes\mathcal T_{\bm{\mathcal G},\bm\Lambda})(\Phi^+)\) is a normalized quantum state. Since \(0\preceq\Phi^+\preceq I_{A'A}\),
\begin{equation}
F_+(\bm{\mathcal G};\bm\Lambda) =\operatorname{Tr}(\Phi^+\rho_{\bm{\mathcal G},\bm\Lambda}^{A'A})\le\operatorname{Tr}(I_{A'A}\rho_{\bm{\mathcal G},\bm\Lambda}^{A'A})=1.
\label{eq:maxent-recovery-unit-bound}
\end{equation}
Substituting \mbox{Eqs.~\eqref{eq:maxent-free-term}} and \eqref{eq:maxent-recovery-unit-bound} into \mbox{Eq.~\eqref{eq:maxent-recovery-linearity-proof}} gives
\begin{equation}
F_+(\bm{\mathcal E};\bm\Lambda) \le(1-w)\frac1{d^2}+w=w+\frac{1-w}{d^2}.
\label{eq:maxent-fixed-recovery-upper-bound}
\end{equation}
Since this bound is independent of the recovery strategy, maximizing over all recovery channels and combining with \mbox{Eq.~\eqref{eq:maxent-recovery-lower-bound-proof}} yields
\begin{equation}
\frac1{d^2}\le F_+(\bm{\mathcal E})
\le W_{\mathrm{int}}(\bm{\mathcal E})
+\frac{1-W_{\mathrm{int}}(\bm{\mathcal E})}{d^2}.
\label{eq:maxent-recovery-bounds-proof}
\end{equation}
For \(d\ge2\), the upper bound further implies
\begin{equation}
d^2F_+(\bm{\mathcal E})\le d^2w+1-w=1+(d^2-1)w.
\label{eq:maxent-weight-rearrangement}
\end{equation}
Therefore, \(d^2F_+(\bm{\mathcal E})-1\le(d^2-1)w\), and hence
\begin{equation}
W_{\mathrm{int}}(\bm{\mathcal E})
\ge\frac{d^2F_+(\bm{\mathcal E})-1}{d^2-1}.
\label{eq:maxent-weight-bound-proof}
\end{equation}
This completes the proof of \hyperref[prop:3]{Proposition 3}. \hfill\ensuremath{\square}

\section{Proof of Proposition 4}\label{app:C}
This appendix proves the average recovery bound for an unknown Haar-random pure state. For a fixed family of recovery channels \(\bm\Lambda=\{\Lambda_a^{B\to A}\}_a,\) define

\begin{equation}
F_{\mathrm{ave}}(\bm{\mathcal E};\bm\Lambda)
=
\int d\psi
\sum_a
\langle\psi|
(\Lambda_a\circ\mathcal E_a)(\rho_\psi)
|\psi\rangle,
\label{eq:haar-performance-proof}
\end{equation}
where \(\rho_\psi=|\psi\rangle\langle\psi|,\) and

\begin{equation}
F_{\mathrm{ave}}(\bm{\mathcal E})
=
\max_{\{\Lambda_a\}_a}
F_{\mathrm{ave}}(\bm{\mathcal E};\bm\Lambda).
\label{eq:haar-optimum-proof}
\end{equation}
\subsection{Non-interactive benchmark}
Let \(\mathcal L_a(X)=\tau_a\operatorname{Tr}X.\) Using the notation introduced in \hyperref[app:B]{Appendix~B}, \(\rho_{\bm\Lambda}^A = \sum_a \Lambda_a(\tau_a),\) with \(\rho_{\bm\Lambda}^A\succeq0,\allowbreak\; \operatorname{Tr}\rho_{\bm\Lambda}^A=1.\) For an arbitrary pure-state input \(\rho_\psi\), \(\operatorname{Tr}\rho_\psi=1\), and therefore

\begin{equation}
\sum_a (\Lambda_a\circ\mathcal L_a)(\rho_\psi) = \sum_a \Lambda_a(\tau_a\operatorname{Tr}\rho_\psi) = \sum_a \Lambda_a(\tau_a) = \rho_{\bm\Lambda}^A.
\label{eq:haar-recovered-ni-state}
\end{equation}
Hence, after any recovery operation, the output of a non-interactive instrument remains independent of the unknown input state \(\rho_\psi\).

Using the normalized first moment of the Haar measure,

\begin{equation}
\int d\psi\,\rho_\psi
=
\frac{I_A}{d},
\label{eq:haar-first-moment}
\end{equation}
one obtains

\begin{equation}
\begin{aligned}
F_{\mathrm{ave}}(\bm{\mathcal L};\bm\Lambda)
&=
\int d\psi\,
\langle\psi|\rho_{\bm\Lambda}^A|\psi\rangle
\\
&=
\int d\psi\,
\operatorname{Tr}(\rho_\psi\rho_{\bm\Lambda}^A)
\\
&=
\operatorname{Tr}
\left[
\left(
\int d\psi\,\rho_\psi
\right)
\rho_{\bm\Lambda}^A
\right]
\\
&=
\operatorname{Tr}
\left(
\frac{I_A}{d}\rho_{\bm\Lambda}^A
\right)
\\
&=
\frac{1}{d}.
\end{aligned}
\label{eq:haar-ni-benchmark-calculation}
\end{equation}
Therefore,

\begin{equation}
F_{\mathrm{ave}}(\bm{\mathcal L};\bm\Lambda)
=
\frac{1}{d}
\label{eq:haar-ni-benchmark-proof}
\end{equation}
for every \(\bm{\mathcal L}\in\mathbf{NI}\) and every recovery family.

\subsection{Proof of Proposition 4}
The general lower bound is established first. Fix an arbitrary density operator \(\rho_{\mathrm{fix}}^A\), and define \(\Lambda_a(Y) = \rho_{\mathrm{fix}}^A\operatorname{Tr}Y\) for every \(a\). For an arbitrary pure-state input \(\rho_\psi\),

\begin{equation}
\sum_a (\Lambda_a\circ\mathcal E_a)(\rho_\psi) = \rho_{\mathrm{fix}}^A \sum_a \operatorname{Tr}[\mathcal E_a(\rho_\psi)] = \rho_{\mathrm{fix}}^A \operatorname{Tr} \left[ \left( \sum_a\mathcal E_a \right)(\rho_\psi) \right] = \rho_{\mathrm{fix}}^A.
\label{eq:haar-constant-recovery}
\end{equation}
It follows that

\begin{equation}
F_{\mathrm{ave}}(\bm{\mathcal E};\bm\Lambda) = \int d\psi\, \langle\psi|\rho_{\mathrm{fix}}^A|\psi\rangle = \operatorname{Tr} \left[ \rho_{\mathrm{fix}}^A \int d\psi\,\rho_\psi \right] = \frac{\operatorname{Tr}\rho_{\mathrm{fix}}^A}{d} = \frac{1}{d}.
\label{eq:haar-constant-recovery-fidelity}
\end{equation}
Hence,

\begin{equation}
F_{\mathrm{ave}}(\bm{\mathcal E})
\ge
\frac{1}{d}.
\label{eq:haar-recovery-lower-bound-proof}
\end{equation}
The upper bound is established next. Let \(w:=W_{\mathrm{int}}(\bm{\mathcal E}),\) and choose an optimal weight decomposition \(\bm{\mathcal E} = (1-w)\bm{\mathcal L}+w\bm{\mathcal G},\allowbreak\; \bm{\mathcal L}\in\mathbf{NI}.\) For any fixed recovery family \(\bm\Lambda\),

{\small
\begin{equation}
\begin{aligned}
F_{\mathrm{ave}}(\bm{\mathcal E};\bm\Lambda)&=\int d\psi\sum_a\langle\psi|\Lambda_a\circ\bigl[(1-w)\mathcal L_a+w\mathcal G_a\bigr](\rho_\psi)|\psi\rangle\\
&=\int d\psi\sum_a\langle\psi|\Bigl[(1-w)(\Lambda_a\circ\mathcal L_a)(\rho_\psi)
+w(\Lambda_a\circ\mathcal G_a)(\rho_\psi)\Bigr]|\psi\rangle\\
&=(1-w)\int d\psi\sum_a\langle\psi|(\Lambda_a\circ\mathcal L_a)(\rho_\psi)|\psi\rangle+w\int d\psi\sum_a\langle\psi|(\Lambda_a\circ\mathcal G_a)(\rho_\psi)|\psi\rangle\\
&=(1-w)F_{\mathrm{ave}}(\bm{\mathcal L};\bm\Lambda)+wF_{\mathrm{ave}}(\bm{\mathcal G};\bm\Lambda).
\end{aligned}
\label{eq:haar-recovery-linearity-proof}
\end{equation}
}
Equation \eqref{eq:haar-ni-benchmark-proof} gives \(F_{\mathrm{ave}}(\bm{\mathcal L};\bm\Lambda) = \frac{1}{d}.\) On the other hand, define \(\mathcal T_{\bm{\mathcal G},\bm\Lambda} = \sum_a \Lambda_a\circ\mathcal G_a.\) This map is a CPTP channel. Hence, for every pure state \(\rho_\psi\), \(\rho_{\psi,\mathrm{out}}^A := \mathcal T_{\bm{\mathcal G},\bm\Lambda}(\rho_\psi)\) is a density operator. Since \(0\preceq\rho_\psi\preceq I,\) it follows that

\begin{equation}
\langle\psi|\rho_{\psi,\mathrm{out}}^A|\psi\rangle = \operatorname{Tr}(\rho_\psi\rho_{\psi,\mathrm{out}}^A) \le \operatorname{Tr}(I\rho_{\psi,\mathrm{out}}^A) = 1.
\label{eq:haar-recovery-unit-bound}
\end{equation}
Integration over the Haar measure therefore gives \(F_{\mathrm{ave}}(\bm{\mathcal G};\bm\Lambda) \le 1.\) Equation \eqref{eq:haar-recovery-linearity-proof} then implies

\begin{equation}
F_{\mathrm{ave}}(\bm{\mathcal E};\bm\Lambda) \le (1-w)\frac{1}{d}+w = w+\frac{1-w}{d}.
\label{eq:haar-fixed-recovery-upper-bound}
\end{equation}
The right-hand side is again independent of the recovery strategy. Therefore,

\begin{equation}
\frac{1}{d}
\le
F_{\mathrm{ave}}(\bm{\mathcal E})
\le
W_{\mathrm{int}}(\bm{\mathcal E})
+
\frac{1-W_{\mathrm{int}}(\bm{\mathcal E})}{d}.
\label{eq:haar-recovery-bounds-proof}
\end{equation}
For \(d\ge2\), \(F_{\mathrm{ave}}(\bm{\mathcal E}) \le w+\frac{1-w}{d}\) is equivalent to \(dF_{\mathrm{ave}}(\bm{\mathcal E}) \le dw+1-w = 1+(d-1)w.\) Hence, \(dF_{\mathrm{ave}}(\bm{\mathcal E})-1 \le (d-1)w,\) and therefore

\begin{equation}
W_{\mathrm{int}}(\bm{\mathcal E})
\ge
\frac{
dF_{\mathrm{ave}}(\bm{\mathcal E})-1
}{
d-1
}.
\label{eq:haar-weight-bound-proof}
\end{equation}
This proves \hyperref[prop:4]{Proposition 4}. \hfill\ensuremath{\square}

\section{Proof of the exact characterization by positive linear loss games}\label{app:D}
This appendix provides the complete proof of \hyperref[prop:5]{Proposition~5}. To this end, we first establish the Choi-linear representation of the average loss.

In the main text, we introduced the positive linear loss game in \mbox{Eq.~\eqref{eq:game-def}} and defined its average loss \(P(\bm{\mathcal E},g)\) and corresponding Choi loss operators \(H_a^g\) in \mbox{Eqs.~\eqref{eq:game-average-loss}} and \eqref{eq:game-choi-loss-operators}, respectively. To explicitly establish the Choi representation of the average loss, we apply the normalized inverse Choi formula in \mbox{Eq.~\eqref{eq:instrument-inverse-choi}}. For each \(a,i,j\), we obtain
\begin{equation}
\begin{aligned}
\operatorname{Tr}[\mathcal E_a(\rho_i)M_j]
&=d\,\operatorname{Tr}_B\left[
M_j\operatorname{Tr}_{A'}\left(
(\rho_i^T\otimes I_B)J_a\right)\right]\\
&=d\,\operatorname{Tr}_{A'B}\left[
(I_{A'}\otimes M_j)(\rho_i^T\otimes I_B)J_a\right]\\
&=d\,\operatorname{Tr}\left[(\rho_i^T\otimes M_j)J_a\right].
\end{aligned}
\label{eq:game-probability-choi-expansion}
\end{equation}
Substituting this expression into the average loss defined in \mbox{Eq.~\eqref{eq:game-average-loss}} and using the definition of \(H_a^g\) in \mbox{Eq.~\eqref{eq:game-choi-loss-operators}}, we obtain
\begin{equation}
\begin{aligned}
P(\bm{\mathcal E},g)
&=d\sum_{a,i,j}p_i\alpha_{aij}
\operatorname{Tr}\left[(\rho_i^T\otimes M_j)J_a\right]\\
&=\sum_a\operatorname{Tr}\left[
\left(d\sum_{i,j}p_i\alpha_{aij}\rho_i^T\otimes M_j\right)J_a\right]\\
&=\sum_a\operatorname{Tr}(H_a^gJ_a).
\end{aligned}
\label{eq:game-average-loss-choi-proof}
\end{equation}
This establishes the Choi-linear representation of the average loss given in \mbox{Eq.~\eqref{eq:game-choi-loss-representation}} of the main text. The resulting expression provides the basis for deriving the non-interactive benchmark and establishing the connection between admissible positive linear loss games and the dual SDP characterization of the maximal non-interactive content.

Based on the above relations, we now prove the exact operational characterization stated in \hyperref[prop:5]{Proposition~5}.

\medskip
\noindent\textbf{\hyperref[prop:5]{Proposition 5.}} For any finite-dimensional quantum instrument \(\bm{\mathcal E}\), one has \(1-W_{\mathrm{int}}(\bm{\mathcal E})=\allowbreak\inf_{g\in\mathcal G_+}\frac{P(\bm{\mathcal E},g)}{\min_{\bm{\mathcal L}\in\mathbf{NI}}P(\bm{\mathcal L},g)}\).

Before proving this proposition, we need to prove the following two lemmas.

\medskip
\phantomsection\label{lem:D.1}\noindent \textbf{Lemma D.1.} For any game \(g\) satisfying \(H_a^g\succeq0,\) one has

\begin{equation}
b_{\mathbf{NI}}(g) =\min_{\bm{\mathcal L}\in\mathbf{NI}}P(\bm{\mathcal L},g)=\min_a\lambda_{\min}\!\left( \frac{\operatorname{Tr}_{A'}H_a^g}{d} \right),
\label{eq:game-free-benchmark-lemma}
\end{equation}
and the minimum is attained.

\noindent \textbf{Proof.} Consider an arbitrary \(\bm{\mathcal L}\in\mathbf{NI}\), whose Choi branches can be written as \(J_a^{\bm{\mathcal L}} = \frac{I_{A'}}{d} \otimes\tau_a,\) where \(\tau_a\succeq0,\allowbreak\; \sum_a\operatorname{Tr}\tau_a=1.\) Equation \eqref{eq:game-average-loss-choi-proof} gives

\begin{equation}
P(\bm{\mathcal L},g) =\sum_a\operatorname{Tr}\left[ H_a^g\left(\frac{I_{A'}}{d}\otimes\tau_a\right) \right]=\sum_a\operatorname{Tr}\left[ \frac{\operatorname{Tr}_{A'}H_a^g}{d}\tau_a \right].
\label{eq:game-ni-loss-pairing}
\end{equation}
Let \(\gamma := \min_a \lambda_{\min}\!\left(\frac{\operatorname{Tr}_{A'}H_a^g}{d}\right).\) Since \(\frac{\operatorname{Tr}_{A'}H_a^g}{d}\succeq\gamma I_B\) for every \(a\),

\begin{equation}
P(\bm{\mathcal L},g) = \sum_a \operatorname{Tr} \left[\frac{\operatorname{Tr}_{A'}H_a^g}{d}\tau_a\right] \ge \gamma \sum_a \operatorname{Tr}\tau_a = \gamma.
\label{eq:game-free-loss-lower-bound}
\end{equation}
Choose a label \(a^\star\) satisfying \(\lambda_{\min}\!\left(\frac{\operatorname{Tr}_{A'}H_{a^\star}^g}{d}\right) = \gamma,\) and let \(|v^\star\rangle\) be a corresponding normalized eigenvector. Set \(\tau_{a^\star} = |v^\star\rangle\langle v^\star|,\allowbreak\; \tau_a=0 \quad \forall a\neq a^\star.\) This family defines a valid non-interactive instrument, and

\begin{equation}
P(\bm{\mathcal L},g) = \operatorname{Tr} \left[ \frac{\operatorname{Tr}_{A'}H_{a^\star}^g}{d} |v^\star\rangle\langle v^\star| \right] = \left\langle v^\star\middle| \frac{\operatorname{Tr}_{A'}H_{a^\star}^g}{d} \middle|v^\star\right\rangle = \gamma.
\label{eq:game-free-loss-attainment}
\end{equation}
Hence, the lower bound in \mbox{Eq.~\eqref{eq:game-free-loss-lower-bound}} is attained, which proves the lemma. \hfill\ensuremath{\square}

In particular, for any real number \(t\), \hyperref[lem:D.1]{Lemma D.1} gives

\begin{equation}
b_{\mathbf{NI}}(g)\ge t
\iff
\operatorname{Tr}_{A'}H_a^g\succeq dtI_B,
\qquad\forall a.
\label{eq:game-benchmark-spectral-constraint}
\end{equation}
\begingroup
\medskip
\phantomsection\label{lem:D.2}\noindent \textbf{Lemma D.2.}
For any finite family of Hermitian operators \(X_a\in \operatorname{Herm}(A'\otimes B),\) there exist a common finite set of input states \(\{\rho_i\}_i\), a common finite POVM \(\{M_j\}_j\), a strictly positive probability distribution \(\{p_i\}_i\), and real coefficients \(\alpha_{aij}\) such that

\begin{equation}
X_a
=
d
\sum_{i,j}
p_i\alpha_{aij}
\rho_i^T\otimes M_j
\qquad
\forall a.
\label{eq:finite-game-realization-lemma}
\end{equation}
\noindent \textbf{Proof.} Choose a basis \(\{F_\mu\}_{\mu=1}^{d^2-1}\) for the space of traceless Hermitian operators on \(A'\), rescaled such that \(\|F_\mu\|_\infty\le1\). Define the input states \(\rho_0:=I_A/d\) and \(\rho_\mu:=(I_A+\frac12F_\mu^T)/d\), where \(1\le\mu\le d^2-1\). Since \(F_\mu\) is traceless and has operator norm at most one, these operators are density operators. Moreover, \(I_{A'}=d\rho_0^T\) and \(F_\mu=2d(\rho_\mu^T-\rho_0^T)\). Hence, \(\{\rho_i^T\}_i\) spans \(\operatorname{Herm}(A')\) over the real numbers.

Similarly, let \(d_B:=\dim B\), and choose a basis \(\{G_\nu\}_{\nu=1}^{d_B^2-1}\) for the space of traceless Hermitian operators on \(B\), with \(\|G_\nu\|_\infty\le1\). Directly define \(M_0:=I_B/(2d_B^2-1)\) and \(M_{\nu,\pm}:=(I_B\pm\frac12G_\nu)/(2d_B^2-1)\). These operators are positive semidefinite, and
\begin{equation}
M_0+\sum_{\nu=1}^{d_B^2-1}(M_{\nu,+}+M_{\nu,-}) =\frac{1+2(d_B^2-1)}{2d_B^2-1}I_B=I_B.
\label{eq:informationally-complete-povm-normalization}
\end{equation}
Thus, denoting them collectively by \(\{M_j\}_j\), we obtain a finite POVM. Moreover, \(I_B=(2d_B^2-1)M_0\) and \(G_\nu=(2d_B^2-1)(M_{\nu,+}-M_{\nu,-})\). Hence, \(\{M_j\}_j\) spans \(\operatorname{Herm}(B)\) over the real numbers.

Consequently, the tensor-product family \(\{\rho_i^T\otimes M_j\}_{i,j}\) spans \(\operatorname{Herm}(A'\otimes B)\) over the real numbers. Hence, for every \(a\), there exist real coefficients \(c_{aij}\) such that \(X_a=\sum_{i,j}c_{aij}\rho_i^T\otimes M_j\). Choose any strictly positive probability distribution \(\{p_i\}_i\), with \(p_i>0\) and \(\sum_i p_i=1\), and define \(\alpha_{aij}:=c_{aij}/(dp_i)\). Then
\begin{equation}
d\sum_{i,j}p_i\alpha_{aij}\rho_i^T\otimes M_j =d\sum_{i,j}p_i\frac{c_{aij}}{dp_i}\rho_i^T\otimes M_j=\sum_{i,j}c_{aij}\rho_i^T\otimes M_j=X_a.
\label{eq:finite-game-coefficient-construction}
\end{equation}
This proves \mbox{Eq.~\eqref{eq:finite-game-realization-lemma}}. The input states, probability distribution, and POVM are common to all branches \(a\). The branch dependence enters only through the real coefficients \(\alpha_{aij}\). \hfill\ensuremath{\square}

Next, we provide the proof of \hyperref[prop:5]{Proposition~5}:\par
\endgroup

\noindent\textbf{Proof of \hyperref[prop:5]{Proposition 5}.}
Let \(\lambda := \lambda_{\mathbf{NI}}(\bm{\mathcal E}) = 1-W_{\mathrm{int}}(\bm{\mathcal E}).\) From the dual SDP in \hyperref[app:A]{Appendix~A},

\begin{equation}
\lambda
=
\inf_{\substack{
\Omega_a\succeq0\\
\operatorname{Tr}_{A'}\Omega_a\succeq dI_B,\ \forall a
}}
\sum_a
\operatorname{Tr}
(\Omega_aJ_a).
\label{eq:game-dual-content-proof}
\end{equation}
First, it is shown that the normalized loss of every admissible game is no smaller than \(\lambda\). Fix \(g\in\mathcal G_+,\) let \(b:=b_{\mathbf{NI}}(g)>0\), and define \(\Omega_a:=H_a^g/b.\) By \hyperref[lem:D.1]{Lemma D.1} and \mbox{Eq.~\eqref{eq:game-benchmark-spectral-constraint}},

\begin{equation}
\Omega_a\succeq0,\operatorname{Tr}_{A'}\Omega_a =\frac{1}{b}\operatorname{Tr}_{A'}H_a^g \succeq dI_B,\qquad\forall a.
\label{eq:game-normalized-witness-feasibility}
\end{equation}
Thus, \(\{\Omega_a\}_a\) is feasible for the dual problem in \mbox{Eq.~\eqref{eq:game-dual-content-proof}}. Using \mbox{Eq.~\eqref{eq:game-average-loss-choi-proof}},

\begin{equation}
\frac{ P(\bm{\mathcal E},g) }{ b_{\mathbf{NI}}(g) } = \frac{1}{b} \sum_a \operatorname{Tr}(H_a^gJ_a) = \sum_a \operatorname{Tr}(\Omega_aJ_a) \ge \lambda.
\label{eq:game-normalized-loss-lower-bound-proof}
\end{equation}
Taking the infimum over all \(g\in\mathcal G_+\) gives

\begin{equation}
\inf_{g\in\mathcal G_+}
\frac{
P(\bm{\mathcal E},g)
}{
b_{\mathbf{NI}}(g)
}
\ge
\lambda.
\label{eq:game-infimum-lower-bound-proof}
\end{equation}
The reverse inequality is established next. Fix an arbitrary \(\epsilon>0.\) By \mbox{Eq.~\eqref{eq:game-dual-content-proof}}, there exists a dual-feasible family \(\{\Omega_a\}_a\) satisfying \(\Omega_a\succeq0,\allowbreak\; \operatorname{Tr}_{A'}\Omega_a \succeq dI_B \quad \forall a,\) and

\begin{equation}
\sum_a
\operatorname{Tr}(\Omega_aJ_a)
\le
\lambda+\epsilon.
\label{eq:game-epsilon-dual-objective}
\end{equation}
Dual feasibility implies \(\operatorname{Tr}_{A'}\Omega_a/d\succeq I_B\) for every \(a\), so each of these operators has minimum eigenvalue at least one. Define \(\delta:=\min_a\lambda_{\min}\!\left(\frac{\operatorname{Tr}_{A'}\Omega_a}{d}\right)\ge1\) and \(\widetilde{\Omega}_a:=\frac{\Omega_a}{\delta}\).
Then \(\widetilde{\Omega}_a\succeq0\), and
\begin{equation}
\min_a\lambda_{\min}\!\left(
\frac{\operatorname{Tr}_{A'}\widetilde{\Omega}_a}{d}
\right)=\frac{1}{\delta}\min_a\lambda_{\min}\!\left(
\frac{\operatorname{Tr}_{A'}\Omega_a}{d}\right)=1.
\label{eq:game-rescaled-witness-benchmark}
\end{equation}
Applying \hyperref[lem:D.2]{Lemma D.2} with \(X_a=\widetilde{\Omega}_a\) produces a finite set of input states, a common finite POVM, a strictly positive probability distribution, and real coefficients such that the corresponding game \(g\) satisfies

\begin{equation}
H_a^g
=
d
\sum_{i,j}
p_i\alpha_{aij}
\rho_i^T\otimes M_j
=
\widetilde{\Omega}_a
\qquad
\forall a.
\label{eq:game-rescaled-witness-realization}
\end{equation}
Hence, \(H_a^g\succeq0.\) By \hyperref[lem:D.1]{Lemma D.1} and \mbox{Eq.~\eqref{eq:game-rescaled-witness-benchmark}},

\begin{equation}
b_{\mathbf{NI}}(g) =\min_a\lambda_{\min}\!\left(\frac{\operatorname{Tr}_{A'}H_a^g}{d}\right)=\min_a\lambda_{\min}\!\left(\frac{\operatorname{Tr}_{A'}\widetilde{\Omega}_a}{d}\right)=1.
\label{eq:game-unit-free-benchmark}
\end{equation}
Therefore, the constructed game belongs to \(\mathcal G_+\). Its normalized loss satisfies

\begin{equation}
\begin{aligned}
\frac{P(\bm{\mathcal E},g)}{b_{\mathbf{NI}}(g)}
&=\sum_a\operatorname{Tr}(\widetilde{\Omega}_aJ_a)\\
&=\frac{1}{\delta}\sum_a\operatorname{Tr}(\Omega_aJ_a)\\
&\le\frac{\lambda+\epsilon}{\delta}\\
&\le\lambda+\epsilon.
\end{aligned}
\label{eq:game-normalized-loss-upper-bound-proof}
\end{equation}
where the last inequality follows from \(\delta\ge1\) and \(\lambda\ge0\). Since \(\epsilon>0\) is arbitrary, taking the infimum over games and then letting \(\epsilon\to0\) gives

\begin{equation}
\inf_{g\in\mathcal G_+}
\frac{
P(\bm{\mathcal E},g)
}{
b_{\mathbf{NI}}(g)
}
\le
\lambda.
\label{eq:game-infimum-upper-bound-proof}
\end{equation}
Combining \mbox{Eqs.~\eqref{eq:game-infimum-lower-bound-proof}} and \eqref{eq:game-infimum-upper-bound-proof} yields

\begin{equation}
1-W_{\mathrm{int}}(\bm{\mathcal E})
=
\inf_{g\in\mathcal G_+}
\frac{
P(\bm{\mathcal E},g)
}{
\displaystyle
\min_{\bm{\mathcal L}\in\mathbf{NI}}
P(\bm{\mathcal L},g)
}.
\label{eq:game-exact-characterization-proof}
\end{equation}
This proves \hyperref[prop:5]{Proposition 5}. \hfill\ensuremath{\square}

\section{Proof of the environment-assisted input-blind extraction}\label{app:E}

\noindent\textbf{Proposition 6.} For any finite-dimensional quantum instrument \(\bm{\mathcal E}\), if the environmental assistant has access to the minimal Stinespring environment of the complete cq channel \(\widetilde{\mathcal E}\), then \(P_{\mathrm{blind}}(\bm{\mathcal E})=1-W_{\mathrm{int}}(\bm{\mathcal E})\). Moreover, the optimal success probability is attained by an actual environmental effect \(M_{\mathrm{succ}}^{E,\star}\).

Before proving this proposition, we need to prove the following two lemmas.

\medskip
\phantomsection\label{lem:E.1}\noindent \textbf{Lemma E.1.} 
Let a linear map \(\mathcal S:A\to D\) satisfy

\begin{equation}
\mathcal S(X)
=
Y^D\operatorname{Tr}X,
\qquad
\forall X\in\mathcal B(A).
\label{eq:replacement-map-lemma}
\end{equation}
Then, for every finite-dimensional reference system \(R\) and every \(X^{RA}\in\mathcal B(R\otimes A),\) one has

\begin{equation}
(\operatorname{id}_R\otimes\mathcal S)(X^{RA})
=
\operatorname{Tr}_A X^{RA}
\otimes
Y^D.
\label{eq:replacement-map-decoupling}
\end{equation}
\noindent \textbf{Proof.} In a fixed orthonormal basis of \(A\), write \(X^{RA} = \sum_{i,j} X_{ij}^R \otimes |i\rangle\langle j|^A.\) Using \mbox{Eq.~\eqref{eq:replacement-map-lemma}},

\begin{equation}
\begin{aligned}
(\operatorname{id}_R\otimes\mathcal S)(X^{RA})
&=
\sum_{i,j}
X_{ij}^R
\otimes
Y^D
\operatorname{Tr}
(|i\rangle\langle j|)
\\
&=
\sum_i
X_{ii}^R
\otimes
Y^D
\\
&=
\operatorname{Tr}_A X^{RA}
\otimes
Y^D.
\end{aligned}
\label{eq:replacement-map-decoupling-calculation}
\end{equation}
\hfill\ensuremath{\square}

For the input-blind successful map introduced in the main text, take \(Y^{CB} = \sum_a |a\rangle\langle a|^C \otimes \kappa_a^B.\) Equation \eqref{eq:replacement-map-decoupling} then gives

\begin{equation}
(\operatorname{id}_R\otimes\mathcal S_M)(\rho^{RA})
=
\operatorname{Tr}_A\rho^{RA}
\otimes
\sum_a
|a\rangle\langle a|^C
\otimes
\kappa_a^B.
\label{eq:successful-map-decoupling-proof}
\end{equation}
Thus, the fully input-blind condition not only removes all information about the local input \(A\), but also completely decouples the successful public output from any pre-existing \(R\)-\(A\) correlations.

\phantomsection\label{lem:E.2}\noindent \textbf{Lemma E.2~\cite{belavkin1986,raginsky2003}.} Let a finite-dimensional CP map \(\mathcal T:A\to D\) have a minimal Kraus representation

\begin{equation}
\mathcal T(X)
=
\sum_{i=1}^{r}
K_i X K_i^\dagger,
\label{eq:minimal-kraus-representation}
\end{equation}
where \(\{K_i\}_{i=1}^{r}\) is linearly independent, and fix the corresponding minimal Stinespring realization

\begin{equation}
V
=
\sum_{i=1}^{r}
K_i\otimes|i\rangle^E.
\label{eq:kraus-stinespring-operator}
\end{equation}
For any CP map \(\mathcal S:A\to D,\) the following statements are equivalent:

(1)
\begin{equation}
0
\preceq_{\mathrm{CP}}
\mathcal S
\preceq_{\mathrm{CP}}
\mathcal T.
\label{eq:radon-nikodym-cp-order}
\end{equation}

(2) The existence of a unique environmental effect \(
0
\preceq
M_{\mathrm{succ}}^E
\preceq
I_E
\) such that


\begin{equation}
\mathcal S(X)
=
\operatorname{Tr}_E
\!\left[
(I_D\otimes \sqrt{M_{\mathrm{succ}}^E})
VXV^\dagger
(I_D\otimes \sqrt{M_{\mathrm{succ}}^E})
\right].
\label{eq:radon-nikodym-square-root-effect}
\end{equation}
If \(\mathcal T\) is trace preserving, then \(V^\dagger V=I_A,\) so \(V\) is a minimal Stinespring isometry.

\noindent \textbf{Proof.} The normalized Choi convention is used. Let the vectorization of the \(i\)-th Kraus operator be \(|k_i\rangle\rangle := (I_{A'}\otimes K_i)|\Phi^+\rangle.\) Define the column operator

\begin{equation}
\mathsf K
:=
\bigl(
|k_1\rangle\rangle
\;
|k_2\rangle\rangle
\;
\cdots
\;
|k_r\rangle\rangle
\bigr).
\label{eq:kraus-vectorization-matrix}
\end{equation}
By the Choi definition,

\begin{equation}
J^{\mathcal T}
=
\sum_{i=1}^{r}
|k_i\rangle\rangle
\langle\langle k_i|
=
\mathsf K\mathsf K^\dagger.
\label{eq:kraus-choi-factorization}
\end{equation}
The map \(K \mapsto (I_{A'}\otimes K)|\Phi^+\rangle\) is linearly injective. Therefore, linear independence of the minimal Kraus family \(\{K_i\}_i\) implies linear independence of \(\{|k_i\rangle\rangle\}_i\). Hence, \(\mathsf K\) has full column rank. It follows that \(\mathsf K^\dagger \mathsf K\) is positive definite and invertible, so the left inverse

\begin{equation}
\mathsf L
:=
(\mathsf K^\dagger \mathsf K)^{-1}\mathsf K^\dagger,
\qquad
\mathsf L\mathsf K=I_r,
\label{eq:kraus-matrix-left-inverse}
\end{equation}
is well defined.

We first prove that \mbox{Eq.~\eqref{eq:radon-nikodym-cp-order}} implies statement (2) of Lemma E.2. Since the Choi isomorphism preserves the CP order,

\begin{equation}
0
\preceq
J^{\mathcal S}
\preceq
J^{\mathcal T}
=
\mathsf K\mathsf K^\dagger.
\label{eq:dominated-choi-operator}
\end{equation}
The positive-semidefinite order implies

\begin{equation}
\operatorname{supp}J^{\mathcal S}
\subseteq
\operatorname{supp}J^{\mathcal T}
=
\operatorname{Ran}\mathsf K.
\label{eq:dominated-choi-support}
\end{equation}
Define \(\mathsf P := \mathsf K\mathsf L = \mathsf K(\mathsf K^\dagger \mathsf K)^{-1}\mathsf K^\dagger.\) Since \(\mathsf K\) has full column rank, \(\mathsf P\) is the orthogonal projector onto \(\operatorname{Ran}\mathsf K\). Equation \eqref{eq:dominated-choi-support} therefore gives

\begin{equation}
\mathsf PJ^{\mathcal S}
=
J^{\mathcal S}\mathsf P
=
J^{\mathcal S}.
\label{eq:choi-support-projection}
\end{equation}
Define \(\mathsf C := \mathsf LJ^{\mathcal S}\mathsf L^\dagger.\) Since \(J^{\mathcal S}\succeq0,\) one has \(\mathsf C\succeq0.\) Moreover, \(\mathsf LJ^{\mathcal T}\mathsf L^\dagger = \mathsf L\mathsf K\mathsf K^\dagger \mathsf L^\dagger = I_r,\) and therefore \(I_r-\mathsf C = \mathsf L(J^{\mathcal T}-J^{\mathcal S})\mathsf L^\dagger \succeq0.\) Thus,

\begin{equation}
0
\preceq
\mathsf C
\preceq
I_r.
\label{eq:radon-nikodym-coefficient-interval}
\end{equation}
Using \mbox{Eq.~\eqref{eq:choi-support-projection}},

\begin{equation}
\mathsf K\mathsf C\mathsf K^\dagger
=
\mathsf K\mathsf LJ^{\mathcal S}\mathsf L^\dagger \mathsf K^\dagger
=
\mathsf PJ^{\mathcal S}\mathsf P
=
J^{\mathcal S}.
\label{eq:dominated-choi-factorization}
\end{equation}
Expanding this expression in terms of the columns of \(\mathsf K\) gives

\begin{equation}
J^{\mathcal S}
=
\sum_{i,j=1}^{r}
\mathsf C_{ij}
|k_i\rangle\rangle
\langle\langle k_j|.
\label{eq:dominated-choi-expansion}
\end{equation}
The Choi operator of the linear map \(X\mapsto K_i X K_j^\dagger\) is precisely \(|k_i\rangle\rangle \langle\langle k_j|.\) By linearity and injectivity of the Choi map,

\begin{equation}
\mathcal S(X)
=
\sum_{i,j=1}^{r}
\mathsf C_{ij}
K_i X K_j^\dagger.
\label{eq:dominated-kraus-expansion}
\end{equation}
In the environmental basis \(\{|i\rangle^E\}\), define \(M_{\mathrm{succ}}^E := \mathsf C^T.\) Equation \eqref{eq:radon-nikodym-coefficient-interval}, together with preservation of the positive-semidefinite order under transposition, gives \(0 \preceq M_{\mathrm{succ}}^E \preceq I_E.\) On the other hand, \mbox{Eq.~\eqref{eq:kraus-stinespring-operator}} gives \(VXV^\dagger = \sum_{i,j=1}^{r} K_iXK_j^\dagger \otimes |i\rangle\langle j|^E.\) Hence,

\begin{equation}
\begin{aligned}
\operatorname{Tr}_E
\!\left[
(I_D\otimes M_{\mathrm{succ}}^E)VXV^\dagger
\right]
&=
\sum_{i,j}
\langle j|M_{\mathrm{succ}}^E|i\rangle
K_iXK_j^\dagger
\\
&=
\sum_{i,j}
\mathsf C_{ij}
K_iXK_j^\dagger
\\
&=
\mathcal S(X),
\end{aligned}
\label{eq:transposed-effect-realization}
\end{equation}
where \(M_{\mathrm{succ}}^E=\mathsf C^T,\allowbreak\; \langle j|M_{\mathrm{succ}}^E|i\rangle=\mathsf C_{ij}\) has been used. This proves the existence of the environmental effect.

The converse direction is proved next. Suppose an effect \(M_{\mathrm{succ}}^E\) satisfies \(0\preceq M_{\mathrm{succ}}^E\preceq I_E\) and realizes \(\mathcal S\) through \mbox{Eq.~\eqref{eq:radon-nikodym-square-root-effect}}. Since \(M_{\mathrm{succ}}^E\succeq0,\) the positive square root \(\sqrt{M_{\mathrm{succ}}^E}\) exists. By cyclicity of the partial trace over the environment,

\begin{equation}
\mathcal S(X)
=
\operatorname{Tr}_E
\!\left[
(I_D\otimes \sqrt{M_{\mathrm{succ}}^E})
VXV^\dagger
(I_D\otimes \sqrt{M_{\mathrm{succ}}^E})
\right].
\label{eq:effect-submap-complete-positivity}
\end{equation}
This expression is a composition of the completely positive maps \(X\mapsto VXV^\dagger,\) \(Y\mapsto (I_D\otimes \sqrt{M_{\mathrm{succ}}^E}) Y (I_D\otimes \sqrt{M_{\mathrm{succ}}^E}),\) and \(Y\mapsto\operatorname{Tr}_E Y.\) Therefore, \(\mathcal S\) is completely positive.

Similarly, since \(I_E-M_{\mathrm{succ}}^E\succeq0,\) define

\begin{equation}
\mathcal S_{I-M}(X)
:=\operatorname{Tr}_E\Bigl[
(I_D\otimes(I_E-M_{\mathrm{succ}}^E)^{1/2})VXV^\dagger
(I_D\otimes(I_E-M_{\mathrm{succ}}^E)^{1/2})\Bigr].
\label{eq:complementary-effect-submap}
\end{equation}
Then \(\mathcal S_{I-M}\) is also completely positive. Using cyclicity and linearity of the partial trace,

\begin{equation}
\begin{aligned}
\mathcal S(X)+\mathcal S_{I-M}(X)
&=\operatorname{Tr}_E\!\left[(I_D\otimes M_{\mathrm{succ}}^E)VXV^\dagger\right]+\operatorname{Tr}_E\!\left[(I_D\!\otimes(I_E-M_{\mathrm{succ}}^E))VXV^\dagger\right]\\
&=\operatorname{Tr}_E\!\left[VXV^\dagger\right]=\mathcal T(X).
\end{aligned}
\label{eq:complementary-submap-sum}
\end{equation}
Hence, \(\mathcal T-\mathcal S = \mathcal S_{I-M}\) is completely positive, and therefore

\begin{equation}
0
\preceq_{\mathrm{CP}}
\mathcal S
\preceq_{\mathrm{CP}}
\mathcal T.
\label{eq:effect-submap-cp-domination}
\end{equation}
It remains to prove uniqueness of the environmental effect. Suppose \(M_{\mathrm{succ}}^E\) and \(\widehat M_{\mathrm{succ}}^E\) generate the same map \(\mathcal S\) on the same minimal Stinespring realization. Equations \eqref{eq:dominated-choi-expansion}--\eqref{eq:transposed-effect-realization} then imply \(\mathsf K((M_{\mathrm{succ}}^E)^T-(\widehat M_{\mathrm{succ}}^E)^T)\mathsf K^\dagger = 0.\) Left multiplication by \(\mathsf L\), right multiplication by \(\mathsf L^\dagger\), and use of \(\mathsf L\mathsf K=I_r\) give \((M_{\mathrm{succ}}^E)^T-(\widehat M_{\mathrm{succ}}^E)^T = 0.\) Therefore, \(M_{\mathrm{succ}}^E=\widehat M_{\mathrm{succ}}^E.\) Finally, if \(\mathcal T\) is trace preserving, its Kraus operators satisfy \(\sum_i K_i^\dagger K_i = I_A.\) Equation \eqref{eq:kraus-stinespring-operator} then gives \(V^\dagger V = \sum_i K_i^\dagger K_i = I_A.\) Thus, \(V\) is a minimal Stinespring isometry. \hfill\ensuremath{\square}

The essential role of minimality in the proof is to guarantee that the Kraus-vectorization matrix \(\mathsf K\) has full column rank. Consequently, \(\mathsf K^\dagger \mathsf K\) is invertible, and the environmental effect is uniquely determined by the dominated Choi operator.

Next, we provide the proof of Proposition~6:

\noindent\textbf{Proof of Proposition 6.} From the definition of the associated cq channel in \mbox{Eq.~\eqref{eq:associated-cq-channel}} of the main text, its normalized Choi operator, with tensor factors ordered as \(CA'B\), takes the block-diagonal form
\begin{equation}
J^{\widetilde{\mathcal E}}_{CA'B}
=\sum_a |a\rangle\langle a|^C\otimes J_a^{\bm{\mathcal E}}.
\label{eq:cq-choi-blocks}
\end{equation}
We use this representation to establish the two inequalities.

We first prove \(P_{\mathrm{blind}}(\bm{\mathcal E}) \le \lambda_{\mathbf{NI}}(\bm{\mathcal E}).\) Consider an arbitrary feasible environmental effect in the optimization defining \(P_{\mathrm{blind}}\), \(0 \preceq M_{\mathrm{succ}}^E \preceq I_E.\) By the fully input-blind condition, there exists a family \(\kappa_a\succeq0\) such that

\begin{equation}
\mathcal S_M(X)
=
\sum_a
|a\rangle\langle a|^C
\otimes
\kappa_a
\operatorname{Tr}X.
\label{eq:extraction-input-blind-submap-proof}
\end{equation}
For any normalized input state, the success probability is \(p_M = \sum_a \operatorname{Tr}\kappa_a.\) By \hyperref[lem:E.2]{Lemma E.2},

\begin{equation}
0
\preceq_{\mathrm{CP}}
\mathcal S_M
\preceq_{\mathrm{CP}}
\widetilde{\mathcal E}.
\label{eq:extraction-cp-domination-proof}
\end{equation}
The normalized Choi operator of a single replacement branch \(X\mapsto\kappa_a\operatorname{Tr}X\) is \(\frac{I_{A'}}{d} \otimes \kappa_a.\) Hence, in the tensor-factor ordering \(CA'B\),

\begin{equation}
J^{\mathcal S_M}
=
\sum_a
|a\rangle\langle a|^C
\otimes
\left(
\frac{I_{A'}}{d}
\otimes
\kappa_a
\right).
\label{eq:input-blind-choi-blocks}
\end{equation}
Using the block-diagonal Choi representation above and the CP-order relation, we obtain

\begin{equation}
J^{\widetilde{\mathcal E}}
-
J^{\mathcal S_M}
=
\sum_a
|a\rangle\langle a|^C
\otimes
\left[
J_a
-
\frac{I_{A'}}{d}
\otimes
\kappa_a
\right]
\succeq0.
\label{eq:extraction-choi-residual-blocks}
\end{equation}
Different classical labels \(a\) correspond to mutually orthogonal direct-sum blocks. Hence, every branch must satisfy \(J_a - \frac{I_{A'}}{d} \otimes \kappa_a \succeq0,\allowbreak\; \forall a,\) or equivalently, \(\frac{I_{A'}}{d} \otimes \kappa_a \preceq J_a,\allowbreak\; \forall a.\) Thus, \(\{\kappa_a\}_a\) is feasible for the primal SDP of \(\lambda_{\mathbf{NI}}(\bm{\mathcal E})\) in \hyperref[prop:1]{Proposition~1}. Therefore, \(p_M = \sum_a \operatorname{Tr}\kappa_a \le \lambda_{\mathbf{NI}}(\bm{\mathcal E}).\) Since this inequality holds for every feasible environmental effect,

\begin{equation}
P_{\mathrm{blind}}(\bm{\mathcal E})
\le
\lambda_{\mathbf{NI}}(\bm{\mathcal E}).
\label{eq:extraction-upper-bound-proof}
\end{equation}

We now prove \(P_{\mathrm{blind}}(\bm{\mathcal E}) \ge \lambda_{\mathbf{NI}}(\bm{\mathcal E}).\) \hyperref[app:A]{Appendix~A} established that the maximum in the primal problem is attained. Hence, there exists an optimal family \(\{\kappa_a^\star\}_a\) satisfying

\begin{equation}
\kappa_a^\star
\succeq0,
\qquad
\frac{I_{A'}}{d}
\otimes
\kappa_a^\star
\preceq
J_a,
\qquad
\forall a,
\label{eq:optimal-primal-extraction-family}
\end{equation}
and \(\sum_a \operatorname{Tr} \kappa_a^\star = \lambda_{\mathbf{NI}}(\bm{\mathcal E}).\) Define the cq map

\begin{equation}
\mathcal S^\star(X)
:=
\sum_a
|a\rangle\langle a|^C
\otimes
\kappa_a^\star
\operatorname{Tr}X.
\label{eq:optimal-replacement-submap-proof}
\end{equation}
Its Choi operator is \(J^{\mathcal S^\star} = \sum_a |a\rangle\langle a|^C \otimes \left( \frac{I_{A'}}{d} \otimes \kappa_a^\star \right) \succeq0,\) where \(\kappa_a^\star\succeq0\) has been used. Hence, \(\mathcal S^\star\) is completely positive.

Equation \eqref{eq:optimal-primal-extraction-family} gives

\begin{equation}
J^{\widetilde{\mathcal E}}
-
J^{\mathcal S^\star}
=
\sum_a
|a\rangle\langle a|^C
\otimes
\left[
J_a
-
\frac{I_{A'}}{d}
\otimes
\kappa_a^\star
\right]
\succeq0.
\label{eq:optimal-extraction-choi-domination}
\end{equation}
Therefore,

\begin{equation}
0
\preceq_{\mathrm{CP}}
\mathcal S^\star
\preceq_{\mathrm{CP}}
\widetilde{\mathcal E}.
\label{eq:optimal-extraction-cp-domination}
\end{equation}
Since \(V\) is the fixed minimal Stinespring realization of \(\widetilde{\mathcal E}\) introduced in \mbox{Eq.~\eqref{eq:minimal-stinespring-realization}} of the main text, \hyperref[lem:E.2]{Lemma E.2} guarantees the existence of an environmental effect \(M_{\mathrm{succ}}^{E,\star}\) that realizes \(\mathcal S^\star\) exactly. This effect satisfies

\begin{equation}
0
\preceq
M_{\mathrm{succ}}^{E,\star}
\preceq
I_E
\label{eq:optimal-environmental-effect}
\end{equation}
such that

\begin{equation}
\mathcal S^\star(X)
=\operatorname{Tr}_E\Bigl[
(I_{CB}\otimes(M_{\mathrm{succ}}^{E,\star})^{1/2})VXV^\dagger
(I_{CB}\otimes(M_{\mathrm{succ}}^{E,\star})^{1/2})\Bigr].
\label{eq:optimal-effect-submap-realization}
\end{equation}
Thus, \(M_{\mathrm{succ}}^{E,\star}\) is a valid environmental success effect in the original physical task. By \mbox{Eq.~\eqref{eq:optimal-replacement-submap-proof}}, the corresponding successful submap is fully input-blind. Its total success probability is

\begin{equation}
p_{M^\star}
=
\sum_a
\operatorname{Tr}\kappa_a^\star
=
\lambda_{\mathbf{NI}}(\bm{\mathcal E}).
\label{eq:optimal-extraction-success-probability}
\end{equation}
Therefore, \(P_{\mathrm{blind}}(\bm{\mathcal E}) \ge \lambda_{\mathbf{NI}}(\bm{\mathcal E}).\) Combining this with \mbox{Eq.~\eqref{eq:extraction-upper-bound-proof}} gives \(P_{\mathrm{blind}}(\bm{\mathcal E}) = \lambda_{\mathbf{NI}}(\bm{\mathcal E}).\) Finally, using \(\lambda_{\mathbf{NI}}(\bm{\mathcal E}) = 1-W_{\mathrm{int}}(\bm{\mathcal E}),\) one obtains

\begin{equation}
P_{\mathrm{blind}}(\bm{\mathcal E})
=
1-W_{\mathrm{int}}(\bm{\mathcal E}).
\label{eq:full-access-extraction-equality-proof}
\end{equation}
At the same time, \mbox{Eqs.~\eqref{eq:optimal-environmental-effect}}--\eqref{eq:optimal-extraction-success-probability} show that the optimal success probability is attained by an actual environmental effect \(M_{\mathrm{succ}}^{E,\star}\), rather than merely being a supremum. This completes the proof of \hyperref[prop:6]{Proposition 6}. \hfill\ensuremath{\square}

\section{Proof of the exact multiplicativity under parallel composition}\label{app:F}
This appendix proves \hyperref[prop:2]{Proposition 2}. Let \(\bm{\mathcal E}^{A_1\to B_1}=\{\mathcal E_a\}_a,\allowbreak\; \bm{\mathcal F}^{A_2\to B_2}=\{\mathcal F_b\}_b,\) and introduce the shorthand notation \(\lambda_{\bm{\mathcal E}}:=\lambda_{\mathbf{NI}}(\bm{\mathcal E}),\allowbreak\; \lambda_{\bm{\mathcal F}}:=\lambda_{\mathbf{NI}}(\bm{\mathcal F}).\) Let \(A_1'\simeq A_1\) and \(A_2'\simeq A_2\) be reference systems, and denote \(d_{A_1}:=\dim A_1,\allowbreak\; d_{A_2}:=\dim A_2.\) To fix the tensor-factor ordering, define the swap unitary \(U_{\mathrm{swap}}: A_1'\otimes B_1\otimes A_2'\otimes B_2 \longrightarrow A_1'\otimes A_2'\otimes B_1\otimes B_2\) by

\begin{equation}
U_{\mathrm{swap}}
\bigl(
|x\rangle_{A_1'}
|y\rangle_{B_1}
|z\rangle_{A_2'}
|w\rangle_{B_2}
\bigr)
=
|x\rangle_{A_1'}
|z\rangle_{A_2'}
|y\rangle_{B_1}
|w\rangle_{B_2}.
\label{eq:tensor-factor-swap}
\end{equation}
By the normalized Choi definition, the product branches satisfy

\begin{equation}
J_{a,b}^{\bm{\mathcal E}\otimes\bm{\mathcal F}}
=
U_{\mathrm{swap}}
\left(
J_a^{\bm{\mathcal E}}\otimes J_b^{\bm{\mathcal F}}
\right)
U_{\mathrm{swap}}^\dagger.
\label{eq:product-instrument-choi}
\end{equation}
We first record two relations used in the proof. For positive operators satisfying \(0\preceq X\preceq X',\allowbreak\; 0\preceq Y\preceq Y',\) one has

\begin{equation}
X\otimes Y
\preceq
X'\otimes Y'.
\label{eq:positive-tensor-order}
\end{equation}
Indeed, \(X'\otimes Y'-X\otimes Y = (X'-X)\otimes Y' + X\otimes(Y'-Y) \succeq0.\) In addition, for arbitrary operators \(X^{A_1'B_1}\) and \(Y^{A_2'B_2}\),

\begin{equation}
\operatorname{Tr}_{A_1'A_2'}
\left[
U_{\mathrm{swap}}(X\otimes Y)U_{\mathrm{swap}}^\dagger
\right]
=
\operatorname{Tr}_{A_1'}X
\otimes
\operatorname{Tr}_{A_2'}Y.
\label{eq:swapped-partial-trace}
\end{equation}
To verify \mbox{Eq.~\eqref{eq:swapped-partial-trace}}, take arbitrary test operators \(H^{B_1}\) and \(K^{B_2}\). Then

\begin{equation}
\begin{aligned}
&
\operatorname{Tr}
\left[
(H\otimes K)
\operatorname{Tr}_{A_1'A_2'}
\left(
U_{\mathrm{swap}}(X\otimes Y)U_{\mathrm{swap}}^\dagger
\right)
\right]
\\
&=
\operatorname{Tr}
\left[
(I_{A_1'A_2'}\otimes H\otimes K)
U_{\mathrm{swap}}(X\otimes Y)U_{\mathrm{swap}}^\dagger
\right]
\\
&=
\operatorname{Tr}
\left[
(I_{A_1'}\otimes H\otimes I_{A_2'}\otimes K)
(X\otimes Y)
\right]
\\
&=
\operatorname{Tr}
\left[
H\operatorname{Tr}_{A_1'}X
\right]
\operatorname{Tr}
\left[
K\operatorname{Tr}_{A_2'}Y
\right]
\\
&=
\operatorname{Tr}
\left[
(H\otimes K)
\left(
\operatorname{Tr}_{A_1'}X
\otimes
\operatorname{Tr}_{A_2'}Y
\right)
\right].
\end{aligned}
\label{eq:swapped-partial-trace-calculation}
\end{equation}
Since product test operators linearly span \(\mathcal B(B_1\otimes B_2)\), \mbox{Eq.~\eqref{eq:swapped-partial-trace}} follows.

We now prove the multiplicative law. The lower bound is established first:

\begin{equation}
\lambda_{\mathbf{NI}}
(\bm{\mathcal E}\otimes\bm{\mathcal F})
\ge
\lambda_{\bm{\mathcal E}}\lambda_{\bm{\mathcal F}}.
\label{eq:multiplicativity-lower-bound}
\end{equation}
By attainment of the primal SDP optimum in \hyperref[app:A]{Appendix~A}, there exist optimal families \(\{\kappa_a\}_a\) and \(\{\eta_b\}_b\) such that \(\kappa_a\succeq0,\allowbreak\; \frac{I_{A_1'}}{d_{A_1}}\otimes\kappa_a \preceq J_a^{\bm{\mathcal E}},\allowbreak\; \sum_a\operatorname{Tr}\kappa_a = \lambda_{\bm{\mathcal E}},\) and \(\eta_b\succeq0,\allowbreak\; \frac{I_{A_2'}}{d_{A_2}}\otimes\eta_b \preceq J_b^{\bm{\mathcal F}},\allowbreak\; \sum_b\operatorname{Tr}\eta_b = \lambda_{\bm{\mathcal F}}.\) By \mbox{Eq.~\eqref{eq:positive-tensor-order}},

\begin{equation}
\left(
\frac{I_{A_1'}}{d_{A_1}}\otimes\kappa_a
\right)
\otimes
\left(
\frac{I_{A_2'}}{d_{A_2}}\otimes\eta_b
\right)
\preceq
J_a^{\bm{\mathcal E}}
\otimes
J_b^{\bm{\mathcal F}}.
\label{eq:primal-product-domination}
\end{equation}
Applying \(U_{\mathrm{swap}}(\cdot)U_{\mathrm{swap}}^\dagger\) to both sides and using \mbox{Eq.~\eqref{eq:product-instrument-choi}} gives

\begin{equation}
\frac{I_{A_1'A_2'}}{d_{A_1}d_{A_2}}
\otimes
(\kappa_a\otimes\eta_b)
\preceq
J_{a,b}^{\bm{\mathcal E}\otimes\bm{\mathcal F}}.
\label{eq:product-primal-feasibility}
\end{equation}
Since \(\kappa_a\otimes\eta_b\succeq0,\) the family \(\{\kappa_a\otimes\eta_b\}_{a,b}\) is feasible for the primal SDP of \(\bm{\mathcal E}\otimes\bm{\mathcal F}\). Its objective value is

\begin{equation}
\sum_{a,b} \operatorname{Tr} (\kappa_a\otimes\eta_b) = \sum_{a,b} (\operatorname{Tr}\kappa_a) (\operatorname{Tr}\eta_b) = \left( \sum_a \operatorname{Tr}\kappa_a \right) \left( \sum_b \operatorname{Tr}\eta_b \right) = \lambda_{\bm{\mathcal E}} \lambda_{\bm{\mathcal F}}.
\label{eq:product-primal-objective}
\end{equation}
Thus, \mbox{Eq.~\eqref{eq:multiplicativity-lower-bound}} follows.

We next prove \(\lambda_{\mathbf{NI}} (\bm{\mathcal E}\otimes\bm{\mathcal F}) \le \lambda_{\bm{\mathcal E}}\lambda_{\bm{\mathcal F}}.\) No assumption concerning attainment of the dual optimum for either individual instrument is required. Fix an arbitrary \(\epsilon>0.\) From the dual representation in \hyperref[app:A]{Appendix~A}, dual-feasible families \(\{\Omega_a\}_a\) and \(\{\Theta_b\}_b\) can be chosen such that \(\Omega_a\succeq0,\allowbreak\; \operatorname{Tr}_{A_1'}\Omega_a \succeq d_{A_1}I_{B_1},\) \(\Theta_b\succeq0,\allowbreak\; \operatorname{Tr}_{A_2'}\Theta_b \succeq d_{A_2}I_{B_2},\) and \(\sum_a \operatorname{Tr} (\Omega_aJ_a^{\bm{\mathcal E}}) \le \lambda_{\bm{\mathcal E}}+\epsilon,\) \(\sum_b \operatorname{Tr} (\Theta_bJ_b^{\bm{\mathcal F}}) \le \lambda_{\bm{\mathcal F}}+\epsilon.\) Define

\begin{equation}
\Xi_{a,b}
:=
U_{\mathrm{swap}}(\Omega_a\otimes\Theta_b)U_{\mathrm{swap}}^\dagger.
\label{eq:product-dual-witness}
\end{equation}
Clearly, \(\Xi_{a,b}\succeq0.\) Using \mbox{Eq.~\eqref{eq:swapped-partial-trace}},

\begin{equation}
\begin{aligned}
\operatorname{Tr}_{A_1'A_2'}\Xi_{a,b}
&=
\operatorname{Tr}_{A_1'}\Omega_a
\otimes
\operatorname{Tr}_{A_2'}\Theta_b
\\
&\succeq
(d_{A_1}I_{B_1})\otimes(d_{A_2}I_{B_2})
\\
&=
d_{A_1}d_{A_2}I_{B_1B_2}.
\end{aligned}
\label{eq:product-dual-feasibility}
\end{equation}
Hence, \(\{\Xi_{a,b}\}_{a,b}\) is a feasible family for the dual SDP of the product instrument. Therefore,

\begin{equation}
\lambda_{\mathbf{NI}}
(\bm{\mathcal E}\otimes\bm{\mathcal F})
\le
\sum_{a,b}
\operatorname{Tr}
\left(
\Xi_{a,b}
J_{a,b}^{\bm{\mathcal E}\otimes\bm{\mathcal F}}
\right).
\label{eq:product-dual-objective-bound}
\end{equation}
Using \mbox{Eqs.~\eqref{eq:product-instrument-choi}} and \eqref{eq:product-dual-witness}, together with invariance of the trace under unitary conjugation,

\begin{equation}
\operatorname{Tr}\left(\Xi_{a,b}J_{a,b}^{\bm{\mathcal E}\otimes\bm{\mathcal F}}\right)=\operatorname{Tr}\left[(\Omega_a\otimes\Theta_b)(J_a^{\bm{\mathcal E}}\otimes J_b^{\bm{\mathcal F}})\right]=\operatorname{Tr}(\Omega_aJ_a^{\bm{\mathcal E}})\operatorname{Tr}(\Theta_bJ_b^{\bm{\mathcal F}}).
\label{eq:product-witness-trace-factorization}
\end{equation}
Consequently,

\begin{equation}
\begin{aligned}
\lambda_{\mathbf{NI}}
(\bm{\mathcal E}\otimes\bm{\mathcal F})
&\le
\left[
\sum_a
\operatorname{Tr}
(\Omega_aJ_a^{\bm{\mathcal E}})
\right]
\left[
\sum_b
\operatorname{Tr}
(\Theta_bJ_b^{\bm{\mathcal F}})
\right]
\\
&\le
(\lambda_{\bm{\mathcal E}}+\epsilon)
(\lambda_{\bm{\mathcal F}}+\epsilon).
\end{aligned}
\label{eq:multiplicativity-epsilon-bound}
\end{equation}
Each term \(\operatorname{Tr} (\Omega_aJ_a^{\bm{\mathcal E}})\) and \(\operatorname{Tr} (\Theta_bJ_b^{\bm{\mathcal F}})\) is nonnegative, since both operators in each trace pairing are positive semidefinite.

Because \mbox{Eq.~\eqref{eq:multiplicativity-epsilon-bound}} holds for every \(\epsilon>0\), taking \(\epsilon\to0\) gives

\begin{equation}
\lambda_{\mathbf{NI}}
(\bm{\mathcal E}\otimes\bm{\mathcal F})
\le
\lambda_{\bm{\mathcal E}}\lambda_{\bm{\mathcal F}}.
\label{eq:multiplicativity-upper-bound}
\end{equation}
Combining \mbox{Eqs.~\eqref{eq:multiplicativity-lower-bound}} and \eqref{eq:multiplicativity-upper-bound} yields

\begin{equation}
\lambda_{\mathbf{NI}}
(\bm{\mathcal E}\otimes\bm{\mathcal F})
=
\lambda_{\mathbf{NI}}(\bm{\mathcal E})
\lambda_{\mathbf{NI}}(\bm{\mathcal F}).
\label{eq:ni-content-multiplicativity-proof}
\end{equation}
Finally, substituting \(\lambda_{\mathbf{NI}}(\bm{\mathcal X}) = 1-W_{\mathrm{int}}(\bm{\mathcal X})\) gives

\begin{equation}
1-W_{\mathrm{int}}
(\bm{\mathcal E}\otimes\bm{\mathcal F})
=
\left[
1-W_{\mathrm{int}}(\bm{\mathcal E})
\right]
\left[
1-W_{\mathrm{int}}(\bm{\mathcal F})
\right],
\label{eq:complementary-weight-product-proof}
\end{equation}
and

\begin{equation}
W_{\mathrm{int}}
(\bm{\mathcal E}\otimes\bm{\mathcal F})
=
W_{\mathrm{int}}(\bm{\mathcal E})
+
W_{\mathrm{int}}(\bm{\mathcal F})
-
W_{\mathrm{int}}(\bm{\mathcal E})
W_{\mathrm{int}}(\bm{\mathcal F}).
\label{eq:weight-parallel-composition-proof}
\end{equation}
This completes the proof of \hyperref[prop:2]{Proposition 2}. \hfill\ensuremath{\square}

\section{Environment-assisted input-blind extraction under restricted environmental access: Framework and proofs}
\label{app:restricted-access}
\subsection{Framework for restricted environmental access}

We consider the associated cq channel \(\widetilde{\mathcal E}^{A\to CB}\) and its fixed minimal Stinespring realization \(V:A\to CBE\), as introduced in \hyperref[interpretation:4]{Interpretation~4}. To describe restricted environmental access, let \(\mathcal A:\mathcal B(E)\to\mathcal B(Z)\) be a CPTP channel, where \(Z\) is the system accessible to the environmental assistant. The assistant has access only to \(Z\), with no additional access to the public register \(C\). Any information about the native outcome \(a\) contained in \(Z\) may be used through the environmental measurement, while the original public output \(CB\) remains unmodified.

The public--accessible extension associated with the fixed realization \(V\) and access channel \(\mathcal A\) is the map \(\mathcal T^{V,\mathcal A}:\mathcal B(A)\to\mathcal B(CBZ)\), defined by

\begin{equation}
\mathcal T^{V,\mathcal A}(X)
:=
(\operatorname{id}_{CB}\otimes\mathcal A)
(VXV^\dagger).
\label{eq:restricted-extension-proof}
\end{equation}

Because \(\mathcal A\) is trace preserving, \(\operatorname{Tr}_Z\mathcal T^{V,\mathcal A}(X)=\widetilde{\mathcal E}(X).\) Thus, discarding the accessible environment recovers the original public channel. No additional operation is applied to the public output \(CB\). The joint extension and a general postselected output need not be block diagonal in \(C\); the cq replacement form is required only for fully input-blind successful submaps. The assistant performs a binary POVM on \(Z\), with success effect \(0\preceq M_{\mathrm{succ}}^Z\preceq I_Z\). The corresponding successful public submap is

\begin{equation}
\begin{aligned}
\mathcal S_{V,\mathcal A,M_{\mathrm{succ}}^Z}(X)
:&=\operatorname{Tr}_Z\Big[
(I_{CB}\otimes\sqrt{M_{\mathrm{succ}}^Z})
\cdot\mathcal T^{V,\mathcal A}(X)
(I_{CB}\otimes\sqrt{M_{\mathrm{succ}}^Z})\Big]\\
&=\operatorname{Tr}_Z\!\left[(I_{CB}\otimes M_{\mathrm{succ}}^Z)
\mathcal T^{V,\mathcal A}(X)\right],
\end{aligned}
\label{eq:restricted-successful-submap-proof}
\end{equation}
the first expression also makes the complete positivity of the successful submap explicit.

The adjoint map \(\mathcal A^\dagger\) is characterized by \(\operatorname{Tr}[Y^Z\mathcal A(X^E)] \allowbreak{}=\allowbreak{}\operatorname{Tr}[\mathcal A^\dagger(Y^Z)X^E].\) As stated in \hyperref[rem:2]{Remark 2}, the success effect on \(Z\) induces the effective environmental effect \(M_{\mathrm{eff}}^E :\allowbreak{}=\allowbreak{} \mathcal A^\dagger(M_{\mathrm{succ}}^Z).\) Since \(\mathcal A^\dagger\) is completely positive and unital,

\begin{equation}
M_{\mathrm{eff}}^E\succeq0,\quad I_E-M_{\mathrm{eff}}^E=\mathcal A^\dagger(I_Z-M_{\mathrm{succ}}^Z)\succeq0.
\label{eq:restricted-effective-effect-validity-proof}
\end{equation}
Thus, \(M_{\mathrm{eff}}^E\) is a valid effect on the original environment.

To establish the equivalence between the restricted measurement and its pulled-back effect, let \(Y^{CB}\) be an arbitrary test operator. Using \mbox{Eq.~\eqref{eq:restricted-successful-submap-proof}} and the adjoint relation, we obtain

\begin{equation}
\begin{aligned}
&\operatorname{Tr}\!\left[
Y^{CB}\mathcal S_{V,\mathcal A,M_{\mathrm{succ}}^Z}(X)
\right]\\
&=\operatorname{Tr}\!\left[
(Y^{CB}\otimes M_{\mathrm{succ}}^Z)
\mathcal T^{V,\mathcal A}(X)
\right]\\
&=\operatorname{Tr}\!\left[
(Y^{CB}\otimes M_{\mathrm{eff}}^E)VXV^\dagger
\right].
\end{aligned}
\label{eq:restricted-trace-pullback}
\end{equation} 

\noindent Since this holds for every \(Y^{CB}\), nondegeneracy of the trace pairing and cyclicity of the partial trace give the equivalent representation

\begin{equation}
\mathcal S_{V,\mathcal A,M_{\mathrm{succ}}^Z}(X)
=\operatorname{Tr}_E\Big[
(I_{CB}\otimes\sqrt{M_{\mathrm{eff}}^E})VXV^\dagger
\cdot(I_{CB}\otimes\sqrt{M_{\mathrm{eff}}^E})\Big].
\label{eq:restricted-effect-pullback-proof}
\end{equation}

Therefore, the strategy under restricted environmental access is fully determined by the effective environmental effect \(M_{\mathrm{eff}}^E=\mathcal A^\dagger(M_{\mathrm{succ}}^Z)\). Comparing \mbox{Eq.~\eqref{eq:restricted-effect-pullback-proof}} with \mbox{Eq.~\eqref{eq:environmental-success-submap}}, the resulting submap is exactly the full-environment successful submap with \(M_{\mathrm{succ}}^E\) replaced by \(M_{\mathrm{eff}}^E\). The input-blind condition is imposed on this submap itself, without any additional processing of \(CB\).

Following the fully input-blind condition introduced in \hyperref[interpretation:4]{Interpretation~4}, a restricted successful submap is admissible if there exist positive operators \(\kappa_a^B\succeq0\) such that \(\mathcal S_{V,\mathcal A,M_{\mathrm{succ}}^Z}(X) \quad=\sum_a |a\rangle\langle a|^C \otimes\kappa_a^B\operatorname{Tr}X,\quad\forall X.\) For every normalized input state, the corresponding success probability is \(p_M=\sum_a\operatorname{Tr}\kappa_a^B\), which is independent of the input. We therefore define the optimal probability of environment-assisted input-blind extraction under restricted environmental access as

\begin{equation}
\begin{aligned}
P_{\mathrm{blind}}(\bm{\mathcal E};V,\mathcal A)
&=\sup_{M_{\mathrm{succ}}^Z,\{\kappa_a^B\}_a}
\sum_a\operatorname{Tr}\kappa_a^B\\
\text{subject to}\quad
&0\preceq M_{\mathrm{succ}}^Z\preceq I_Z,\qquad
\kappa_a^B\succeq0,\quad\forall a,\\
&\mathcal S_{V,\mathcal A,M_{\mathrm{succ}}^Z}(X)
=\sum_a|a\rangle\langle a|^C
\otimes\kappa_a^B\operatorname{Tr}X,\quad\forall X.
\end{aligned}
\label{eq:restricted-optimization-proof}
\end{equation}

By the second expression in \mbox{Eq.~\eqref{eq:restricted-successful-submap-proof}}, the successful submap is linear in \(M_{\mathrm{succ}}^Z\). In finite dimensions, the map equality in \mbox{Eq.~\eqref{eq:restricted-optimization-proof}} can be imposed on a finite operator basis of \(\mathcal B(A)\), giving finitely many linear equality constraints. Together with the positive-semidefinite constraints, this makes \mbox{Eq.~\eqref{eq:restricted-optimization-proof}} a semidefinite program.

The feasible set is nonempty because the zero effect is always admissible. Moreover, the interval of effects \(0\preceq M_{\mathrm{succ}}^Z\preceq I_Z\) is compact in finite dimensions. For any feasible strategy, the operators \(\kappa_a^B\) are uniquely determined by the classical blocks of \(\mathcal S_{V,\mathcal A,M_{\mathrm{succ}}^Z}(I_A/d_A)\) and hence depend continuously on the effect. Since the input-blind constraints are closed, the feasible set is compact and the maximum in \mbox{Eq.~\eqref{eq:restricted-optimization-proof}} is attained.

In particular, when \(Z=E\) and \(\mathcal A=\operatorname{id}_E\), every full-environment effect is accessible and the successful submap and input-blind condition coincide with \mbox{Eqs.~\eqref{eq:environmental-success-submap}} and \eqref{eq:fully-input-blind-condition}. Hence, by \hyperref[prop:6]{Proposition~6}, \(P_{\mathrm{blind}}(\bm{\mathcal E};V,\operatorname{id}_E) =P_{\mathrm{blind}}(\bm{\mathcal E}) =1-W_{\mathrm{int}}(\bm{\mathcal E}).\)

\subsection{Restricted environmental access: universal bound and access data-processing inequality}

The two conclusions stated in \hyperref[rem:2]{Remark 2} are proved separately below. We first establish the universal bound in \hyperref[rem:2]{Remark 2}.

\medskip
\phantomsection\label{prop:G2.1}\noindent\textbf{Proposition G2.1.} For any finite-dimensional quantum instrument \(\bm{\mathcal E}\), any fixed minimal Stinespring realization \(V\), and any CPTP environmental access channel \(\mathcal A^{E\to Z}\), the optimal probability of environment-assisted input-blind extraction under restricted environmental access satisfies \(0\le P_{\mathrm{blind}}(\bm{\mathcal E};V,\mathcal A)\le\lambda_{\mathbf{NI}}(\bm{\mathcal E})=1-W_{\mathrm{int}}(\bm{\mathcal E})\).

\medskip
\noindent\textbf{Proof.} For any admissible success effect \(M_{\mathrm{succ}}^Z\), the complementary effect \(I_Z-M_{\mathrm{succ}}^Z\) induces another completely positive submap. By \mbox{Eq.~\eqref{eq:restricted-successful-submap-proof}} and trace preservation of the access channel, the two submaps satisfy \(\mathcal S_{V,\mathcal A,M_{\mathrm{succ}}^Z} + \mathcal S_{V,\mathcal A,I_Z-M_{\mathrm{succ}}^Z}\) \(= \widetilde{\mathcal E}.\) Consequently,

\begin{equation}
0
\preceq_{\mathrm{CP}}
\mathcal S_{V,\mathcal A,M_{\mathrm{succ}}^Z}
\preceq_{\mathrm{CP}}
\widetilde{\mathcal E}.
\label{eq:restricted-cp-domination-proof}
\end{equation}

Suppose that the successful submap is fully input-blind, with positive output operators \(\{\kappa_a^B\}_a\). Using the normalized Choi representation, \mbox{Eq.~\eqref{eq:restricted-cp-domination-proof}} implies, in the tensor-factor ordering \(CA'B\),

\begin{equation}
J^{\widetilde{\mathcal E}}-J^{\mathcal S_{V,\mathcal A,M_{\mathrm{succ}}^Z}}
=\sum_a|a\rangle\langle a|^C\otimes
\left[J_a^{\bm{\mathcal E}}
-\frac{I_{A'}}{d_A}\otimes\kappa_a^B\right]\succeq0.
\label{eq:restricted-choi-domination-proof}
\end{equation}

\noindent The classical outcome labels correspond to mutually orthogonal blocks. Hence, each block must be positive semidefinite, giving \(\frac{I_{A'}}{d_A}\otimes\kappa_a^B\preceq J_a^{\bm{\mathcal E}}\) for every \(a\). Therefore, \(\{\kappa_a^B\}_a\) is feasible for the primal SDP defining \(\lambda_{\mathbf{NI}}(\bm{\mathcal E})\). Its total success probability consequently satisfies \(p_M = \sum_a\operatorname{Tr}\kappa_a^B \le \lambda_{\mathbf{NI}}(\bm{\mathcal E}).\) Since the inequality holds for every admissible restricted strategy, maximizing over all such strategies and using \hyperref[prop:6]{Proposition 6} yields

\begin{equation}
0\le P_{\mathrm{blind}}(\bm{\mathcal E};V,\mathcal A)
\le\lambda_{\mathbf{NI}}(\bm{\mathcal E})
=1-W_{\mathrm{int}}(\bm{\mathcal E}).
\label{eq:restricted-intrinsic-bound-proof}
\end{equation}

Thus, restricted environmental access cannot yield a larger probability of environment-assisted input-blind extraction of a fully input-blind subprocess than the intrinsic maximal non-interactive content. \hfill\ensuremath{\square}

\medskip
We next consider the effect of further processing the accessible environment and prove the access data-processing inequality.

\medskip
\phantomsection\label{prop:G2.2}\noindent\textbf{Proposition G2.2.} Let \(\mathcal A_1^{E\to Z_1}\) and \(\mathcal A_2^{E\to Z_2}\) be two environmental access channels acting on the same fixed Stinespring realization \(V\). If \(\mathcal A_2=\mathcal Q\circ\mathcal A_1\) for a CPTP channel \(\mathcal Q^{Z_1\to Z_2}\), then \(P_{\mathrm{blind}}(\bm{\mathcal E};V,\mathcal A_2)\le P_{\mathrm{blind}}(\bm{\mathcal E};V,\mathcal A_1)\).

\medskip
\noindent\textbf{Proof.} For a fixed access channel \(\mathcal A^{E\to Z}\), define the set of environmental effects implementable through this access as

\begin{equation}
\mathcal F_{\mathcal A}
:=
\left\{
\mathcal A^\dagger(M^Z):
0\preceq M^Z\preceq I_Z
\right\}.
\label{eq:restricted-accessible-effects-proof}
\end{equation}

Consider two access channels \(\mathcal A_1^{E\to Z_1}\) and \(\mathcal A_2^{E\to Z_2}\) acting on the same fixed realization \(V\), with the same public output \(CB\) and the same fully input-blind condition. Suppose that \(\mathcal A_2 = \mathcal Q\circ\mathcal A_1,\) where \(\mathcal Q^{Z_1\to Z_2}\) is CPTP. For any admissible success effect \(0\preceq M_{\mathrm{succ}}^{Z_2}\preceq I_{Z_2}\), define \(M_{\mathrm{succ}}^{Z_1} := \mathcal Q^\dagger(M_{\mathrm{succ}}^{Z_2}).\) Because \(\mathcal Q^\dagger\) is completely positive and unital,

\begin{equation}
M_{\mathrm{succ}}^{Z_1}\succeq0,\quad I_{Z_1}-M_{\mathrm{succ}}^{Z_1}=\mathcal Q^\dagger(I_{Z_2}-M_{\mathrm{succ}}^{Z_2})\succeq0.
\label{eq:restricted-postprocessed-effect-validity-proof}
\end{equation}

\noindent Hence, \(M_{\mathrm{succ}}^{Z_1}\) is an admissible effect for the first access model. Furthermore, the adjoint of the channel composition satisfies

\begin{equation}
\mathcal A_1^\dagger(M_{\mathrm{succ}}^{Z_1})=\mathcal A_1^\dagger\left[\mathcal Q^\dagger(M_{\mathrm{succ}}^{Z_2})\right]=(\mathcal Q\circ\mathcal A_1)^\dagger(M_{\mathrm{succ}}^{Z_2})=\mathcal A_2^\dagger(M_{\mathrm{succ}}^{Z_2}).
\label{eq:restricted-adjoint-composition-proof}
\end{equation}

\noindent It follows that \(\mathcal F_{\mathcal A_2} \subseteq \mathcal F_{\mathcal A_1}.\) By \mbox{Eq.~\eqref{eq:restricted-effect-pullback-proof}}, two strategies with the same pulled-back environmental effect induce exactly the same successful public submap. Therefore, every fully input-blind strategy feasible under \(\mathcal A_2\) can also be implemented under \(\mathcal A_1\), with the same successful output and success probability. Taking the maximum over all admissible strategies gives

\begin{equation}
P_{\mathrm{blind}}(\bm{\mathcal E};V,\mathcal A_2)
\le
P_{\mathrm{blind}}(\bm{\mathcal E};V,\mathcal A_1).
\label{eq:restricted-data-processing-proof}
\end{equation}

This establishes the access data-processing inequality. Together with \mbox{Eq.~\eqref{eq:restricted-intrinsic-bound-proof}}, it shows that the maximal non-interactive content sets an intrinsic upper bound on environment-assisted input-blind extraction, while the achievable probability depends on the environmental effects accessible through the specified physical implementation and access channel. \hfill\ensuremath{\square}

\end{document}